\documentclass[review,3p,authoryear,times]{elsarticle}
\usepackage{multirow,setspace,times,amssymb,amsmath,graphicx,color,rotating,subfigure,url}

\usepackage{lscape}
\usepackage[colorlinks, citecolor=blue]{hyperref}

\usepackage{caption}
\journal{}

\begin{document}
\UseRawInputEncoding

\begin{frontmatter}

\title{Horse race of weekly idiosyncratic momentum strategies with respect to various risk metrics: Evidence from the Chinese stock market}

\author[DF]{Huai-Long Shi}

\author[BS,RCE,SS]{Wei-Xing Zhou\corref{cor3}}
\ead{wxzhou@ecust.edu.cn}
\cortext[cor3]{Corresponding author. Address: East China University of Science and Technology, 130 Meilong Road, Shanghai 200237, China}

\address[DF]{School of Management Science and Engineering, Nanjing University of Information Science \& Technology, Nanjing 210044, China}
\address[BS]{Department of Finance, School of Business, East China University of Science and Technology, Shanghai 200237, China}
\address[SS]{Department of Mathematics, School of Science, East China University of Science and Technology, Shanghai 200237, China}
\address[RCE]{Research Center for Econophysics, East China University of Science and Technology, Shanghai 200237, China}

\begin{abstract}
This paper focuses on the horse race of weekly idiosyncratic momentum (IMOM) with respect to various idiosyncratic risk metrics. Using the A-share individual stocks in the Chinese market from January 1997 to December 2017, we first evaluate the performance of the weekly momentum and idiosyncratic momentum based on raw returns and idiosyncratic returns, respectively. After that the univariate portfolio analysis is conducted to investigate the return predictability with respect to various idiosyncratic risk metrics. Further, we perform a comparative study on the performance of the IMOM portfolios with respect to various risk metrics. At last, we explore the possible explanations to the IMOM as well as risk-based IMOM portfolios. We find that 1) there is a prevailing contrarian effect and a IMOM effect for the whole sample; 2) a negative relation exists between most of the idiosyncratic risk metrics and the cross-sectional returns, and better performance is found that is linked to idiosyncratic volatility (IVol) and maximum drawdowns (IMDs); 3) additionally, the IVol-based and IMD-based IMOM portfolios exhibit a better explanatory power to the IMOM portfolios with respect to other risk metrics; 4) finally, higher profitability of the IMOM as well as IVol-based and IMD-based IMOM portfolios is found to be related to upside market states, high levels of liquidity and high levels of investor sentiment.
\end{abstract}

\end{frontmatter}

\section{Introduction}
\label{S1:Intro}

Among the market anomalies, the cross-sectional momentum (MOM) effect is one of the most wildly-studied phenomena. In fact, it has been found ubiquitous in almost every stock market and and asset class \citep{Asness-Moskowitz-Pedersen-2013-JF}.\footnote{Literature regarding the momentum (contrarian) effect are abundant, and
\cite{Asness-Moskowitz-Pedersen-2013-JF} conduct the comprehensive research on cross-sectional momentum effect.}
It exerts the big challenge to the Efficient Markets Hypothesis (EMH) \citep{Fama-1970-JF,Fama-1991-JF} since its introduction in the seminal work of \cite{Jegadeesh-Titman-1993-JF}.
Typically, the momentum effect exists in the intermediate term (mainly from 3 months to 12 months), while the contrarian effect exists in the short term (1 month) and long term (more than 12 months).

Despite of its popularity, MOM still suffers dramatic drawdowns during financial crises \citep{Daniel-Moskowitz-2016-JFE}, and it is more sensitive to common factors' performance in the future \citep{Grundy-Martin-2001-RFS,Blitz-Huij-Martens-2011-JEF,Blitz-Hanauer-Vidojevic-2020-IRFA}.
Accordingly, scholars have been attempting to explore more refined versions of momentum that maintains the profitability while reduces downside risks.

In light of classical pricing models, the premium is compensated for systematic risk only. In other words, unsystematic risks or idiosyncratic risks, would not be related to asset returns \citep{Sharpe-1964-JF,Lintner-1965-JF}. However, a surge of literature provide the evidence suggesting the linkage between the idiosyncratic risk and asset performance
\citep{Ang-Hodrick-Xing-Zhang-2006-JF,Ang-Hodrick-Xing-Zhang-2009-JFE,Fu-2009-JFE,Stambaugh-Yu-Yuan-2015-JF,Gu-Kang-Xu-2018-JBF}, although some of their findings are mixed. The relation between idiosyncratic risk and expected returns, if any, is expected to be positive in classical paradigm, while a growing body of literature have recognized a negative relation between them \citep{Stambaugh-Yu-Yuan-2015-JF,Atilgan-Bali-Demirtas-Gunaydin-2020-JFE}.
And this may provide new thought or solution to the enhancement of profitability related to the combination of different firm-specific characteristics.

Therefore, the residual or idiosyncratic momentum (IMOM) comes out,\footnote{Note that the abbreviation IMOM in our work merely refers to as the momentum portfolios constructed by sorting on idiosyncratic return that is not adjusted by the risk, i.e., pure IMOM.}
which has been receiving scholars' attention in recent years
\citep{Gutierrez-Prinsky-2007-JFinM,Blitz-Huij-Martens-2011-JEF,Blitz-Hanauer-Vidojevic-2020-IRFA,Chang-Ko-Nakano-Rhee-2018-JEF}. IMOM portfolios are constructed by sorting on idiosyncratic return that is usually adjusted by the idiosyncratic volatility, and it is documented to give rise to more robust and better performance compared to the MOM, especially in the US stock market \citep{Gutierrez-Prinsky-2007-JFinM,Blitz-Hanauer-Vidojevic-2020-IRFA}.
On the other hand, although more and more idiosyncratic risk metrics have been developed, few attentions have been paid to how the various risk-adjustment methods affect the momentum performance.

Our work contributes to the literature on momentum effects from the following aspects. First, our work fills in the gap of the research on the short-term idiosyncratic momentum for the Chinese stock market.
Compared with other mature financial markets, the Chinese stock market is relatively young. Some phenomena characterize the Chinese market, including the less transparent information environment
at the market and firm levels and a larger proportion of irrational individual investors, etc.
These further result in remarkably different performance associated with price trend anomalies \citep{Shi-Jiang-Zhou-2015-PLoS1}.
In this regard, examining anomalies over different periods is meaningful for both academic scholars as well as practitioners.
For the MOM, although majority of researches conduct the studies on the monthly basis, there is a few studies paying attention on the short-term MOM, but in the earlier days \citep{Kang-Liu-Ni-2002-PBFJ, Pan-Tang-Xu-2013-PBFJ}.
By comparison, scarce attention has been paid to the IMOM, especially in emerging markets
\citep{Lu-Lu-2018-SSRN,Lin-2019-FRL}.
And to the best of our knowledge, the short-term idiosyncratic momentum has remained unexplored in the Chinese stock market yet.
\textcolor{black}{
In this regard, our paper contributes to the literature by providing the evidence regarding the performance of the short-term IMOM at weekly basis, which has been usually ignored in previous studies.
}

Second and more importantly, we attempt to conduct a comparative study on how different risk-adjustment methods are related to the idiosyncratic momentum in the Chinese stock market.
In fact, Chinese investors mainly invest in real estate and equities. And according to the data published by China Securities Depository and Clearing Corporation Limited (CSDC), each of the retail investors holds only 1.4 stocks as of June 2018, which are obviously under-diversified. As such, the aggregated effect of under-diversified portfolios would result in more idiosyncratic risks in the market \citep{Fu-2009-JFE}. Hence, we expect a better performance of the IMOM strategies with respect to various risk-adjustment methods in China.

Conventionally, idiosyncratic risk always refers to as idiosyncratic volatility in most of the literature \citep{Ang-Hodrick-Xing-Zhang-2006-JF,Ang-Hodrick-Xing-Zhang-2009-JFE,Blitz-Huij-Martens-2011-JEF,Blitz-Hanauer-Vidojevic-2020-IRFA}.
However, more idiosyncratic risk metrics have been developed to capture the idiosyncratic risk more accurately.
We attempt to explore if these risk metrics can predict cross-sectional returns and if they can lead to better performance of risk-adjusted idiosyncratic momentum.

We take several risk metrics associated with idiosyncratic return into consideration. Among them, idiosyncratic volatility (IVol, hereafter) is one of the most frequently used risk metrics in the literature, which has been shown to function well in predicting the cross-sectional stock returns in the US market. A quite feasible theoretical explanation is proposed by \cite{Stambaugh-Yu-Yuan-2015-JF}, documenting that the IVol viewed as arbitrage risk, accompanied with arbitrage asymmetry, leads to the negative relation between IVol and expected return. The Chinese market still experiences costly short-selling, despite of the launch of margin trading on March 31, 2010.
Thus, the negative relation between the IVol and the cross-sectional stock return is expected to be more pronounced in the Chinese market.
Besides the IVol, we also consider the skewness and kurtosis associated with idiosyncratic returns, i.e., ISkew and IKurt. Some of the investors possess lottery-like demands, resulting in the positive skewness of historical stock returns but with very small probability of persistent outperformance in the future \citep{Boyer-Mitton-Vorkink-2009-RFS,Bali-Cakici-Whitelaw-2011-JFE}. Additionally, a few studies also find the negative relation between IKurt and cross-sectional return, while a solid theoretical basis is still needed.

We also consider several tail risk metrics with respect to the idiosyncratic return. As argued in \cite{Pontiff-JAE-2006},  the risk with respect to idiosyncratic return, including its tail risks, could be viewed as part of ``holding costs'', thus contributing to the arbitrage cost. Thereby following the explanation from \cite{Stambaugh-Yu-Yuan-2015-JF}, there also would be the negative relation between the idiosyncratic tail risk and cross-sectional return, due to  the arbitrage asymmetry.
From another perspective, the arguments associated with tail risk anomalies by \cite{Atilgan-Bali-Demirtas-Gunaydin-2020-JFE} also shed light on the idiosyncratic tail risk metrics adopted in our work:
Investors exhibit stronger underreaction to bad news, which provides a negative return drift in the future, and this further leads to the negative relation between tail risk and stock returns.
Correspondingly, stronger underreaction to \textit{idiosyncratically} bad news is expected to result in negatively abnormal return drifts, thus seemingly causing the assets more overpriced. This also implies the negative relation between the idiosyncratic tail risk and cross-sectional stock return.

Following \cite{Atilgan-Bali-Demirtas-Gunaydin-2020-JFE}, we consider the idiosyncratic VaR and idiosyncratic expected shortfall as the proxies of idiosyncratic tail risks (IVaR and IES, hereafter) in our work. Additionally, it is straightforward to consider the idiosyncratic maximum drawdown (IMD) as one of the idiosyncratic tail risk metrics for the Chinese stock market.
The price limit trading rules implemented currently in China became effective since December 1996, requiring that the maximum daily price fluctuation in terms of the last closing price is $\pm10\%$ for common stocks.
Apparently, it slows down investors' reactions to the bad (good) news associated with firms via limiting the magnitude of price slumping (climbing). Therefore, as to some extreme events, the ``slow-down'' effect would be more pronounced in this situation and the price reflection is expected to last for a much longer period of time.
Additionally and more importantly, price limits also serve as the alleviator of the pessimism (enthusiasm) of investors, incurred by massive selling (buying). Specifically, after hitting the limits caused by bad or good news, the price would stop reflecting the information until the upcoming news completely offset the influence posed by the preceding news.
Therefore, the suspension of price reflection essentially ``shortens'' the reaction time of last news, which contributes to the ``speed-up'' effect in this situation.
Hence, in the first situation, tail risks might not be perfectly captured by the expected shortfall or VaR that functions well in the US market \citep{Atilgan-Bali-Demirtas-Gunaydin-2020-JFE}. Instead, the maximum drawdown does a better job by virtue of capturing the risk of long-term price adjustment.
And, it would be opposite in the second situation.
Therefore, the idiosyncratic tail risk anomaly would depend on which metric is adopted. Thereby, we could examine which effect is dominant via comparing the results regarding the IMD and the IVaR (or the IES): If the ``slow-down" (``speed-up") effect is dominant, higher return magnitudes would be mainly achieved by the short side of portfolio with higher (lower) idiosyncratic tail risks.

The rest of this paper proceeds as follows. Section \ref{S1:DM} describes data and methodology employed in the study. Section \ref{S1:Results} presents the empirical results and conducts further examinations. Section \ref{S1:Conclusion} concludes.

\section{Data and methodology}
\label{S1:DM}

\subsection{Data}

Daily data for the Chinese A-share common stocks are retrieved from the China Stock Market \& Accounting Research (CSMAR) database, covering the period from December 1997 to December 2017. There are a total of 3,726 common stocks listed on Shanghai Stock Exchange (SHSE) and Shenzhen Stock Exchange (SZSE) as of December 2017. Our data set mainly contains adjusted closing prices and returns (for split and dividend). We also retrieve the data of Fama-French 3 and 5 factors as well as risk-free rates from the CSMAR database.

We preprocess the data and exclude the prices and returns of the first months for individual stocks on the ground that abnormal fluctuations usually exist in the IPO month in the Chinese stock market. We also exclude the stock returns after a continuous period of suspension,\footnote{The period is set as ten trading days or two trading weeks in our work.} which is typically related to the abnormal returns caused by major events like M\&A.

\subsection{MOM test}

The classical $J-K$ portfolios based on calendar-time methods are constructed \citep{Jegadeesh-Titman-1993-JF,Jegadeesh-Titman-2001-JF,Gutierrez-Kelley-2008-JF}, which are argued to be more conducive to convey the information regarding the factors \citep{Fama-1998-JFE}. The parameters $J$ and $K$ represent the lengths of the estimation period and the holding period, respectively.
Specifically, we conduct the univariate portfolio analysis according to the following steps:

STEP 1: Rank and divide individual stocks into ten decile groups at the beginning of each week $t$, according to some criteria: For the MOM test, the ranking criteria is cumulatively raw return over the past $J$ weeks from $t-J-1$ to $t-2$, as described below:\footnote{As argued by \cite{Lehmann-1990-QJE}, \cite{Ball-Kothari-Wasley-1995-JPM} and \cite{Conrad-Gultekin-Kaul-1997-JBES}, the bid-ask spread, the non-synchronous trading as well as the lack of liquidity would enlarge the momentum effects. To avoid biased results, we follow the common approach to skip one week between the estimation period $J$ and the holding period $K$.}
\begin{equation}
\textrm{MOM}_{i,t}=\prod_{j=2}^{J+1}(1+r_{i,t-j})-1,
\label{Eq:MOM}
\end{equation}
The winner refers to the decile group with the best performance, while the loser is the group with the worst performance.

STEP 2:  Construct the zero-cost arbitrage portfolio via longing the stocks from the winner portfolio and shorting the stocks from the loser portfolio for each week during the whole sample period.

STEP 3: Evaluate the average performance of the loser, winner and zero-cost arbitrage portfolios over the subsequent $K$ weeks.\footnote{The calendar-time method indicates that, when calculating the average weekly returns for a specific $J-K$ portfolio, the portfolios at the week $t$ should contain the portfolio formed at week $t$ and the portfolios constructed in the former $K-1$ weeks, and the average return at week $t$ is the equal-weighted return of all available portfolios.}
Weekly return can be obtained using the Friday-to-Friday method.\footnote{We compare the results for the Wednesday-to-Wednesday and Friday-to-Friday methods and obtain very similar results. In fact, even more pronounced results are observed based on the Wednesday-to-Wednesday method. It is somewhat counter-intuitive because abnormal fluctuations is expected because of the frequently-issued news in Fridays in China. As a result, Friday-to-Friday method is criticized for overestimating the weekly performance \citep{Pan-Tang-Xu-2013-PBFJ}, which seemingly has not been validated by previous research. We employ the Friday-to-Friday method in this paper.} Considering some short vacations or Lunar holidays in China that usually range from three days to one week, we skip the weeks that contain two trading days or less.
Lastly, \cite{Newey-West-1987-Em}'s $t$-statistics adjusted for autocorrelation and heteroscedasticity is applied to average weekly return series for each $J-K$ portfolio.  For the MOM test,  it would be a momentum (contrarian) effect when the winner portfolios have better (worse) performance than the loser portfolios.

\subsection{IMOM test}

As for the IMOM test, we first resort to popular pricing models to retrieve the idiosyncratic return series for each individual stock. Specifically, at the beginning of each week $t$, we retrieve the daily excess returns of individual stocks over the past $J$ trading weeks from $t-2$ to $t-J-1$, and we conduct the Fama-French 5-factor (FF5F, hereafter) regressions to obtain the idiosyncratic return series $\hat{\varepsilon}_{t}^{\rm{FF5F}}$, as described by
\begin{equation}
ER_{t}=\alpha+\beta_{\rm{MKT}}R_{{\rm{MKT}},t}+\beta_{\rm{SMB}}R_{{\rm{SMB}},t}+\beta_{\rm{HML}}R_{{\rm{HML}},t}+\beta_{\rm{RMW}}R_{{\rm{RMW}},t}+\beta_{\rm{CMA}}R_{{\rm{CMA}},t}+\varepsilon_{t}^{\rm{FF5F}}, ~~~\varepsilon_{t}^{FF5F} \thicksim N(0,\sigma^{2}_{\varepsilon})
\label{Eq:FF5F}
\end{equation}
and
\begin{equation}
\hat{\varepsilon}_{t}^{\rm{FF5F}}=ER_{t}-\hat{\alpha}-\hat{\beta}_{\rm{MKT}}R_{{\rm{MKT}},t}-\hat{\beta}_{\rm{SMB}}R_{{\rm{SMB}},t}-\hat{\beta}_{\rm{HML}}R_{{\rm{HML}},t}-\hat{\beta}_{\rm{RMW}}R_{{\rm{RMW}},t}-\hat{\beta}_{\rm{CMA}}R_{{\rm{CMA}},t},
\label{Eq:FF5F:res}
\end{equation}
where $ER_{t}$ and $R_{\rm{MKT}}$ denote daily stock returns and value-weighted market portfolio returns, respectively, in excess of risk-free rate; $R_{\rm{SMB}}$, $R_{\rm{HML}}$, $R_{\rm{RMW}}$,  and $R_{\rm{CMA}}$ constructed following \cite{Fama-French-2015-JFE}, represent size, value, profitability and investment factors, respectively.
\footnote{It is noted that our aim is to retrieve the idiosyncratic component when controlling for the systematic influence as much as possible. Hence, Fama-French 5-factor model serves a more favorable purpose, since it allows for more factors capturing systematic components.
In fact, existing evidence suggests the similar performance regarding the IMOM based on various pricing models \citep{Lin-2020-EFM,Hovmark-2020-SSRN}.
We additionally conduct the study based on the Fama-French 3-factor model \citep{Fama-French-1993-JFE} (hereafter, FF3F) for the robustness. In comparison, the results based on the FF3F exhibit more pronounced findings, which is consistent with those of the FF5F.}
The IMOM test shares the same test procedures as that of the MOM test, except that in STEP 1, the ranking criteria changes to the cumulatively idiosyncratic return over the past $J$ weeks from $t-J-1$ to $t-2$, as described below
\begin{equation}
\centering
\textrm{IMOM}_{i,t}=\prod_{j=2}^{J+1}(1+\hat{\varepsilon}_{i,t-j}^{\rm{FF5F}})-1.
\label{Eq:IMOM}
\end{equation}

\subsection{Risk metrics}

Risk-adjusted IMOM is argued to provide more robust and favorable performance, which enlightens us to explore optimal risk-adjusted IMOM based upon various risk metrics, as presented in \autoref{TB:risks}.
We first attempt to examine the predictability of stock returns with respect to each risk metric.
Specifically, each of the risk metrics at time $t$ is calculated according to the idiosyncratic returns, $\hat{\varepsilon}_{t}^{\rm{FF5F}}$, over past 130 trading days or 26 trading weeks.
Further, we conduct the univariate portfolio analysis. The ranking criteria would be based on each of the idiosyncratic risk metrics and the winner (loser) corresponds to the stocks with the lowest (highest) risk.

\begin{table}
\small
\centering
  \caption{Idiosyncratic risk metrics adopted in study.}
  \vspace{-3mm}
  \begin{tabular}{lll}
    \hline\hline
    Metric  & Description\\
    \hline
    Idiosyncratic Volatility (IVol) &  Standard deviation of idiosyncratic return\\
    Idiosyncratic Skewness (ISkew)  & Skewness of idiosyncratic return\\
    Idiosyncratic Kurtosis (IKurt) & Kurtosis of idiosyncratic return\\
    Idiosyncratic Maximum Drawdown (IMD) &  Maximum drawdown of idiosyncratic return\\
    Idiosyncratic Expected Shortfall (IES1, IES5)
    &  Expected shortfall of idiosyncratic return at the 1\% and 5\% confidence level\\
    Idiosyncratic VaR (IVaR1, IVaR5) & VaR of idiosyncratic return at the 1\% and 5\% confidence level\\
    \hline\hline
  \end{tabular}
  \label{TB:risks}
\end{table}

\subsection{Risk metrics and IMOM}

As for the combination of idiosyncratic return and risk metric, firstly, we follow the literature on risk-adjusted IMOM
\citep{Gutierrez-Prinsky-2007-JFinM,Blitz-Huij-Martens-2011-JEF,Blitz-Hanauer-Vidojevic-2020-IRFA,Chang-Ko-Nakano-Rhee-2018-JEF}
and construct ranking factor as follows
\begin{equation}
\textrm{Risk-adjusted~IMOM}_{i,t}=\frac{\prod_{j=2}^{J+1}\left(1+\hat{\varepsilon}_{i,t-j}^{\rm{FF5F}}\right)-1}{{\rm{risk~metric~value}}_{i,t}},
\label{Eq:Risk-IMOM}
\end{equation}
which could be viewed as a directly adjusted procedure.
Then, we conduct the univariate portfolio analysis following STEP 1 to STEP 3.

Secondly, we adopt the indirectly adjusted procedure.
Specifically, we conduct the bivariate portfolios analysis and form the intersected groups by double sorting. At first, the individual stocks are divided into ten decile groups according to the idiosyncratic return and specific risk metric, respectively. The intersected winner (loser) group is formed through picking out the stocks from groups with the highest (lowest) return and the lowest (highest) risk. The rest of the test process is also the same with the univariate portfolio analysis.
This method further results in relatively fewer stocks in the portfolios, which is more practical in the investment.

For the convenience, we use the following abbreviations to denote the momentum portfolios constructed through the double sorting on idiosyncratic return and each specific risk metric: IVol-IMOM, ISkew-IMOM, IKurt-IMOM, IMD-IMOM, IES-IMOM, and IVaR-IMOM.

\section{Empirical results}
\label{S1:Results}

\subsection{Univariate portfolio analysis based on raw returns}

We first evaluate the performance of weekly momentum expressed in Eq.~(\ref{Eq:MOM}) based on cumulatively raw returns. We adopt various pairs of $J$ and $K$, where $J \in \{2,3,4,8,13,26,52\}$ weeks whilst $K \in \{1,2,3,4,8,13,26,52\}$ weeks. Given the immature price limits mechanism implemented before 1997 and the frequently abnormal fluctuations happened during the same period, our data period is from January 1997 to December 2017.

In view of the prevailing contrarian effect in the Chinese market
\citep{Shi-Jiang-Zhou-2015-PLoS1,Shi-Zhou-2017a-PA,Shi-Zhou-2017b-PA}, the time series of the MOM returns are multiplied by -1 so that the Sharpe ratios and maximum drawdowns could be further obtained, and the positive results indicate the higher (lower) performance of the loser (winner) portfolios. This is confirmed by the significant returns ranging from 5.2\% to 13\% annually in Panel A of \autoref{TB:MOM:basic} for each $J-K$ portfolio, which is also in accordance with previous studies \citep{Kang-Liu-Ni-2002-PBFJ,Zhu-Wu-Wang-2003-cnJWE,Pan-Tang-Xu-2013-PBFJ}. Except that the significant momentum return is achieved only when $J=2$ and $K =1$, we can readily observe the prevailingly and significantly contrarian effects when $J$ and $K$ are less than 26 weeks. The FF5F-$\alpha$ values in Panel B remain statistically significant as well, and most of their magnitudes are larger than those of the raw returns. The maximum drawdowns also vary in a wide range.

\begin{table}[!ht]
\small
\caption{Basic results for momentum portfolios. This table presents basic results of momentum portfolios for various $J-K$ portfolios. The average weekly returns, FF5F-$\alpha$ values, annualized Sharpe ratios and maximum drawdowns are presented in Panel A, B, C and D, respectively. The whole sample period is January 1997 to December 2017. \cite{Newey-West-1987-Em}'s $t$-statistics are obtained and the superscripts * and ** denote the significance at 5\% and 1\% levels, respectively.}
\centering
\vspace{-3mm}
   \begin{tabular}{ccccccccc}
   \hline\hline
     $J$ & $K=1$ & $2$ & $3$ & $4$ &  $8$ & $13$ & $26$ & $52$ \\
   \hline
 \multicolumn{9}{l}{Panel A: Raw return} \\
   2 &  {\textbf{ -0.0020$^{*~}$}} & { -0.0006$^{~~}$}  & { 0.0006$^{~~}$} & {\textbf{ 0.0013$^{*~}$}} & {\textbf{ 0.0011$^{**}$}} & {\textbf{ 0.0010$^{**}$}} & { 0.0002$^{~~}$} & { 0.0000$^{~~}$}  \\
   3 &  { -0.0002$^{~~}$} & { 0.0010$^{~~}$}  & {\textbf{ 0.0019$^{**}$}} & {\textbf{ 0.0022$^{**}$}} & {\textbf{ 0.0016$^{**}$}} & {\textbf{ 0.0013$^{**}$}} & { 0.0003$^{~~}$} & { 0.0001$^{~~}$}  \\
   4 &  { 0.0007$^{~~}$} & {\textbf{ 0.0019$^{*~}$}}  & {\textbf{ 0.0023$^{**}$}} & {\textbf{ 0.0025$^{**}$}} & {\textbf{ 0.0017$^{**}$}} & {\textbf{ 0.0013$^{**}$}} & { 0.0003$^{~~}$} & { 0.0001$^{~~}$}  \\
   8 &  { 0.0018$^{~~}$} & {\textbf{ 0.0019$^{*~}$}}  & {\textbf{ 0.0021$^{*~}$}} & {\textbf{ 0.0022$^{**}$}} & {\textbf{ 0.0018$^{**}$}} & {\textbf{ 0.0012$^{*~}$}} & { 0.0001$^{~~}$} & { 0.0001$^{~~}$}  \\
   13 &  {\textbf{ 0.0022$^{*~}$}} & {\textbf{ 0.0024$^{*~}$}}  & {\textbf{ 0.0023$^{*~}$}} & {\textbf{ 0.0022$^{*~}$}} & { 0.0014$^{~~}$} & { 0.0009$^{~~}$} & { -0.0002$^{~~}$} & { 0.0001$^{~~}$}  \\
   26 &  { 0.0011$^{~~}$} & { 0.0009$^{~~}$}  & { 0.0009$^{~~}$} & { 0.0008$^{~~}$} & { 0.0004$^{~~}$} & { 0.0000$^{~~}$} & { -0.0004$^{~~}$} & { 0.0003$^{~~}$}  \\
   52 &  { 0.0005$^{~~}$} & { 0.0005$^{~~}$}  & { 0.0005$^{~~}$} & { 0.0006$^{~~}$} & { 0.0004$^{~~}$} & { 0.0004$^{~~}$} & { 0.0003$^{~~}$} & { 0.0006$^{~~}$}
       \smallskip\\
 \multicolumn{9}{l}{Panel B: FF5F-$\alpha$} \\
   2 &  {\textbf{ -0.0018$^{*~}$}} & { -0.0004$^{~~}$}  & { 0.0008$^{~~}$} & {\textbf{ 0.0015$^{**}$}} & {\textbf{ 0.0013$^{**}$}} & {\textbf{ 0.0011$^{**}$}} & { 0.0003$^{~~}$} & { 0.0001$^{~~}$}  \\
   3 &  { 0.0001$^{~~}$} & { 0.0013$^{~~}$}  & {\textbf{ 0.0021$^{**}$}} & {\textbf{ 0.0025$^{**}$}} & {\textbf{ 0.0019$^{**}$}} & {\textbf{ 0.0015$^{**}$}} & { 0.0004$^{~~}$} & { 0.0002$^{~~}$}  \\
   4 &  { 0.0010$^{~~}$} & {\textbf{ 0.0022$^{**}$}}  & {\textbf{ 0.0026$^{**}$}} & {\textbf{ 0.0028$^{**}$}} & {\textbf{ 0.0020$^{**}$}} & {\textbf{ 0.0015$^{**}$}} & { 0.0004$^{~~}$} & { 0.0002$^{~~}$}  \\
   8 &  {\textbf{ 0.0024$^{**}$}} & {\textbf{ 0.0024$^{**}$}}  & {\textbf{ 0.0025$^{**}$}} & {\textbf{ 0.0026$^{**}$}} & {\textbf{ 0.0022$^{**}$}} & {\textbf{ 0.0014$^{*~}$}} & { 0.0002$^{~~}$} & { 0.0001$^{~~}$}  \\
   13 &  {\textbf{ 0.0029$^{**}$}} & {\textbf{ 0.0030$^{**}$}}  & {\textbf{ 0.0028$^{**}$}} & {\textbf{ 0.0026$^{**}$}} & {\textbf{ 0.0018$^{*~}$}} & { 0.0011$^{~~}$} & { -0.0000$^{~~}$} & { 0.0001$^{~~}$}  \\
   26 &  { 0.0018$^{~~}$} & { 0.0016$^{~~}$}  & { 0.0015$^{~~}$} & { 0.0014$^{~~}$} & { 0.0010$^{~~}$} & { 0.0004$^{~~}$} & { -0.0001$^{~~}$} & { 0.0005$^{~~}$}  \\
   52 &  { 0.0007$^{~~}$} & { 0.0007$^{~~}$}  & { 0.0008$^{~~}$} & { 0.0008$^{~~}$} & { 0.0007$^{~~}$} & { 0.0005$^{~~}$} & { 0.0004$^{~~}$} & { 0.0006$^{~~}$}
     \smallskip\\
 \multicolumn{9}{l}{Panel C: Annualized Sharpe ratio} \\
   2 &  { -0.6013$^{~~}$} & { -0.2071$^{~~}$}  & { 0.2443$^{~~}$} & { 0.5845$^{~~}$} & { 0.6460$^{~~}$} & { 0.7414$^{~~}$} & { 0.2159$^{~~}$} & { 0.0322$^{~~}$}  \\
   3 &  { -0.0489$^{~~}$} & { 0.3226$^{~~}$}  & { 0.6752$^{~~}$} & { 0.8731$^{~~}$} & { 0.8076$^{~~}$} & { 0.8150$^{~~}$} & { 0.3073$^{~~}$} & { 0.1289$^{~~}$}  \\
   4 &  { 0.2085$^{~~}$} & { 0.5792$^{~~}$}  & { 0.7842$^{~~}$} & { 0.8689$^{~~}$} & { 0.7801$^{~~}$} & { 0.7431$^{~~}$} & { 0.2232$^{~~}$} & { 0.1075$^{~~}$}  \\
   8 &  { 0.4641$^{~~}$} & { 0.5387$^{~~}$}  & { 0.6060$^{~~}$} & { 0.6890$^{~~}$} & { 0.6478$^{~~}$} & { 0.5401$^{~~}$} & { 0.0549$^{~~}$} & { 0.0459$^{~~}$}  \\
   13 &  { 0.5716$^{~~}$} & { 0.6399$^{~~}$}  & { 0.6385$^{~~}$} & { 0.6201$^{~~}$} & { 0.4643$^{~~}$} & { 0.3488$^{~~}$} & { -0.0977$^{~~}$} & { 0.0362$^{~~}$}  \\
   26 &  { 0.2859$^{~~}$} & { 0.2478$^{~~}$}  & { 0.2387$^{~~}$} & { 0.2286$^{~~}$} & { 0.1316$^{~~}$} & { 0.0123$^{~~}$} & { -0.1498$^{~~}$} & { 0.1603$^{~~}$}  \\
   52 &  { 0.1142$^{~~}$} & { 0.1175$^{~~}$}  & { 0.1397$^{~~}$} & { 0.1526$^{~~}$} & { 0.1213$^{~~}$} & { 0.1266$^{~~}$} & { 0.1000$^{~~}$} & { 0.2724$^{~~}$}
       \smallskip\\
 \multicolumn{9}{l}{Panel D: Maxium drawdown} \\
   2 &  { 0.9606$^{~~}$} & { 0.8287$^{~~}$}  & { 0.5401$^{~~}$} & { 0.3940$^{~~}$} & { 0.1301$^{~~}$} & { 0.1202$^{~~}$} & { 0.1910$^{~~}$} & { 0.1498$^{~~}$}  \\
   3 &  { 0.8338$^{~~}$} & { 0.6490$^{~~}$}  & { 0.4823$^{~~}$} & { 0.3554$^{~~}$} & { 0.1513$^{~~}$} & { 0.1317$^{~~}$} & { 0.2119$^{~~}$} & { 0.1839$^{~~}$}  \\
   4 &  { 0.7382$^{~~}$} & { 0.5887$^{~~}$}  & { 0.4522$^{~~}$} & { 0.3255$^{~~}$} & { 0.1653$^{~~}$} & { 0.1595$^{~~}$} & { 0.2702$^{~~}$} & { 0.2133$^{~~}$}  \\
   8 &  { 0.4425$^{~~}$} & { 0.3387$^{~~}$}  & { 0.2775$^{~~}$} & { 0.2573$^{~~}$} & { 0.2520$^{~~}$} & { 0.2387$^{~~}$} & { 0.4223$^{~~}$} & { 0.3221$^{~~}$}  \\
   13 &  { 0.4089$^{~~}$} & { 0.3838$^{~~}$}  & { 0.3903$^{~~}$} & { 0.3822$^{~~}$} & { 0.3412$^{~~}$} & { 0.3813$^{~~}$} & { 0.5429$^{~~}$} & { 0.3853$^{~~}$}  \\
   26 &  { 0.7833$^{~~}$} & { 0.7502$^{~~}$}  & { 0.7197$^{~~}$} & { 0.6967$^{~~}$} & { 0.6823$^{~~}$} & { 0.6757$^{~~}$} & { 0.6526$^{~~}$} & { 0.4770$^{~~}$}  \\
   52 &  { 0.7998$^{~~}$} & { 0.7519$^{~~}$}  & { 0.7205$^{~~}$} & { 0.6907$^{~~}$} & { 0.6596$^{~~}$} & { 0.6460$^{~~}$} & { 0.6229$^{~~}$} & { 0.5174$^{~~}$}  \\
   \hline\hline
   \end{tabular}
   \label{TB:MOM:basic}
\end{table}

We additionally evaluate the MOM performance in subperiods as reported in \autoref{TB:MOM:subperiod},
which are consistent with existing literature \citep{Kang-Liu-Ni-2002-PBFJ,Zhu-Wu-Wang-2003-cnJWE,Pan-Tang-Xu-2013-PBFJ}: Weekly momentum mainly exist for the period from 1993 to 2000, and gradually fade away after 2000 \citep{Zhu-Wu-Wang-2003-cnJWE}, as indicated by our results in Panel B. In comparison, weekly contrarian effects are much more pronounced for the period from 2000 to 2017 (in Panel C and D) \citep{Pan-Tang-Xu-2013-PBFJ}.

\subsection{Univariate portfolio analysis based on idiosyncratic return}

The results of the univariate portfolios analysis based on the cumulatively idiosyncratic returns are reported in \autoref{TB:IMOM:detail:F5}. Compared to the results based on the cumulatively raw returns, \autoref{TB:IMOM:detail:F5} tells a completely different story. We observe that all the portfolios achieve statistically positive profits. Specifically, the annual returns range from 3.64\% ($0.0007 \times 52 \approx 3.64\%$ when $J= 2$ weeks and $ K= 52$ weeks) to 19.76\% ($0.0038 \times 52 \approx 19.76\%$ when $J=4$ weeks and $K = 1$ week).

\begin{table}[!ht]
\small
\caption{Basic results for idiosyncratic momentum portfolios. This table presents basic results of idiosyncratic momentum portfolios for various $J-K$ portfolios. The average weekly returns, FF5F-$\alpha$ values, annualized Sharpe ratios and maximum drawdowns are presented in Panel A, B, C and D, respectively. The whole sample period is January 1997 to December 2017. \cite{Newey-West-1987-Em}'s $t$-statistics are obtained and the superscripts * and ** denote the significance at 5\% and 1\% levels, respectively.}
\centering
\vspace{-3mm}
   \begin{tabular}{ccccccccc}
   \hline\hline
     $J$ & $K=1$ & $2$ & $3$ & $4$ &  $8$ & $13$ & $26$ & $52$ \\
   \hline
 \multicolumn{9}{l}{Panel A: Raw return} \\
   2 &  {\textbf{ 0.0031$^{**}$}} & {\textbf{ 0.0029$^{**}$}}  & {\textbf{ 0.0028$^{**}$}} & {\textbf{ 0.0025$^{**}$}} & {\textbf{ 0.0016$^{**}$}} & {\textbf{ 0.0013$^{**}$}} & {\textbf{ 0.0009$^{**}$}} & {\textbf{ 0.0007$^{**}$}}  \\
   3 &  {\textbf{ 0.0036$^{**}$}} & {\textbf{ 0.0034$^{**}$}}  & {\textbf{ 0.0031$^{**}$}} & {\textbf{ 0.0028$^{**}$}} & {\textbf{ 0.0019$^{**}$}} & {\textbf{ 0.0016$^{**}$}} & {\textbf{ 0.0010$^{**}$}} & {\textbf{ 0.0008$^{*~}$}}  \\
   4 &  {\textbf{ 0.0038$^{**}$}} & {\textbf{ 0.0034$^{**}$}}  & {\textbf{ 0.0031$^{**}$}} & {\textbf{ 0.0028$^{**}$}} & {\textbf{ 0.0020$^{**}$}} & {\textbf{ 0.0017$^{**}$}} & {\textbf{ 0.0010$^{*~}$}} & {\textbf{ 0.0008$^{*~}$}}  \\
   8 &  {\textbf{ 0.0033$^{**}$}} & {\textbf{ 0.0030$^{**}$}}  & {\textbf{ 0.0028$^{**}$}} & {\textbf{ 0.0026$^{**}$}} & {\textbf{ 0.0021$^{**}$}} & {\textbf{ 0.0017$^{**}$}} & {\textbf{ 0.0012$^{*~}$}} & {\textbf{ 0.0010$^{*~}$}}  \\
   13 &  {\textbf{ 0.0030$^{**}$}} & {\textbf{ 0.0027$^{**}$}}  & {\textbf{ 0.0025$^{**}$}} & {\textbf{ 0.0024$^{**}$}} & {\textbf{ 0.0020$^{**}$}} & {\textbf{ 0.0017$^{**}$}} & {\textbf{ 0.0013$^{*~}$}} & {\textbf{ 0.0011$^{*~}$}}  \\
   26 &  {\textbf{ 0.0024$^{**}$}} & {\textbf{ 0.0022$^{**}$}}  & {\textbf{ 0.0021$^{**}$}} & {\textbf{ 0.0020$^{**}$}} & {\textbf{ 0.0018$^{**}$}} & {\textbf{ 0.0017$^{**}$}} & {\textbf{ 0.0014$^{*~}$}} & {\textbf{ 0.0012$^{*~}$}}  \\
   52 &  {\textbf{ 0.0019$^{**}$}} & {\textbf{ 0.0018$^{*~}$}}  & {\textbf{ 0.0017$^{*~}$}} & {\textbf{ 0.0017$^{*~}$}} & {\textbf{ 0.0016$^{*~}$}} & {\textbf{ 0.0016$^{*~}$}} & {\textbf{ 0.0013$^{*~}$}} & {\textbf{ 0.0011$^{*~}$}}
       \smallskip\\
 \multicolumn{9}{l}{Panel B: FF5F-$\alpha$} \\
   2 &  {\textbf{ 0.0038$^{**}$}} & {\textbf{ 0.0036$^{**}$}}  & {\textbf{ 0.0034$^{**}$}} & {\textbf{ 0.0031$^{**}$}} & {\textbf{ 0.0021$^{**}$}} & {\textbf{ 0.0017$^{**}$}} & {\textbf{ 0.0011$^{**}$}} & {\textbf{ 0.0009$^{**}$}}  \\
   3 &  {\textbf{ 0.0045$^{**}$}} & {\textbf{ 0.0043$^{**}$}}  & {\textbf{ 0.0040$^{**}$}} & {\textbf{ 0.0036$^{**}$}} & {\textbf{ 0.0025$^{**}$}} & {\textbf{ 0.0020$^{**}$}} & {\textbf{ 0.0013$^{**}$}} & {\textbf{ 0.0010$^{**}$}}  \\
   4 &  {\textbf{ 0.0048$^{**}$}} & {\textbf{ 0.0044$^{**}$}}  & {\textbf{ 0.0040$^{**}$}} & {\textbf{ 0.0036$^{**}$}} & {\textbf{ 0.0027$^{**}$}} & {\textbf{ 0.0022$^{**}$}} & {\textbf{ 0.0014$^{**}$}} & {\textbf{ 0.0011$^{**}$}}  \\
   8 &  {\textbf{ 0.0044$^{**}$}} & {\textbf{ 0.0040$^{**}$}}  & {\textbf{ 0.0036$^{**}$}} & {\textbf{ 0.0034$^{**}$}} & {\textbf{ 0.0027$^{**}$}} & {\textbf{ 0.0023$^{**}$}} & {\textbf{ 0.0016$^{**}$}} & {\textbf{ 0.0012$^{**}$}}  \\
   13 &  {\textbf{ 0.0041$^{**}$}} & {\textbf{ 0.0037$^{**}$}}  & {\textbf{ 0.0034$^{**}$}} & {\textbf{ 0.0032$^{**}$}} & {\textbf{ 0.0027$^{**}$}} & {\textbf{ 0.0023$^{**}$}} & {\textbf{ 0.0017$^{**}$}} & {\textbf{ 0.0014$^{**}$}}  \\
   26 &  {\textbf{ 0.0033$^{**}$}} & {\textbf{ 0.0031$^{**}$}}  & {\textbf{ 0.0029$^{**}$}} & {\textbf{ 0.0027$^{**}$}} & {\textbf{ 0.0024$^{**}$}} & {\textbf{ 0.0021$^{**}$}} & {\textbf{ 0.0017$^{**}$}} & {\textbf{ 0.0014$^{**}$}}  \\
   52 &  {\textbf{ 0.0025$^{**}$}} & {\textbf{ 0.0023$^{**}$}}  & {\textbf{ 0.0022$^{**}$}} & {\textbf{ 0.0021$^{**}$}} & {\textbf{ 0.0019$^{**}$}} & {\textbf{ 0.0018$^{**}$}} & {\textbf{ 0.0014$^{**}$}} & {\textbf{ 0.0012$^{**}$}}
       \smallskip\\
 \multicolumn{9}{l}{Panel C: Annualized Sharpe ratio} \\
   2 &  { 1.3824$^{~~}$} & { 1.4113$^{~~}$}  & { 1.4296$^{~~}$} & { 1.3555$^{~~}$} & { 1.0187$^{~~}$} & { 0.9648$^{~~}$} & { 0.8175$^{~~}$} & { 0.8825$^{~~}$}  \\
   3 &  { 1.3492$^{~~}$} & { 1.3625$^{~~}$}  & { 1.3158$^{~~}$} & { 1.2255$^{~~}$} & { 0.9248$^{~~}$} & { 0.8654$^{~~}$} & { 0.6882$^{~~}$} & { 0.6686$^{~~}$}  \\
   4 &  { 1.3392$^{~~}$} & { 1.2835$^{~~}$}  & { 1.2211$^{~~}$} & { 1.1353$^{~~}$} & { 0.9211$^{~~}$} & { 0.8380$^{~~}$} & { 0.6409$^{~~}$} & { 0.6335$^{~~}$}  \\
   8 &  { 1.0735$^{~~}$} & { 1.0127$^{~~}$}  & { 0.9715$^{~~}$} & { 0.9350$^{~~}$} & { 0.8112$^{~~}$} & { 0.7508$^{~~}$} & { 0.6422$^{~~}$} & { 0.6651$^{~~}$}  \\
   13 &  { 0.9333$^{~~}$} & { 0.8857$^{~~}$}  & { 0.8404$^{~~}$} & { 0.8235$^{~~}$} & { 0.7258$^{~~}$} & { 0.6948$^{~~}$} & { 0.6186$^{~~}$} & { 0.6741$^{~~}$}  \\
   26 &  { 0.7339$^{~~}$} & { 0.7060$^{~~}$}  & { 0.6780$^{~~}$} & { 0.6567$^{~~}$} & { 0.6342$^{~~}$} & { 0.6492$^{~~}$} & { 0.6517$^{~~}$} & { 0.6951$^{~~}$}  \\
   52 &  { 0.6154$^{~~}$} & { 0.5793$^{~~}$}  & { 0.5820$^{~~}$} & { 0.5735$^{~~}$} & { 0.5776$^{~~}$} & { 0.6102$^{~~}$} & { 0.5975$^{~~}$} & { 0.5906$^{~~}$}
       \smallskip\\
 \multicolumn{9}{l}{Panel D: Maximum drawdown} \\
   2 &  { 0.2397$^{~~}$} & { 0.1828$^{~~}$}  & { 0.1590$^{~~}$} & { 0.1665$^{~~}$} & { 0.1226$^{~~}$} & { 0.1250$^{~~}$} & { 0.1375$^{~~}$} & { 0.1173$^{~~}$}  \\
   3 &  { 0.2673$^{~~}$} & { 0.2058$^{~~}$}  & { 0.1849$^{~~}$} & { 0.1682$^{~~}$} & { 0.1491$^{~~}$} & { 0.1499$^{~~}$} & { 0.1648$^{~~}$} & { 0.1457$^{~~}$}  \\
   4 &  { 0.2756$^{~~}$} & { 0.2317$^{~~}$}  & { 0.1988$^{~~}$} & { 0.1773$^{~~}$} & { 0.1528$^{~~}$} & { 0.1749$^{~~}$} & { 0.1868$^{~~}$} & { 0.1641$^{~~}$}  \\
   8 &  { 0.2340$^{~~}$} & { 0.2269$^{~~}$}  & { 0.2312$^{~~}$} & { 0.2186$^{~~}$} & { 0.1853$^{~~}$} & { 0.1990$^{~~}$} & { 0.2086$^{~~}$} & { 0.1835$^{~~}$}  \\
   13 &  { 0.2950$^{~~}$} & { 0.3011$^{~~}$}  & { 0.2636$^{~~}$} & { 0.2333$^{~~}$} & { 0.2374$^{~~}$} & { 0.2515$^{~~}$} & { 0.2267$^{~~}$} & { 0.1873$^{~~}$}  \\
   26 &  { 0.3411$^{~~}$} & { 0.3481$^{~~}$}  & { 0.3166$^{~~}$} & { 0.2911$^{~~}$} & { 0.2845$^{~~}$} & { 0.2654$^{~~}$} & { 0.2224$^{~~}$} & { 0.2299$^{~~}$}  \\
   52 &  { 0.2678$^{~~}$} & { 0.2798$^{~~}$}  & { 0.2620$^{~~}$} & { 0.2624$^{~~}$} & { 0.2735$^{~~}$} & { 0.2541$^{~~}$} & { 0.2539$^{~~}$} & { 0.2599$^{~~}$}  \\
   \hline\hline
   \end{tabular}
   \label{TB:IMOM:detail:F5}
\end{table}

Furthermore, the results are sensitive to the values of $J$ and $K$. The return magnitude decreases with increasing $J$, whilst there exists an inverted ``U'' shape relation between the return magnitude and $K$, with the superior performance emerging when $J =4$ weeks.
All of the portfolios are statistically significant at the level of 1\%, which mainly appear in the short term and inter-medium term when $J$ and $K$ are less than 13 weeks. With increasing $J$ and $K$, the significance for the IMOM portfolios decreases as well, as suggested by the results that are significant mostly at the level of 5\% when $J$ and $K$ are more than 26 weeks.

Similarly, all of the FF5F-$\alpha$ values remain statistically significant and the magnitudes are larger than that of raw returns.
Furthermore, high Sharpe ratios are obtained in Panel C, mainly ascribing to the sorting on the cumulatively idiosyncratic returns, which avoid the influence of time-varying common risk factors to a large extent \citep{Blitz-Hanauer-Vidojevic-2020-IRFA}. And the higher Sharpe ratios are achieved when $J$ and $K$ are less than four weeks.
Accordingly, the maximum drawdowns in Panel D range from 11.73\% (when $J= 2$ weeks and $K = 52$ weeks) to 27.56\% (when $J = 4$ weeks and $K =1$ week).

\textcolor{black}{It is noted that our results can be roughly comparable to those of \cite{Lin-2019-FRL}, which evaluates the IMOM performance at monthly basis in the Chinese stock market. The author reports statistically significant IMOM profits when the estimation period is from lagged 12 months through lagged two months and the holding period is one month. Specifically, the excess return and FF5F-$\alpha$ are 0.882\% and 0.931\% per month, respectively.
Correspondingly, according to our results when $K = 4$ weeks, the IMOM portfolio generates excess return of 0.68\% ($\approx 4 \times 0.17\%$) and FF5F-$\alpha$ of 0.84\% ($\approx 4 \times 0.21\%$), both of which are lower than those in \cite{Lin-2019-FRL}.
We believe that the small gaps in the results between \cite{Lin-2019-FRL} and our work, can be mainly explained by the difference in time-skip schemes and the data samples adopted in our two studies. }

We additionally evaluate the IMOM performance in subperiods. The results are reported in \autoref{TB:IMOM:subperiod:F5}. Apparently, almost no idiosyncratic momentum profits exist during the first subperiod from 1997 to 2003 in Panel B, although all the portfolios achieve positive returns. In comparison, as for the second and third subperiods in Panels C and D, we observe the pronounced short-term idiosyncratic momentum, ranging from one week to 13 weeks. Meanwhile, the profitability of the IMOM seemingly has strengthened over time, especially in most recent periods, with the return magnitudes being higher than those for the whole sample period (in Panel A) by a range from 5 to 10 basis points.

\subsection{Univariate portfolio analysis based on idiosyncratic risk metrics}

The relation between lagged risk metrics and cross-sectional returns should be identified before we perform the risk-adjustment to the IMOM. Therefore, we conduct the univariate portfolio analysis based on different risk metrics.

Unsurprisingly, the majority of the portfolios achieve positive profits, indicating the negative relation between the idiosyncratic risk metric and cross-sectional returns, as indicated in  \autoref{TB:Risk:F5}. Additionally, as predicted, most results indicate the negative abnormal returns (measured by the FF5F-$\alpha$ values) achieved by the stock group with the highest level of risk metrics, contributing much more to the total abnormal return of arbitrage portfolios.\footnote{ We also examine the results for the long sides (with lowest risk) and the short sides (with highest risk) with respect to the portfolios. As indicated by \autoref{TB:Risk:F5:long} and \autoref{TB:Risk:F5:short}, the results are pronounced in the short term when $K$ is less than four weeks. }

We find that higher profitability is achieved by the portfolios with respect to the IVol, which is presented in Panel A. 
The empirical results imply that the return series of IMOM and IVol might be highly correlated, which would be confirmed in the following tests.
In comparison, the portfolios based on the ISkew and the IKurt in Panel B and C exhibit much worse performance.
It should be noted that lagged ISkew is argued to be a good estimator for the expected skewness but not for the expected idiosyncratic skewness \citep{Boyer-Mitton-Vorkink-2009-RFS}, thereby leading to the biased results regarding the ISkew, while the explanation of lottery-like demand is applicable to both cases.
In addition, it is stressed that our results do not imply the weak lottery-like demand in the Chinese stock market. On the contrary, Chinese market sees the strong lottery-like demand, which are better captured by several indicators jointly \citep{Zheng-Sun-2013-cnERJ}.
As for the IKurt, no results are statistically significant, and moreover the return magnitudes remain very low as well. This is no surprise, however, since no solid theoretical evidence has been provided to support the existence of the relation between the idiosyncratic kurtosis and cross-sectional stock returns.

Moving to the results for idiosyncratic tail risk,
we find more superior performance with respect to the IMD.
In comparison, the returns of portfolios with respect to other tail risk metrics exhibit lower magnitudes and statistical significance.
Although our results suggest that IMD-based portfolios outperform those based on other tail risk metrics, we cannot conclude that a more prevailing slow-down effect of price limits takes place in the Chinese stock market. Conversely, as suggested by the larger magnitude of negatively abnormal returns of the short sides of the portfolios (\autoref{TB:Risk:F5:short}), it would be a speed-up effect instead.
In \autoref{TB:Risk:F5}, we additionally observe the stark difference between the results for IVaR5 (or IES5) and IVaR1 (or IES1). Nearly half of the portfolios are insignificant in IES1 or IVaR1, while all portfolios are significant in IES5 and IVaR5. This is probably because the IVaR5 and the IES5 convey more information about the bad news than IVaR1 (IES1), as indicated by the larger drift of negative abnormal returns in loser portfolios.

We additionally observe the decreasing profitability with increasing holding period $K$ in results for most of the risk metrics.
\cite{Atilgan-Bali-Demirtas-Gunaydin-2020-JFE} also document similar findings but associated with left tail risks. Both cases are in accordance with the argument of the stronger underreaction to bad news in the first phase and the correction of overreaction to bad news in the second phase \citep{Hong-Stein-1999-JF,Atilgan-Bali-Demirtas-Gunaydin-2020-JFE}.


\subsection{Spanning test for idiosyncratic risk metrics}

We further employ the results from \autoref{TB:Risk:F5} to conduct the spanning test
\citep{NovyMarx-2012-JFE,Blitz-Hanauer-Vidojevic-2020-IRFA},
so as to examine if the predictability of one specific risk metric, along with Fama-French 5 factors, could explain those of other risk metrics. The results are presented in \autoref{TB:Risk:F5:spanning:p1}.
In each panel, a total of 56 augmented FF5F time-series regression are performed. Significance level of intercepts ($\alpha$ values) implies the explanatory power of risk metrics. And an insignificant intercept would suggest the test factor is inside the span of the explanatory factor.

We evaluate their performance according to the number of statistically significant $\alpha$ values associated with different risk metrics. For instance, there is a total of 20 $\alpha$ values that remain statistically significant after the augmented FF5F regression in first panel for IVol.
Accordingly, we can sort the idiosyncratic risk metrics based upon their respective numbers of unexplained intercepts: IVol (20), IES5 (22), IMD (23), IES1 (24), IVaR1 (28), IVaR5 (32), ISkew (48), IKurt (51); what present in parentheses are numbers of statistically significant $\alpha$ values.
Therefore, the IVol-based portfolios display more powerful explanation to others, while the IKurt-based portfolios have the weakest power of explanation. In fact, risk metrics with smaller number of significant $\alpha$ values indeed exhibit better explanatory power than others, such as IVol, IES5, IMD, and IES1.

Additionally, we also find the risk metrics ``clustering'' in the results. To be specific, the tail risk metrics including IES1, IES5, IVaR1 and IVaR5 function well in explaining each other.
Generally, the portfolios with respect to tail risk metrics exhibit limited explanatory power to the IMD-based as well as IVol-based portfolios.
In comparison, ISkew and IKurt could offer better explanation to each other, meanwhile they are not able to explain the portfolios based on other risk metrics at all, with significant $\alpha$ values much higher than those in \autoref{TB:Risk:F5}.
And portfolios based on other risk metrics also exhibit weak explanatory power to the portfolios based on ISkew and IKurt.

Apparently, IMD explains the majority of the portfolios associated with tail risk metrics as well as most portfolios with respect to the IVol. In comparison, the IVol-based portfolios seemingly have better explanatory power to those with respect to most risk metrics, except for ISkew, IKurt, and IVaR5, and the insignificant $\alpha$ values approach zero as well.
Accompanied with the previous evidence, we conclude that IVol and IMD outperform the rest of the risk metrics.

\subsection{Risk-adjusted IMOM}

Preceding tests confirm the negative relation between the idiosyncratic risk metrics and the cross-sectional stock returns, which paves the way to the combination of the idiosyncratic risk metrics and the IMOM.
We attempt to explore more profitable risk-adjusted IMOM portfolios.
Likewise, we conduct the study by forming $J-K$ portfolios with varying $K$, while we fix $J$ at 26 weeks, which is coherent with the length of the estimation period (130 trading days) for idiosyncratic risk metrics.
As mentioned in previous sections, we construct risk-adjusted IMOM portfolios from two perspectives for the robustness of results.

Firstly, we follow the directly adjusted IMOM procedure, as described in Eq.~(\ref{Eq:Risk-IMOM}). Given that the negative values in ISkew lead to sorting disorder, we consider only  other seven risk metrics in this method. The results are reported in \autoref{TB:Res2Risk:F5:d}.
Generally, the results are very similar to those of the univariate portfolio analysis with respect to the risk metrics in \autoref{TB:Risk:F5}.
In fact, some of the portfolios achieve higher returns by only 1 basis point.
The Sharpe ratios for most portfolios are indeed improved but very slightly, especially in the results based on idiosyncratic tail risk metrics.
However, the maximum drawdowns of most portfolios are even worse than those in \autoref{TB:Risk:F5}.
In comparison, FF5F-$\alpha$ is the only indicator that gets remarkably improved.
We also obtain some similar findings. For instance, the IVol-IMOM portfolios gain higher profitability than other portfolios.

As for the Sharpe ratios, we observe the outperformance of the IVol-IMOMs and the IMD-IMOMs. In particular, the IVol-IMOMs achieve higher Sharpe ratios when $K$ is less than three weeks, ranging from 0.67 to 0.75, and the IMD-IMOMs achieve higher Sharpe ratios when $K$ is more than three weeks, ranging from 0.69 to 0.87, which are similar to the findings in \autoref{TB:Risk:F5}.
As for the maximum drawdowns, we find that the IES1-IMOMs, the IMD-IMOMs and the IVol-IMOMs outperform the others.
In addition, through the scrutiny of the results, we find that the IVol-IMOM shares the similar performance as those with respect to the IVol and the IMOM, further suggesting that IVol and IMOM share the similar ranking process to thereby construct the zero-cost arbitrage portfolios with similar performance.

\begin{table}[!ht]
\small
\caption{The performance of risk-adjusted IMOM portfolios with the direct adjustment procedure. This table reports the results associated with risk-adjusted IMOM portfolios based on Eq.~(\ref{Eq:Risk-IMOM}), including average weekly raw returns (Raw), FF5F-$\alpha$ values ($\alpha$), annualized Sharpe ratios (SR) and maximum drawdown (MD). The sample period is from January 1997 to December 2017. At the beginning of each week, the risk-adjusted IMOM factor is constructed according to Eq.~(\ref{Eq:Risk-IMOM}), where idiosyncratic return and its risk metrics can be calculated using idiosyncratic returns over past 26 weeks and 130 trading days for each of individual stocks. And they can be sorted into ten decile groups according to the risk-adjusted IMOM factor. Zero-cost arbitrage portfolios are constructed by buying stocks from the group with the highest risk-adjusted idiosyncratic return and selling the stocks from the group with the lowest risk-adjusted idiosyncratic return. The portfolios would be held for $K$ weeks, and calendar-time method is applied to obtain the average weekly return. \cite{Newey-West-1987-Em}'s $t$-statistics are obtained and the superscripts * and ** denote the significance at 5\% and 1\% levels, respectively. }
\centering
\vspace{-3mm}
   \begin{tabular}{ccccccccc}
   \hline\hline
      & $K=1$ & $2$ & $3$ & $4$ &  $8$ & $13$ & $26$ & $52$ \\
   \hline
 \multicolumn{9}{l}{Panel A: Idiosyncratic volatility} \\
   Raw &  {\textbf{ 0.0024$^{**}$}} & {\textbf{ 0.0022$^{**}$}}  & {\textbf{ 0.0021$^{**}$}} & {\textbf{ 0.0019$^{**}$}} & {\textbf{ 0.0018$^{**}$}} & {\textbf{ 0.0017$^{**}$}} & {\textbf{ 0.0014$^{*~}$}} & {\textbf{ 0.0012$^{*~}$}}  \\
   $\alpha$ &  {\textbf{ 0.0033$^{**}$}} & {\textbf{ 0.0031$^{**}$}}  & {\textbf{ 0.0029$^{**}$}} & {\textbf{ 0.0027$^{**}$}} & {\textbf{ 0.0024$^{**}$}} & {\textbf{ 0.0021$^{**}$}} & {\textbf{ 0.0017$^{**}$}} & {\textbf{ 0.0014$^{**}$}}  \\
   SR &  { 0.7419$^{~~}$} & { 0.7060$^{~~}$}  & { 0.6775$^{~~}$} & { 0.6544$^{~~}$} & { 0.6436$^{~~}$} & { 0.6568$^{~~}$} & { 0.6553$^{~~}$} & { 0.6954$^{~~}$}  \\
   MD &  { 0.3169$^{~~}$} & { 0.3259$^{~~}$}  & { 0.3020$^{~~}$} & { 0.2860$^{~~}$} & { 0.2796$^{~~}$} & { 0.2601$^{~~}$} & { 0.2259$^{~~}$} & { 0.2323$^{~~}$}
       \smallskip\\
 \multicolumn{9}{l}{Panel B: Idiosyncratic kurtosis} \\
   Raw &  { 0.0009$^{~~}$} & { 0.0008$^{~~}$}  & { 0.0007$^{~~}$} & { 0.0006$^{~~}$} & { 0.0003$^{~~}$} & { 0.0001$^{~~}$} & { -0.0001$^{~~}$} & { -0.0002$^{~~}$}  \\
   $\alpha$ &  {\textbf{ 0.0010$^{*~}$}} & {\textbf{ 0.0009$^{*~}$}}  & {\textbf{ 0.0008$^{*~}$}} & { 0.0007$^{~~}$} & { 0.0004$^{~~}$} & { 0.0002$^{~~}$} & { -0.0001$^{~~}$} & { -0.0002$^{~~}$}  \\
   SR &  { 0.4664$^{~~}$} & { 0.4520$^{~~}$}  & { 0.4089$^{~~}$} & { 0.3865$^{~~}$} & { 0.1764$^{~~}$} & { 0.0828$^{~~}$} & { -0.0627$^{~~}$} & { -0.2306$^{~~}$}  \\
   MD &  { 0.4108$^{~~}$} & { 0.3838$^{~~}$}  & { 0.3618$^{~~}$} & { 0.3525$^{~~}$} & { 0.3973$^{~~}$} & { 0.3840$^{~~}$} & { 0.3970$^{~~}$} & { 0.4400$^{~~}$}
       \smallskip\\
 \multicolumn{9}{l}{Panel C: Idiosyncratic maximum drawdown} \\
   Raw &  {\textbf{ 0.0016$^{*~}$}} & {\textbf{ 0.0018$^{**}$}}  & {\textbf{ 0.0018$^{**}$}} & {\textbf{ 0.0018$^{**}$}} & {\textbf{ 0.0017$^{**}$}} & {\textbf{ 0.0016$^{**}$}} & {\textbf{ 0.0015$^{**}$}} & {\textbf{ 0.0012$^{**}$}}  \\
   $\alpha$ &  {\textbf{ 0.0022$^{**}$}} & {\textbf{ 0.0023$^{**}$}}  & {\textbf{ 0.0023$^{**}$}} & {\textbf{ 0.0022$^{**}$}} & {\textbf{ 0.0021$^{**}$}} & {\textbf{ 0.0019$^{**}$}} & {\textbf{ 0.0016$^{**}$}} & {\textbf{ 0.0014$^{**}$}}  \\
   SR &  { 0.5851$^{~~}$} & { 0.6634$^{~~}$}  & { 0.6793$^{~~}$} & { 0.7024$^{~~}$} & { 0.7339$^{~~}$} & { 0.7709$^{~~}$} & { 0.8157$^{~~}$} & { 0.8745$^{~~}$}  \\
   MD &  { 0.3110$^{~~}$} & { 0.3187$^{~~}$}  & { 0.3124$^{~~}$} & { 0.3038$^{~~}$} & { 0.2855$^{~~}$} & { 0.2561$^{~~}$} & { 0.2177$^{~~}$} & { 0.2177$^{~~}$}
       \smallskip\\
 \multicolumn{9}{l}{Panel D: Idiosyncratic ES (5\%)} \\
   Raw &  {\textbf{ 0.0019$^{**}$}} & {\textbf{ 0.0017$^{*~}$}}  & {\textbf{ 0.0016$^{*~}$}} & {\textbf{ 0.0015$^{*~}$}} & {\textbf{ 0.0014$^{*~}$}} & {\textbf{ 0.0014$^{*~}$}} & {\textbf{ 0.0012$^{*~}$}} & {\textbf{ 0.0010$^{*~}$}}  \\
   $\alpha$ &  {\textbf{ 0.0029$^{**}$}} & {\textbf{ 0.0026$^{**}$}}  & {\textbf{ 0.0024$^{**}$}} & {\textbf{ 0.0023$^{**}$}} & {\textbf{ 0.0020$^{**}$}} & {\textbf{ 0.0019$^{**}$}} & {\textbf{ 0.0015$^{**}$}} & {\textbf{ 0.0012$^{**}$}}  \\
   SR &  { 0.6196$^{~~}$} & { 0.5655$^{~~}$}  & { 0.5364$^{~~}$} & { 0.5348$^{~~}$} & { 0.5324$^{~~}$} & { 0.5755$^{~~}$} & { 0.5998$^{~~}$} & { 0.6001$^{~~}$}  \\
   MD &  { 0.3250$^{~~}$} & { 0.3455$^{~~}$}  & { 0.3158$^{~~}$} & { 0.2939$^{~~}$} & { 0.2885$^{~~}$} & { 0.2712$^{~~}$} & { 0.2169$^{~~}$} & { 0.2283$^{~~}$}
       \smallskip\\
 \multicolumn{9}{l}{Panel E: Idiosyncratic VaR (5\%)} \\
   Raw &  {\textbf{ 0.0018$^{*~}$}} & {\textbf{ 0.0016$^{*~}$}}  & {\textbf{ 0.0015$^{*~}$}} & {\textbf{ 0.0015$^{*~}$}} & {\textbf{ 0.0014$^{*~}$}} & {\textbf{ 0.0014$^{*~}$}} & {\textbf{ 0.0013$^{*~}$}} & {\textbf{ 0.0011$^{*~}$}}  \\
   $\alpha$ &  {\textbf{ 0.0026$^{**}$}} & {\textbf{ 0.0024$^{**}$}}  & {\textbf{ 0.0023$^{**}$}} & {\textbf{ 0.0022$^{**}$}} & {\textbf{ 0.0020$^{**}$}} & {\textbf{ 0.0018$^{**}$}} & {\textbf{ 0.0015$^{**}$}} & {\textbf{ 0.0012$^{**}$}}  \\
   SR &  { 0.5631$^{~~}$} & { 0.5376$^{~~}$}  & { 0.5218$^{~~}$} & { 0.5166$^{~~}$} & { 0.5149$^{~~}$} & { 0.5609$^{~~}$} & { 0.5969$^{~~}$} & { 0.6165$^{~~}$}  \\
   MD &  { 0.3510$^{~~}$} & { 0.3666$^{~~}$}  & { 0.3403$^{~~}$} & { 0.3161$^{~~}$} & { 0.2997$^{~~}$} & { 0.2748$^{~~}$} & { 0.2298$^{~~}$} & { 0.2387$^{~~}$}
       \smallskip\\
 \multicolumn{9}{l}{Panel F: Idiosyncratic ES (1\%)} \\
   Raw &  {\textbf{ 0.0017$^{**}$}} & {\textbf{ 0.0014$^{*~}$}}  & {\textbf{ 0.0013$^{*~}$}} & {\textbf{ 0.0013$^{*~}$}} & {\textbf{ 0.0012$^{*~}$}} & {\textbf{ 0.0011$^{*~}$}} & {\textbf{ 0.0010$^{*~}$}} & { 0.0008$^{~~}$}  \\
   $\alpha$ &  {\textbf{ 0.0025$^{**}$}} & {\textbf{ 0.0022$^{**}$}}  & {\textbf{ 0.0021$^{**}$}} & {\textbf{ 0.0019$^{**}$}} & {\textbf{ 0.0017$^{**}$}} & {\textbf{ 0.0016$^{**}$}} & {\textbf{ 0.0012$^{**}$}} & {\textbf{ 0.0009$^{**}$}}  \\
   SR &  { 0.6497$^{~~}$} & { 0.5623$^{~~}$}  & { 0.5448$^{~~}$} & { 0.5265$^{~~}$} & { 0.5195$^{~~}$} & { 0.5444$^{~~}$} & { 0.5436$^{~~}$} & { 0.5608$^{~~}$}  \\
   MD &  { 0.2981$^{~~}$} & { 0.3102$^{~~}$}  & { 0.2913$^{~~}$} & { 0.2730$^{~~}$} & { 0.2594$^{~~}$} & { 0.2477$^{~~}$} & { 0.2008$^{~~}$} & { 0.2059$^{~~}$}
       \smallskip\\
 \multicolumn{9}{l}{Panel G: Idiosyncratic VaR (1\%)} \\
   Raw &  {\textbf{ 0.0017$^{*~}$}} & {\textbf{ 0.0014$^{*~}$}}  & {\textbf{ 0.0013$^{*~}$}} & { 0.0012$^{~~}$} & { 0.0011$^{~~}$} & { 0.0011$^{~~}$} & { 0.0011$^{~~}$} & { 0.0008$^{~~}$}  \\
   $\alpha$ &  {\textbf{ 0.0026$^{**}$}} & {\textbf{ 0.0023$^{**}$}}  & {\textbf{ 0.0021$^{**}$}} & {\textbf{ 0.0019$^{**}$}} & {\textbf{ 0.0017$^{**}$}} & {\textbf{ 0.0016$^{**}$}} & {\textbf{ 0.0013$^{**}$}} & {\textbf{ 0.0010$^{*~}$}}  \\
   SR &  { 0.5849$^{~~}$} & { 0.5140$^{~~}$}  & { 0.4819$^{~~}$} & { 0.4653$^{~~}$} & { 0.4549$^{~~}$} & { 0.5039$^{~~}$} & { 0.5466$^{~~}$} & { 0.5540$^{~~}$}  \\
   MD &  { 0.2900$^{~~}$} & { 0.3150$^{~~}$}  & { 0.2896$^{~~}$} & { 0.2705$^{~~}$} & { 0.2650$^{~~}$} & { 0.2534$^{~~}$} & { 0.2256$^{~~}$} & { 0.2236$^{~~}$}  \\
   \hline\hline
   \end{tabular}
   \label{TB:Res2Risk:F5:d}
\end{table}

\begin{table}[!ht]
\small
\caption{The performance of risk-adjusted IMOM portfolios with the indirect adjustment procedure. This table reports the results associated with the bivariate portfolio analysis, including average weekly raw returns (Raw), FF5F-$\alpha$ values ($\alpha$), annualized Sharpe ratios (SR) and maximum drawdowns (MD). The sample period is January 1997 to December 2017. At the beginning of each week, the idiosyncratic return and its risk metrics can be calculated using idiosyncratic returns over past 26 weeks and 130 trading days for individual stocks, by which the stocks can be separately sorted into decile groups according to cumulatively idiosyncratic return and specific risk metric, respectively. The zero-cost arbitrage portfolios are constructed by buying stocks from intersected decile groups with lowest risk and highest cumulatively idiosyncratic returns and selling the stocks from the intersected decile groups with highest risk and lowest cumulatively idiosyncratic returns. Portfolios would be held for $K$ weeks, and calendar-time method is applied to obtain the average weekly return. \cite{Newey-West-1987-Em}'s $t$-statistics are obtained and the superscripts * and ** denote the significance at 5\% and 1\% levels, respectively. }
\centering
\vspace{-3mm}
   \begin{tabular}{ccccccccc}
   \hline\hline
      & $K=1$ & $2$ & $3$ & $4$ &  $8$ & $13$ & $26$ & $52$ \\
   \hline
 \multicolumn{9}{l}{Panel A: Idiosyncratic volatility} \\
   Raw &  {\textbf{ 0.0024$^{**}$}} & {\textbf{ 0.0022$^{**}$}}  & {\textbf{ 0.0021$^{**}$}} & {\textbf{ 0.0020$^{**}$}} & {\textbf{ 0.0018$^{**}$}} & {\textbf{ 0.0017$^{**}$}} & {\textbf{ 0.0015$^{*~}$}} & {\textbf{ 0.0012$^{*~}$}}  \\
   $\alpha$ &  {\textbf{ 0.0034$^{**}$}} & {\textbf{ 0.0031$^{**}$}}  & {\textbf{ 0.0029$^{**}$}} & {\textbf{ 0.0028$^{**}$}} & {\textbf{ 0.0024$^{**}$}} & {\textbf{ 0.0022$^{**}$}} & {\textbf{ 0.0017$^{**}$}} & {\textbf{ 0.0014$^{**}$}}  \\
   SR &  { 0.7410$^{~~}$} & { 0.7051$^{~~}$}  & { 0.6761$^{~~}$} & { 0.6564$^{~~}$} & { 0.6409$^{~~}$} & { 0.6586$^{~~}$} & { 0.6584$^{~~}$} & { 0.7010$^{~~}$}  \\
   MD &  { 0.3341$^{~~}$} & { 0.3449$^{~~}$}  & { 0.3143$^{~~}$} & { 0.2902$^{~~}$} & { 0.2857$^{~~}$} & { 0.2664$^{~~}$} & { 0.2233$^{~~}$} & { 0.2312$^{~~}$}     \smallskip\\
 \multicolumn{9}{l}{Panel B: Idiosyncratic skewness} \\
   Raw &  {\textbf{ 0.0027$^{*~}$}} & {\textbf{ 0.0032$^{**}$}}  & {\textbf{ 0.0041$^{**}$}} & {\textbf{ 0.0034$^{**}$}} & {\textbf{ 0.0019$^{*~}$}} & { 0.0009$^{~~}$} & { 0.0008$^{~~}$} & { 0.0006$^{~~}$}  \\
   $\alpha$ &  {\textbf{ 0.0035$^{**}$}} & {\textbf{ 0.0038$^{**}$}}  & {\textbf{ 0.0046$^{**}$}} & {\textbf{ 0.0038$^{**}$}} & {\textbf{ 0.0023$^{**}$}} & { 0.0012$^{~~}$} & { 0.0010$^{~~}$} & { 0.0008$^{~~}$}  \\
   SR &  { 0.4572$^{~~}$} & { 0.5509$^{~~}$}  & { 0.7894$^{~~}$} & { 0.6729$^{~~}$} & { 0.4500$^{~~}$} & { 0.2201$^{~~}$} & { 0.2478$^{~~}$} & { 0.2969$^{~~}$}  \\
   MD &  { 0.6946$^{~~}$} & { 0.5153$^{~~}$}  & { 0.3912$^{~~}$} & { 0.5875$^{~~}$} & { 0.5374$^{~~}$} & { 0.8075$^{~~}$} & { 0.6315$^{~~}$} & { 0.2775$^{~~}$}  \smallskip\\
 \multicolumn{9}{l}{Panel C: Idiosyncratic maximum drawdown} \\
   Raw &  {\textbf{ 0.0024$^{**}$}} & {\textbf{ 0.0024$^{**}$}}  & {\textbf{ 0.0023$^{**}$}} & {\textbf{ 0.0022$^{**}$}} & {\textbf{ 0.0020$^{**}$}} & {\textbf{ 0.0019$^{**}$}} & {\textbf{ 0.0017$^{**}$}} & {\textbf{ 0.0014$^{**}$}}  \\
   $\alpha$ &  {\textbf{ 0.0033$^{**}$}} & {\textbf{ 0.0032$^{**}$}}  & {\textbf{ 0.0031$^{**}$}} & {\textbf{ 0.0030$^{**}$}} & {\textbf{ 0.0026$^{**}$}} & {\textbf{ 0.0024$^{**}$}} & {\textbf{ 0.0020$^{**}$}} & {\textbf{ 0.0016$^{**}$}}  \\
   SR &  { 0.6555$^{~~}$} & { 0.6746$^{~~}$}  & { 0.6706$^{~~}$} & { 0.6766$^{~~}$} & { 0.6710$^{~~}$} & { 0.6933$^{~~}$} & { 0.7090$^{~~}$} & { 0.7656$^{~~}$}  \\
   MD &  { 0.3920$^{~~}$} & { 0.3992$^{~~}$}  & { 0.3759$^{~~}$} & { 0.3682$^{~~}$} & { 0.3484$^{~~}$} & { 0.3099$^{~~}$} & { 0.2394$^{~~}$} & { 0.2438$^{~~}$}   \smallskip\\
 \multicolumn{9}{l}{Panel D: Idiosyncratic ES (5\%)} \\
   Raw &  {\textbf{ 0.0023$^{**}$}} & {\textbf{ 0.0020$^{*~}$}}  & {\textbf{ 0.0019$^{*~}$}} & {\textbf{ 0.0018$^{*~}$}} & {\textbf{ 0.0016$^{*~}$}} & {\textbf{ 0.0016$^{*~}$}} & {\textbf{ 0.0014$^{*~}$}} & {\textbf{ 0.0012$^{*~}$}}  \\
   $\alpha$ &  {\textbf{ 0.0034$^{**}$}} & {\textbf{ 0.0031$^{**}$}}  & {\textbf{ 0.0029$^{**}$}} & {\textbf{ 0.0027$^{**}$}} & {\textbf{ 0.0024$^{**}$}} & {\textbf{ 0.0022$^{**}$}} & {\textbf{ 0.0018$^{**}$}} & {\textbf{ 0.0014$^{**}$}}  \\
   SR &  { 0.6268$^{~~}$} & { 0.5830$^{~~}$}  & { 0.5678$^{~~}$} & { 0.5536$^{~~}$} & { 0.5377$^{~~}$} & { 0.5712$^{~~}$} & { 0.6037$^{~~}$} & { 0.6143$^{~~}$}  \\
   MD &  { 0.3781$^{~~}$} & { 0.3899$^{~~}$}  & { 0.3538$^{~~}$} & { 0.3402$^{~~}$} & { 0.3315$^{~~}$} & { 0.3114$^{~~}$} & { 0.2438$^{~~}$} & { 0.2436$^{~~}$}  \smallskip\\
 \multicolumn{9}{l}{Panel E: Idiosyncratic VaR (5\%)} \\
   Raw &  {\textbf{ 0.0024$^{**}$}} & {\textbf{ 0.0022$^{**}$}}  & {\textbf{ 0.0021$^{**}$}} & {\textbf{ 0.0020$^{*~}$}} & {\textbf{ 0.0019$^{*~}$}} & {\textbf{ 0.0018$^{*~}$}} & {\textbf{ 0.0016$^{*~}$}} & {\textbf{ 0.0013$^{*~}$}}  \\
   $\alpha$ &  {\textbf{ 0.0034$^{**}$}} & {\textbf{ 0.0032$^{**}$}}  & {\textbf{ 0.0030$^{**}$}} & {\textbf{ 0.0028$^{**}$}} & {\textbf{ 0.0025$^{**}$}} & {\textbf{ 0.0023$^{**}$}} & {\textbf{ 0.0019$^{**}$}} & {\textbf{ 0.0015$^{**}$}}  \\
   SR &  { 0.6639$^{~~}$} & { 0.6526$^{~~}$}  & { 0.6157$^{~~}$} & { 0.6023$^{~~}$} & { 0.6069$^{~~}$} & { 0.6295$^{~~}$} & { 0.6471$^{~~}$} & { 0.6642$^{~~}$}  \\
   MD &  { 0.3842$^{~~}$} & { 0.3935$^{~~}$}  & { 0.3617$^{~~}$} & { 0.3355$^{~~}$} & { 0.3250$^{~~}$} & { 0.2978$^{~~}$} & { 0.2368$^{~~}$} & { 0.2396$^{~~}$}   \smallskip\\
 \multicolumn{9}{l}{Panel F: Idiosyncratic ES (1\%)} \\
   Raw &  {\textbf{ 0.0024$^{**}$}} & {\textbf{ 0.0022$^{**}$}}  & {\textbf{ 0.0020$^{*~}$}} & {\textbf{ 0.0019$^{*~}$}} & {\textbf{ 0.0017$^{*~}$}} & {\textbf{ 0.0017$^{*~}$}} & {\textbf{ 0.0014$^{*~}$}} & {\textbf{ 0.0011$^{*~}$}}  \\
   $\alpha$ &  {\textbf{ 0.0036$^{**}$}} & {\textbf{ 0.0032$^{**}$}}  & {\textbf{ 0.0030$^{**}$}} & {\textbf{ 0.0028$^{**}$}} & {\textbf{ 0.0025$^{**}$}} & {\textbf{ 0.0022$^{**}$}} & {\textbf{ 0.0017$^{**}$}} & {\textbf{ 0.0014$^{**}$}}  \\
   SR &  { 0.6641$^{~~}$} & { 0.6083$^{~~}$}  & { 0.5848$^{~~}$} & { 0.5780$^{~~}$} & { 0.5659$^{~~}$} & { 0.5845$^{~~}$} & { 0.5784$^{~~}$} & { 0.6052$^{~~}$}  \\
   MD &  { 0.3471$^{~~}$} & { 0.3733$^{~~}$}  & { 0.3448$^{~~}$} & { 0.3311$^{~~}$} & { 0.3301$^{~~}$} & { 0.3159$^{~~}$} & { 0.2537$^{~~}$} & { 0.2406$^{~~}$}   \smallskip\\
 \multicolumn{9}{l}{Panel G: Idiosyncratic VaR (1\%)} \\
   Raw &  {\textbf{ 0.0024$^{**}$}} & {\textbf{ 0.0021$^{**}$}}  & {\textbf{ 0.0019$^{*~}$}} & {\textbf{ 0.0018$^{*~}$}} & {\textbf{ 0.0017$^{*~}$}} & {\textbf{ 0.0017$^{*~}$}} & {\textbf{ 0.0015$^{*~}$}} & {\textbf{ 0.0012$^{*~}$}}  \\
   $\alpha$ &  {\textbf{ 0.0035$^{**}$}} & {\textbf{ 0.0031$^{**}$}}  & {\textbf{ 0.0029$^{**}$}} & {\textbf{ 0.0027$^{**}$}} & {\textbf{ 0.0024$^{**}$}} & {\textbf{ 0.0022$^{**}$}} & {\textbf{ 0.0018$^{**}$}} & {\textbf{ 0.0014$^{**}$}}  \\
   SR &  { 0.6426$^{~~}$} & { 0.5892$^{~~}$}  & { 0.5572$^{~~}$} & { 0.5485$^{~~}$} & { 0.5375$^{~~}$} & { 0.5772$^{~~}$} & { 0.5987$^{~~}$} & { 0.6092$^{~~}$}  \\
   MD &  { 0.3665$^{~~}$} & { 0.3863$^{~~}$}  & { 0.3508$^{~~}$} & { 0.3363$^{~~}$} & { 0.3335$^{~~}$} & { 0.3124$^{~~}$} & { 0.2522$^{~~}$} & { 0.2511$^{~~}$}  \\
   \hline\hline
   \end{tabular}
   \label{TB:Res2Risk:F5:id}
\end{table}

\textcolor{black}{
We additionally adopt the indirectly adjusted IMOM procedure via conducting the bivariate portfolios analysis based on double sorting on idiosyncratic returns and some specific risk metrics.
The results are presented in \autoref{TB:Res2Risk:F5:id}.
Given no ability of return predictability based on the IKurt, which is suggested by the results in \autoref{TB:Risk:F5}, the case of IKurt-IMOMs is not taken into account here.
In general, the IMOMs have become more profitable after the adjustment by most of idiosyncratic risk metrics.
}

\textcolor{black}{The exceptions are IVol-adjusted IMOMs. In fact, we observe similar and unimproved performance, compared to the results for pure IMOM or IVol-based portfolios. This is consistent with our preceding finding that the IVol and idiosyncratic return function very similarly in stock sorting.
This will also be confirmed by the results in the spanning test in the proceeding section, which indicate that pure IMOMs and the IVol-IMOMs can totally explain each other, with $\beta$ even close to one.
On the other hand, more profitable IVol-based portfolios are usually obtained if higher IVol is accompanied with high level of arbitrage asymmetry \citep{Stambaugh-Yu-Yuan-2015-JF}, limit-to-arbitrage \citep{Gu-Kang-Xu-2018-JBF}, lottery-type demand \citep{Kumar-2009-JF,Kumar-Page-2014-JFQA}, etc.
Essentially, the cumulatively idiosyncratic return serves as idiosyncratic shocks at the firm level. Thus, the performance regarding the IVol-IMOMs is not enhanced as much as expected.}

\textcolor{black}{ By comparison, the ISkew-IMOMs generate considerable profits, compared to the results of ISkew-based portfolios. For instance, when the holding period $K= 1$ week, the ISkew-IMOM generates average return of 0.27\%, which is higher than that of ISkew-based portfolio by 18 basis points (Panel B in \autoref{TB:Risk:F5}). Overall, the profits of the ISkew-IMOMs have gained more than 4 times increase on average, compared to those of ISkew-based portfolios.
This is mainly attributed to more accurate description of lottery-like demand.
Our finding suggests the similar role shared by IVol and idiosyncratic return in stock sorting, which helps to explain the considerable performance regarding the ISkew-IMOMs. As in \cite{Kumar-2009-JF} and \cite{Kumar-Page-2014-JFQA}, ISkew, IVol and other variables are jointly employed to capture the lottery-like demand of the investors.
Although the ISkew is not able to select out the lottery-type stocks independently, adding another indicator---IVol--- would be more helpful.
In this vein, we expect that the profitability regarding the ISkew-IMOMs will be mainly contributed by their short legs, because of significant mean-reverting process associated with the lottery-type stocks. In other words, one should observe much larger negative return drifts or FF5F-$\alpha$ values in the results, which are confirmed by our results in \autoref{TB:Res2Risk:F5:id:long} and \autoref{TB:Res2Risk:F5:id:short}.
Especially, when $K$ is less than four weeks, the short sides of the ISkew-IMOMs contribute to the total magnitudes of FF5F-$\alpha$ by at least 30 basis points, while the long sides provide with less than 5 basis points only. }

\textcolor{black}{As with ISkew-IMOMs, the performance of IMOMs have also been strengthened starkly after indirect adjustment of idiosyncratic tail risk metrics.
For instance, weekly short-term raw returns reach up to around 0.24\% for IMD, IES and IVaR, compared to the results of around 0.17\% in \autoref{TB:Res2Risk:F5:d}.
The FF5F-$\alpha$ values have accordingly increased significantly to greater than 0.30\%.
This is mainly due to the investors' underreaction to bad firm-level information \citep{Hong-Stein-1999-JF,Hong-Lim-Stein-2000-JF,Atilgan-Bali-Demirtas-Gunaydin-2020-JFE}.
The short sides of the IMOMs adjusted by idiosyncratic tail risk metrics involve the stocks from the group with the highest specific risk metric as well as the lowest idiosyncratic return (or the highest IVol, implied by our finding above).
In fact, we argue that the stocks with the lowest idiosyncratic return or the highest IVol generally experienced abnormally plummets during the estimation period.
As shown in \autoref{TB:Res2Risk:F5:id}, the improved profitability regarding these double-sorted portfolios are not completely subsumed solely by IMD, IVaR or IES; since their performance would remain unchanged, otherwise.
In this vein, the idiosyncratic tail risk is assumed to be depicted more accurately in the IMOMs adjusted by the IMD, IVaR and IES.
Therefore, the negative return drifts (or FF5F-$\alpha$ values) are expected to be much larger, since investors are prone to underreact to bad news at the firm level.
As a result, the IMOMs adjusted by idiosyncratic tail risk metrics are expected to be more profitable, which are highlighted by the results in \autoref{TB:Res2Risk:F5:id}.
Overall, when $K$ is less than four weeks, the short sides of IMOMs adjusted by idiosyncratic tail risks contribute to much larger proportion of magnitudes of FF5F-$\alpha$ with more than 20 basis points, which is highlighted by the results in \autoref{TB:Res2Risk:F5:id:short}. In comparison, their long sides provide with less than 10 basis points only, as can be observed in \autoref{TB:Res2Risk:F5:id:long}.
Furthermore, it should be noted that the IVaR and the IES would more likely to be substituted by the measure of combination of the lowest idiosyncratic return and the highest IVol, which functions more comprehensively to capture the extremely tail risk regarding idiosyncratic return.
Accordingly, the IMD-IMOMs are related to a more complete picture of idiosyncratic tail risk and we thus expect more profitable IMD-IMOMs that are largely contributed by their short sides, which are confirmed by the results in \autoref{TB:Res2Risk:F5:id:short}.
}

\textcolor{black}{Although the profitability of most aforementioned portfolios is greatly enhanced, their Sharpe ratios and maximum drawdowns are not improved remarkably, and there are even falls for several cases, such as the IMD.
On balance, according to the Sharpe ratios and the maximum drawdowns, we still arrive to a similar conclusion that IVol-IMOMs and IMD-IMOMs outperform the IMOMs with respect to other risk metrics.}

\subsection{Spanning test for risk-adjusted IMOM}

Likewise, this section is to test whether the IMOM portfolios based on various risk metrics could be explained by the IMOM portfolios adjusted by some specific risk metric. To this end, we conduct the spanning tests with augmented FF5F model that incorporates the IMOM based on each specific risk metric.
We employ the results based on double sorting in \autoref{TB:Res2Risk:F5:id}. We additionally consider the pure MOM portfolios and the IMOM portfolios with varying $K$ and fixed $J$ ($=26$ weeks), which are constructed by merely sorting on raw returns and idiosyncratic returns, respectively.
The results are reported in \autoref{TB:Res2Risk:F5:spanning:p1}.

Accordingly, we can rank different versions of momentum portfolios as follows:
IMOM (1), IVol-IMOM (1), IES1-IMOM (6), IMD-IMOM (10), IVaR5-IMOM (11), IES5-IMOM (17), IVaR5-IMOM (17), and MOM (61), and the numbers of insignificant $\alpha$ values are presented in the parentheses.

Then, we take a closer look at the results.
The pure MOM and IMOM portfolios exhibit completely different results. The MOMs could explain nearly no portfolios regarding the IMOMs and its risk-adjusted versions, except for the ISkew- and the IKurt-IMOM. In fact, although most $\alpha$ values remain insignificant after the FF5F regression with ISkew-IMOMs or IKurt-IMOMs, the corresponding $\beta$ values are not significant yet, also indicating the poor explanatory power associated with the MOMs.
Conversely, the MOMs could be explained by any other version of momentum portfolios.
Compared with the MOMs, the IMOMs perform much better in explaining the MOMs as well as other risk-adjusted IMOMs. Except for the ISkew-IMOMs and IKurt-IMOMs, insignificant $\alpha$ values approach zero in results of other risk metrics, indicating the good explanatory power with respect to the IMOMs.
Very similar results are obtained for the IVol-IMOMs, which share almost identical $\alpha$ values and $\beta$ values with those of the IMOMs, in line with our findings mentioned above.
Also, the ISkew-IMOMs and IKurt-IMOMs are still not capable to explain the others, except for the powerful explanation delivered to the MOMs.
As for the IMOMs with respect to the idiosyncratic tail risk metrics, they also demonstrate good explanatory power to each other. Moreover, the IES1-IMOMs and IMD-IMOMs could explain most of the IMOMs and IVol-IMOMs. In particularly, the IES1-IMOMs exhibit such better explanatory power that IMOMs and IVol-IMD could be completely explained, with the only exception of IVol-IMOM when $K =2$ weeks.

In summary, we conclude that IVol-IMOMs, IMD-IMOMs and IES1-IMOMs exhibit better explanatory power to others, which is similar to the finding in the spanning test for idiosyncratic risk metrics.

\subsection{Market state, illiquility and sentiment}

Based upon the results in previous sections, we do find that IVol and IMD adjustments could  give rise to more favorable IMOM performance compared to pure IMOM, especially in the bivariate portfolio analysis. Moreover, IVol, IMD, IVol-IMOMs and IMD-IMOMs perform better in the spanning tests.
We additionally find the superior performance associated with pure IMOM as well as other risk-based portfolios, such as IVol and IMD.
The profitability regarding aforementioned portfolios survived in the FF5F regression, and we further attempt to explore if other factors that are not incorporated in the FF5F, could explain IVol-IMOMs and IMD-IMOMs as well as pure IMOMs. Specifically, we wonder whether their performance differs during the periods featured by the factors associated with certain market environment.
Therefore, we consider following three factors: Market state, market illiquidity and investor sentiment, which have been documented to be closely linked to momentum performance
\citep{Cooper-Gutierrez-Hameed-2004-JF,Moskowitz-Ooi-Pedersen-2012-JFE,Antoniou-Doukas-Subrahmanyam-2013-JFQA,Avramov-Cheng-Hameed-2016-JFQA}.
The data that are employed to construct the three aforementioned factors are also retrieved from the CSMAR database.

Given that the momentum acts as the trend strategy, we first consider the market states. Following \cite{Cooper-Gutierrez-Hameed-2004-JF}, the upside (downside) market state at week $t$ is defined as the periods with the cumulative return of Shanghai Composite Index (SHCI) over the past 26 or 52 weeks being positive (negative).

\begin{landscape}
\begin{table}[!ht]
\setlength\tabcolsep{0pt}
\footnotesize
\caption{Spanning test for risk-adjusted IMOM portfolios. Augmented FF5F-regression is conducted to examine if the profitability of various risk-based IMOM portfolios could be explained by one of others. Variable X in the first column refers to as the profitability of the X incorporated in the FF5F. Column 2 presents $K$'s value. Columns 3 to 8 report $\alpha$ values of explained variables that are IMOMs based on the rest of risk metrics, and $\beta$ values of explanatory variable X. \cite{Newey-West-1987-Em}'s $t$-statistics are presented in parentheses and the superscripts * and ** denote the significance at 5\% and 1\% levels, respectively. }
\centering
\vspace{-3mm}
   \begin{tabular}{cccccccccccccccccccccc}
   \hline\hline
    Explanatory  & \multirow{2}{*}{$K$} & \multirow{2}{*}{$\alpha$(RR)}  &\multirow{2}{*}{$\beta_X$} & \multirow{2}{*}{$\alpha$(IR)}  &\multirow{2}{*}{$\beta_X$} & \multirow{2}{*}{$\alpha$(IVol)}  &\multirow{2}{*}{$\beta_X$}  &  \multirow{2}{*}{$\alpha$(ISkew)}  &\multirow{2}{*}{$\beta_X$} & \multirow{2}{*}{$\alpha$(IKurt)}  &\multirow{2}{*}{$\beta_X$} & \multirow{2}{*}{$\alpha$(IMD) } &\multirow{2}{*}{$\beta_X$} & \multirow{2}{*}{$\alpha$(IES5)} & \multirow{2}{*}{$\beta_X$ } & \multirow{2}{*}{$\alpha$(IVaR5)} & \multirow{2}{*}{$\beta_X$ } & \multirow{2}{*}{$\alpha$(IES1)} & \multirow{2}{*}{$\beta_X$ } &\multirow{2}{*}{$\alpha$(IVaR1)} & \multirow{2}{*}{$\beta_X$} \\
   variable X  & & & & &  & & &  &  &  &  &  & & & & & & & & & \\
   \hline
   MOM &  1 & { $^{~~}$} & { $^{~~}$} & {\textbf{ -0.0030$^{**}$}} & {\textbf{ -0.19$^{**}$}}  & {\textbf{ -0.0031$^{**}$}} & {\textbf{ -0.18$^{**}$}} & {\textbf{ -0.0036$^{**}$}} &  {  0.02$^{~~}$} & { 0.0009$^{~~}$} &  { -0.07$^{~~}$}& {\textbf{ -0.0029$^{**}$}} &  {\textbf{ -0.22$^{**}$}} & {\textbf{ -0.0030$^{**}$}} &  {\textbf{ -0.21$^{**}$}}  & {\textbf{ -0.0030$^{**}$}} & {\textbf{ -0.19$^{**}$}} & {\textbf{ -0.0032$^{**}$}} & {\textbf{ -0.22$^{**}$}} & {\textbf{ -0.0031$^{**}$}} & {\textbf{ -0.23$^{**}$}}  \\
    &  2 & { $^{~~}$} & { $^{~~}$} & {\textbf{ -0.0028$^{**}$}} & {\textbf{ -0.18$^{**}$}}  & {\textbf{ -0.0028$^{**}$}} & {\textbf{ -0.18$^{**}$}} & {\textbf{ -0.0038$^{**}$}} &  { -0.05$^{~~}$} & { 0.0017$^{~~}$} &  { -0.10$^{~~}$}& {\textbf{ -0.0029$^{**}$}} &  {\textbf{ -0.21$^{**}$}} & {\textbf{ -0.0028$^{**}$}} &  {\textbf{ -0.20$^{**}$}}  & {\textbf{ -0.0029$^{**}$}} & {\textbf{ -0.18$^{**}$}} & {\textbf{ -0.0029$^{**}$}} & {\textbf{ -0.21$^{**}$}} & {\textbf{ -0.0028$^{**}$}} & {\textbf{ -0.21$^{**}$}}  \\
    &  3 & { $^{~~}$} & { $^{~~}$} & {\textbf{ -0.0026$^{**}$}} & {\textbf{ -0.18$^{**}$}}  & {\textbf{ -0.0027$^{**}$}} & {\textbf{ -0.18$^{**}$}} & {\textbf{ -0.0045$^{**}$}} &  { -0.05$^{~~}$} & { 0.0015$^{~~}$} &  { -0.09$^{~~}$}& {\textbf{ -0.0027$^{**}$}} &  {\textbf{ -0.21$^{**}$}} & {\textbf{ -0.0026$^{**}$}} &  {\textbf{ -0.19$^{**}$}}  & {\textbf{ -0.0027$^{**}$}} & {\textbf{ -0.18$^{**}$}} & {\textbf{ -0.0027$^{**}$}} & {\textbf{ -0.20$^{**}$}} & {\textbf{ -0.0026$^{**}$}} & {\textbf{ -0.20$^{**}$}}  \\
    &  4 & { $^{~~}$} & { $^{~~}$} & {\textbf{ -0.0025$^{**}$}} & {\textbf{ -0.16$^{**}$}}  & {\textbf{ -0.0025$^{**}$}} & {\textbf{ -0.16$^{**}$}} & {\textbf{ -0.0037$^{**}$}} &  { -0.08$^{~~}$} & { 0.0001$^{~~}$} &  { -0.03$^{~~}$}& {\textbf{ -0.0027$^{**}$}} &  {\textbf{ -0.19$^{**}$}} & {\textbf{ -0.0025$^{**}$}} &  {\textbf{ -0.18$^{**}$}}  & {\textbf{ -0.0026$^{**}$}} & {\textbf{ -0.16$^{**}$}} & {\textbf{ -0.0026$^{**}$}} & {\textbf{ -0.18$^{**}$}} & {\textbf{ -0.0024$^{**}$}} & {\textbf{ -0.19$^{**}$}}  \\
    &  8 & { $^{~~}$} & { $^{~~}$} & {\textbf{ -0.0022$^{**}$}} & {\textbf{ -0.17$^{**}$}}  & {\textbf{ -0.0023$^{**}$}} & {\textbf{ -0.17$^{**}$}} & {\textbf{ -0.0023$^{**}$}} &  { -0.04$^{~~}$} & { 0.0001$^{~~}$} &  {  0.02$^{~~}$}& {\textbf{ -0.0024$^{**}$}} &  {\textbf{ -0.19$^{**}$}} & {\textbf{ -0.0022$^{**}$}} &  {\textbf{ -0.18$^{**}$}}  & {\textbf{ -0.0024$^{**}$}} & {\textbf{ -0.17$^{**}$}} & {\textbf{ -0.0023$^{**}$}} & {\textbf{ -0.17$^{**}$}} & {\textbf{ -0.0022$^{**}$}} & {\textbf{ -0.18$^{**}$}}  \\
    &  13 & { $^{~~}$} & { $^{~~}$} & {\textbf{ -0.0021$^{**}$}} & {\textbf{ -0.16$^{**}$}}  & {\textbf{ -0.0021$^{**}$}} & {\textbf{ -0.16$^{**}$}} & { -0.0012$^{~~}$} &  { -0.05$^{~~}$} & { -0.0001$^{~~}$} &  {  0.09$^{~~}$}& {\textbf{ -0.0023$^{**}$}} &  {\textbf{ -0.18$^{**}$}} & {\textbf{ -0.0021$^{**}$}} &  {\textbf{ -0.16$^{**}$}}  & {\textbf{ -0.0022$^{**}$}} & {\textbf{ -0.16$^{**}$}} & {\textbf{ -0.0022$^{**}$}} & {\textbf{ -0.16$^{**}$}} & {\textbf{ -0.0021$^{**}$}} & {\textbf{ -0.17$^{**}$}}  \\
    &  26 & { $^{~~}$} & { $^{~~}$} & {\textbf{ -0.0017$^{**}$}} & {\textbf{ -0.14$^{**}$}}  & {\textbf{ -0.0018$^{**}$}} & {\textbf{ -0.14$^{**}$}} & { -0.0010$^{~~}$} &  {  0.04$^{~~}$} & { -0.0015$^{~~}$} &  {  0.06$^{~~}$}& {\textbf{ -0.0020$^{**}$}} &  {\textbf{ -0.16$^{**}$}} & {\textbf{ -0.0018$^{**}$}} &  {\textbf{ -0.15$^{*~}$}}  & {\textbf{ -0.0019$^{**}$}} & {\textbf{ -0.16$^{**}$}} & {\textbf{ -0.0017$^{**}$}} & {\textbf{ -0.13$^{*~}$}} & {\textbf{ -0.0018$^{**}$}} & {\textbf{ -0.14$^{*~}$}}  \\
    &  52 & { $^{~~}$} & { $^{~~}$} & {\textbf{ -0.0014$^{**}$}} & { -0.10$^{~~}$}  & {\textbf{ -0.0014$^{**}$}} & { -0.10$^{~~}$} & { -0.0008$^{~~}$} &  {  0.04$^{~~}$} & { -0.0015$^{~~}$} &  {  0.10$^{~~}$}& {\textbf{ -0.0016$^{**}$}} &  {\textbf{ -0.12$^{*~}$}} & {\textbf{ -0.0013$^{**}$}} &  { -0.10$^{~~}$}  & {\textbf{ -0.0014$^{**}$}} & { -0.11$^{~~}$} & {\textbf{ -0.0013$^{**}$}} & { -0.09$^{~~}$} & {\textbf{ -0.0013$^{**}$}} & { -0.09$^{~~}$}  \smallskip\\
   IMOM &  1 & { 0.0004$^{~~}$} & {\textbf{ -0.43$^{**}$}} & { $^{~~}$} & { $^{~~}$}  & { -0.0001$^{~~}$} & {\textbf{  1.01$^{**}$}} & { -0.0012$^{~~}$} &  {\textbf{  0.78$^{**}$}} & { 0.0022$^{~~}$} &  {\textbf{  0.77$^{**}$}}& { 0.0003$^{~~}$} &  {\textbf{  1.09$^{**}$}} & { 0.0001$^{~~}$} &  {\textbf{  1.07$^{**}$}}  & { 0.0002$^{~~}$} & {\textbf{  1.10$^{**}$}} & { -0.0001$^{~~}$} & {\textbf{  1.04$^{**}$}} & { 0.0001$^{~~}$} & {\textbf{  1.08$^{**}$}}  \\
    &  2 & { 0.0004$^{~~}$} & {\textbf{ -0.41$^{**}$}} & { $^{~~}$} & { $^{~~}$}  & { -0.0000$^{~~}$} & {\textbf{  1.01$^{**}$}} & { -0.0017$^{~~}$} &  {\textbf{  0.76$^{**}$}} & { 0.0032$^{~~}$} &  {\textbf{  0.87$^{**}$}}& { 0.0002$^{~~}$} &  {\textbf{  1.09$^{**}$}} & { 0.0002$^{~~}$} &  {\textbf{  1.07$^{**}$}}  & { 0.0002$^{~~}$} & {\textbf{  1.10$^{**}$}} & { -0.0000$^{~~}$} & {\textbf{  1.04$^{**}$}} & { 0.0002$^{~~}$} & {\textbf{  1.08$^{**}$}}  \\
    &  3 & { 0.0004$^{~~}$} & {\textbf{ -0.40$^{**}$}} & { $^{~~}$} & { $^{~~}$}  & { -0.0000$^{~~}$} & {\textbf{  1.01$^{**}$}} & {\textbf{ -0.0023$^{*~}$}} &  {\textbf{  0.83$^{**}$}} & { 0.0030$^{~~}$} &  {\textbf{  0.88$^{**}$}}& { 0.0001$^{~~}$} &  {\textbf{  1.09$^{**}$}} & { 0.0002$^{~~}$} &  {\textbf{  1.07$^{**}$}}  & { 0.0002$^{~~}$} & {\textbf{  1.09$^{**}$}} & { 0.0000$^{~~}$} & {\textbf{  1.04$^{**}$}} & { 0.0003$^{~~}$} & {\textbf{  1.08$^{**}$}}  \\
    &  4 & { 0.0004$^{~~}$} & {\textbf{ -0.37$^{**}$}} & { $^{~~}$} & { $^{~~}$}  & { -0.0000$^{~~}$} & {\textbf{  1.01$^{**}$}} & { -0.0014$^{~~}$} &  {\textbf{  0.82$^{**}$}} & { 0.0016$^{~~}$} &  {\textbf{  0.84$^{**}$}}& { -0.0000$^{~~}$} &  {\textbf{  1.09$^{**}$}} & { 0.0002$^{~~}$} &  {\textbf{  1.07$^{**}$}}  & { 0.0002$^{~~}$} & {\textbf{  1.09$^{**}$}} & { -0.0000$^{~~}$} & {\textbf{  1.04$^{**}$}} & { 0.0003$^{~~}$} & {\textbf{  1.09$^{**}$}}  \\
    &  8 & { 0.0002$^{~~}$} & {\textbf{ -0.34$^{**}$}} & { $^{~~}$} & { $^{~~}$}  & { -0.0000$^{~~}$} & {\textbf{  1.01$^{**}$}} & { -0.0004$^{~~}$} &  {\textbf{  0.74$^{**}$}} & { 0.0014$^{~~}$} &  {\textbf{  0.75$^{**}$}}& { -0.0001$^{~~}$} &  {\textbf{  1.08$^{**}$}} & { 0.0002$^{~~}$} &  {\textbf{  1.06$^{**}$}}  & { 0.0001$^{~~}$} & {\textbf{  1.09$^{**}$}} & { -0.0000$^{~~}$} & {\textbf{  1.03$^{**}$}} & { 0.0002$^{~~}$} & {\textbf{  1.08$^{**}$}}  \\
    &  13 & { -0.0002$^{~~}$} & {\textbf{ -0.30$^{**}$}} & { $^{~~}$} & { $^{~~}$}  & { -0.0000$^{~~}$} & {\textbf{  1.01$^{**}$}} & { 0.0005$^{~~}$} &  {\textbf{  0.78$^{**}$}} & { 0.0013$^{~~}$} &  {\textbf{  0.78$^{**}$}}& { -0.0001$^{~~}$} &  {\textbf{  1.07$^{**}$}} & { 0.0001$^{~~}$} &  {\textbf{  1.07$^{**}$}}  & { 0.0001$^{~~}$} & {\textbf{  1.09$^{**}$}} & { -0.0000$^{~~}$} & {\textbf{  1.04$^{**}$}} & { 0.0001$^{~~}$} & {\textbf{  1.08$^{**}$}}  \\
    &  26 & { -0.0005$^{~~}$} & {\textbf{ -0.23$^{**}$}} & { $^{~~}$} & { $^{~~}$}  & { -0.0000$^{~~}$} & {\textbf{  1.01$^{**}$}} & { 0.0002$^{~~}$} &  {\textbf{  0.69$^{**}$}} & { -0.0003$^{~~}$} &  {\textbf{  0.83$^{**}$}}& { -0.0001$^{~~}$} &  {\textbf{  1.07$^{**}$}} & { 0.0001$^{~~}$} &  {\textbf{  1.08$^{**}$}}  & { 0.0000$^{~~}$} & {\textbf{  1.10$^{**}$}} & { 0.0001$^{~~}$} & {\textbf{  1.06$^{**}$}} & { 0.0001$^{~~}$} & {\textbf{  1.10$^{**}$}}  \\
    &  52 & { 0.0003$^{~~}$} & {\textbf{ -0.13$^{*~}$}} & { $^{~~}$} & { $^{~~}$}  & { -0.0000$^{~~}$} & {\textbf{  1.01$^{**}$}} & { 0.0002$^{~~}$} &  {\textbf{  0.71$^{**}$}} & { -0.0006$^{~~}$} &  {\textbf{  0.63$^{**}$}}& { -0.0001$^{~~}$} &  {\textbf{  1.06$^{**}$}} & { 0.0001$^{~~}$} &  {\textbf{  1.07$^{**}$}}  & { 0.0001$^{~~}$} & {\textbf{  1.09$^{**}$}} & { 0.0001$^{~~}$} & {\textbf{  1.04$^{**}$}} & { 0.0002$^{~~}$} & {\textbf{  1.08$^{**}$}}
      \smallskip\\
   IVol-IMOM &  1 & { 0.0004$^{~~}$} & {\textbf{ -0.42$^{**}$}} & { 0.0001$^{~~}$} & {\textbf{  0.99$^{**}$}}  & { $^{~~}$} & { $^{~~}$} & { -0.0012$^{~~}$} &  {\textbf{  0.77$^{**}$}} & { 0.0021$^{~~}$} &  {\textbf{  0.72$^{**}$}}& { 0.0004$^{~~}$} &  {\textbf{  1.08$^{**}$}} & { 0.0002$^{~~}$} &  {\textbf{  1.06$^{**}$}}  & {\textbf{ 0.0003$^{*~}$}} & {\textbf{  1.09$^{**}$}} & { -0.0001$^{~~}$} & {\textbf{  1.03$^{**}$}} & { 0.0002$^{~~}$} & {\textbf{  1.07$^{**}$}}  \\
    &  2 & { 0.0004$^{~~}$} & {\textbf{ -0.40$^{**}$}} & { 0.0000$^{~~}$} & {\textbf{  0.99$^{**}$}}  & { $^{~~}$} & { $^{~~}$} & { -0.0017$^{~~}$} &  {\textbf{  0.75$^{**}$}} & { 0.0031$^{~~}$} &  {\textbf{  0.85$^{**}$}}& { 0.0002$^{~~}$} &  {\textbf{  1.08$^{**}$}} & { 0.0002$^{~~}$} &  {\textbf{  1.06$^{**}$}}  & { 0.0002$^{~~}$} & {\textbf{  1.09$^{**}$}} & { -0.0000$^{~~}$} & {\textbf{  1.03$^{**}$}} & { 0.0002$^{~~}$} & {\textbf{  1.07$^{**}$}}  \\
    &  3 & { 0.0004$^{~~}$} & {\textbf{ -0.39$^{**}$}} & { -0.0000$^{~~}$} & {\textbf{  0.99$^{**}$}}  & { $^{~~}$} & { $^{~~}$} & {\textbf{ -0.0023$^{*~}$}} &  {\textbf{  0.83$^{**}$}} & { 0.0030$^{~~}$} &  {\textbf{  0.86$^{**}$}}& { 0.0001$^{~~}$} &  {\textbf{  1.08$^{**}$}} & { 0.0002$^{~~}$} &  {\textbf{  1.05$^{**}$}}  & { 0.0002$^{~~}$} & {\textbf{  1.08$^{**}$}} & { 0.0000$^{~~}$} & {\textbf{  1.03$^{**}$}} & { 0.0003$^{~~}$} & {\textbf{  1.07$^{**}$}}  \\
    &  4 & { 0.0004$^{~~}$} & {\textbf{ -0.36$^{**}$}} & { -0.0000$^{~~}$} & {\textbf{  0.99$^{**}$}}  & { $^{~~}$} & { $^{~~}$} & { -0.0014$^{~~}$} &  {\textbf{  0.82$^{**}$}} & { 0.0016$^{~~}$} &  {\textbf{  0.83$^{**}$}}& { 0.0000$^{~~}$} &  {\textbf{  1.07$^{**}$}} & { 0.0002$^{~~}$} &  {\textbf{  1.05$^{**}$}}  & { 0.0002$^{~~}$} & {\textbf{  1.08$^{**}$}} & { -0.0000$^{~~}$} & {\textbf{  1.02$^{**}$}} & { 0.0003$^{~~}$} & {\textbf{  1.07$^{**}$}}  \\
    &  8 & { 0.0002$^{~~}$} & {\textbf{ -0.33$^{**}$}} & { 0.0000$^{~~}$} & {\textbf{  0.99$^{**}$}}  & { $^{~~}$} & { $^{~~}$} & { -0.0004$^{~~}$} &  {\textbf{  0.74$^{**}$}} & { 0.0014$^{~~}$} &  {\textbf{  0.73$^{**}$}}& { -0.0000$^{~~}$} &  {\textbf{  1.07$^{**}$}} & { 0.0002$^{~~}$} &  {\textbf{  1.05$^{**}$}}  & { 0.0001$^{~~}$} & {\textbf{  1.08$^{**}$}} & { -0.0000$^{~~}$} & {\textbf{  1.02$^{**}$}} & { 0.0002$^{~~}$} & {\textbf{  1.07$^{**}$}}  \\
    &  13 & { -0.0002$^{~~}$} & {\textbf{ -0.29$^{**}$}} & { 0.0000$^{~~}$} & {\textbf{  0.99$^{**}$}}  & { $^{~~}$} & { $^{~~}$} & { 0.0005$^{~~}$} &  {\textbf{  0.77$^{**}$}} & { 0.0013$^{~~}$} &  {\textbf{  0.77$^{**}$}}& { -0.0001$^{~~}$} &  {\textbf{  1.06$^{**}$}} & { 0.0001$^{~~}$} &  {\textbf{  1.06$^{**}$}}  & { 0.0001$^{~~}$} & {\textbf{  1.08$^{**}$}} & { -0.0000$^{~~}$} & {\textbf{  1.03$^{**}$}} & { 0.0001$^{~~}$} & {\textbf{  1.07$^{**}$}}  \\
    &  26 & { -0.0005$^{~~}$} & {\textbf{ -0.22$^{**}$}} & { 0.0000$^{~~}$} & {\textbf{  0.99$^{**}$}}  & { $^{~~}$} & { $^{~~}$} & { 0.0002$^{~~}$} &  {\textbf{  0.68$^{**}$}} & { -0.0003$^{~~}$} &  {\textbf{  0.82$^{**}$}}& { -0.0001$^{~~}$} &  {\textbf{  1.06$^{**}$}} & { 0.0001$^{~~}$} &  {\textbf{  1.07$^{**}$}}  & { 0.0000$^{~~}$} & {\textbf{  1.08$^{**}$}} & { 0.0001$^{~~}$} & {\textbf{  1.05$^{**}$}} & { 0.0001$^{~~}$} & {\textbf{  1.09$^{**}$}}  \\
    &  52 & { 0.0003$^{~~}$} & {\textbf{ -0.13$^{*~}$}} & { 0.0000$^{~~}$} & {\textbf{  0.99$^{**}$}}  & { $^{~~}$} & { $^{~~}$} & { 0.0002$^{~~}$} &  {\textbf{  0.71$^{**}$}} & { -0.0006$^{~~}$} &  {\textbf{  0.62$^{**}$}}& { -0.0001$^{~~}$} &  {\textbf{  1.05$^{**}$}} & { 0.0001$^{~~}$} &  {\textbf{  1.06$^{**}$}}  & { 0.0001$^{~~}$} & {\textbf{  1.08$^{**}$}} & { 0.0001$^{~~}$} & {\textbf{  1.02$^{**}$}} & { 0.0002$^{~~}$} & {\textbf{  1.07$^{**}$}}
      \smallskip\\
   ISkew-IMOM &  1 & {\textbf{ 0.0021$^{*~}$}} & {  0.01$^{~~}$} & {\textbf{ -0.0025$^{**}$}} & {\textbf{  0.14$^{**}$}}  & {\textbf{ -0.0026$^{**}$}} & {\textbf{  0.14$^{**}$}} & { $^{~~}$} &  { $^{~~}$} & { 0.0020$^{~~}$} &  {\textbf{  0.73$^{**}$}}& {\textbf{ -0.0023$^{**}$}} &  {\textbf{  0.14$^{**}$}} & {\textbf{ -0.0027$^{**}$}} &  {\textbf{  0.12$^{**}$}}  & {\textbf{ -0.0026$^{**}$}} & {\textbf{  0.13$^{**}$}} & {\textbf{ -0.0029$^{**}$}} & {\textbf{  0.12$^{**}$}} & {\textbf{ -0.0026$^{**}$}} & {\textbf{  0.11$^{**}$}}  \\
    &  2 & { 0.0015$^{~~}$} & { -0.02$^{~~}$} & {\textbf{ -0.0023$^{**}$}} & {\textbf{  0.14$^{**}$}}  & {\textbf{ -0.0023$^{**}$}} & {\textbf{  0.14$^{**}$}} & { $^{~~}$} &  { $^{~~}$} & { 0.0028$^{~~}$} &  {\textbf{  0.71$^{**}$}}& {\textbf{ -0.0023$^{**}$}} &  {\textbf{  0.15$^{**}$}} & {\textbf{ -0.0023$^{**}$}} &  {\textbf{  0.12$^{**}$}}  & {\textbf{ -0.0024$^{**}$}} & {\textbf{  0.13$^{**}$}} & {\textbf{ -0.0024$^{**}$}} & {\textbf{  0.12$^{**}$}} & {\textbf{ -0.0023$^{**}$}} & {\textbf{  0.12$^{**}$}}  \\
    &  3 & { 0.0014$^{~~}$} & { -0.03$^{~~}$} & {\textbf{ -0.0019$^{**}$}} & {\textbf{  0.18$^{**}$}}  & {\textbf{ -0.0019$^{**}$}} & {\textbf{  0.18$^{**}$}} & { $^{~~}$} &  { $^{~~}$} & { 0.0027$^{~~}$} &  {\textbf{  0.76$^{**}$}}& {\textbf{ -0.0020$^{**}$}} &  {\textbf{  0.19$^{**}$}} & {\textbf{ -0.0019$^{**}$}} &  {\textbf{  0.16$^{**}$}}  & {\textbf{ -0.0020$^{**}$}} & {\textbf{  0.17$^{**}$}} & {\textbf{ -0.0020$^{**}$}} & {\textbf{  0.16$^{**}$}} & {\textbf{ -0.0018$^{**}$}} & {\textbf{  0.16$^{**}$}}  \\
    &  4 & { 0.0015$^{~~}$} & { -0.04$^{~~}$} & {\textbf{ -0.0022$^{**}$}} & {\textbf{  0.19$^{**}$}}  & {\textbf{ -0.0022$^{**}$}} & {\textbf{  0.20$^{**}$}} & { $^{~~}$} &  { $^{~~}$} & { 0.0008$^{~~}$} &  {\textbf{  0.68$^{**}$}}& {\textbf{ -0.0024$^{**}$}} &  {\textbf{  0.21$^{**}$}} & {\textbf{ -0.0022$^{**}$}} &  {\textbf{  0.18$^{**}$}}  & {\textbf{ -0.0023$^{**}$}} & {\textbf{  0.19$^{**}$}} & {\textbf{ -0.0023$^{**}$}} & {\textbf{  0.18$^{**}$}} & {\textbf{ -0.0022$^{**}$}} & {\textbf{  0.18$^{**}$}}  \\
    &  8 & { 0.0011$^{~~}$} & { -0.03$^{~~}$} & {\textbf{ -0.0021$^{**}$}} & {\textbf{  0.22$^{**}$}}  & {\textbf{ -0.0021$^{**}$}} & {\textbf{  0.23$^{**}$}} & { $^{~~}$} &  { $^{~~}$} & { 0.0009$^{~~}$} &  {\textbf{  0.68$^{**}$}}& {\textbf{ -0.0023$^{**}$}} &  {\textbf{  0.23$^{**}$}} & {\textbf{ -0.0021$^{**}$}} &  {\textbf{  0.20$^{**}$}}  & {\textbf{ -0.0022$^{**}$}} & {\textbf{  0.22$^{**}$}} & {\textbf{ -0.0022$^{**}$}} & {\textbf{  0.20$^{**}$}} & {\textbf{ -0.0021$^{**}$}} & {\textbf{  0.21$^{**}$}}  \\
    &  13 & { 0.0005$^{~~}$} & { -0.02$^{~~}$} & {\textbf{ -0.0020$^{**}$}} & {\textbf{  0.20$^{**}$}}  & {\textbf{ -0.0020$^{**}$}} & {\textbf{  0.20$^{**}$}} & { $^{~~}$} &  { $^{~~}$} & { 0.0008$^{~~}$} &  {\textbf{  0.65$^{**}$}}& {\textbf{ -0.0022$^{**}$}} &  {\textbf{  0.21$^{**}$}} & {\textbf{ -0.0020$^{**}$}} &  {\textbf{  0.19$^{**}$}}  & {\textbf{ -0.0021$^{**}$}} & {\textbf{  0.20$^{**}$}} & {\textbf{ -0.0021$^{**}$}} & {\textbf{  0.19$^{**}$}} & {\textbf{ -0.0020$^{**}$}} & {\textbf{  0.19$^{**}$}}  \\
    &  26 & { -0.0000$^{~~}$} & {  0.02$^{~~}$} & {\textbf{ -0.0015$^{**}$}} & {\textbf{  0.25$^{**}$}}  & {\textbf{ -0.0015$^{**}$}} & {\textbf{  0.25$^{**}$}} & { $^{~~}$} &  { $^{~~}$} & { -0.0009$^{~~}$} &  {\textbf{  0.67$^{**}$}}& {\textbf{ -0.0017$^{**}$}} &  {\textbf{  0.26$^{**}$}} & {\textbf{ -0.0015$^{**}$}} &  {\textbf{  0.24$^{**}$}}  & {\textbf{ -0.0016$^{**}$}} & {\textbf{  0.25$^{**}$}} & {\textbf{ -0.0015$^{**}$}} & {\textbf{  0.24$^{**}$}} & {\textbf{ -0.0015$^{**}$}} & {\textbf{  0.24$^{**}$}}  \\
    &  52 & { 0.0005$^{~~}$} & {  0.03$^{~~}$} & {\textbf{ -0.0011$^{**}$}} & {\textbf{  0.39$^{**}$}}  & {\textbf{ -0.0011$^{**}$}} & {\textbf{  0.40$^{**}$}} & { $^{~~}$} &  { $^{~~}$} & { -0.0011$^{~~}$} &  {\textbf{  0.65$^{**}$}}& {\textbf{ -0.0013$^{**}$}} &  {\textbf{  0.40$^{**}$}} & {\textbf{ -0.0011$^{**}$}} &  {\textbf{  0.38$^{**}$}}  & {\textbf{ -0.0012$^{**}$}} & {\textbf{  0.40$^{**}$}} & {\textbf{ -0.0011$^{**}$}} & {\textbf{  0.38$^{**}$}} & {\textbf{ -0.0011$^{*~}$}} & {\textbf{  0.38$^{**}$}}
      \smallskip\\
   IKurt-IMOM &  1 & { 0.0005$^{~~}$} & { -0.02$^{~~}$} & {\textbf{ -0.0018$^{**}$}} & {\textbf{  0.07$^{**}$}}  & {\textbf{ -0.0018$^{**}$}} & {\textbf{  0.06$^{**}$}} & { -0.0017$^{~~}$} &  {\textbf{  0.30$^{**}$}} & { $^{~~}$} &  { $^{~~}$}& { -0.0015$^{~~}$} &  {\textbf{  0.06$^{**}$}} & {\textbf{ -0.0017$^{*~}$}} &  {\textbf{  0.07$^{**}$}}  & { -0.0014$^{~~}$} & {\textbf{  0.05$^{*~}$}} & {\textbf{ -0.0017$^{*~}$}} & {\textbf{  0.08$^{**}$}} & {\textbf{ -0.0018$^{*~}$}} & {\textbf{  0.07$^{**}$}}  \\
    &  2 & { 0.0010$^{~~}$} & { -0.03$^{~~}$} & {\textbf{ -0.0019$^{**}$}} & {\textbf{  0.08$^{**}$}}  & {\textbf{ -0.0019$^{**}$}} & {\textbf{  0.08$^{**}$}} & { -0.0022$^{~~}$} &  {\textbf{  0.28$^{**}$}} & { $^{~~}$} &  { $^{~~}$}& {\textbf{ -0.0019$^{*~}$}} &  {\textbf{  0.08$^{**}$}} & {\textbf{ -0.0017$^{*~}$}} &  {\textbf{  0.08$^{**}$}}  & {\textbf{ -0.0016$^{*~}$}} & {\textbf{  0.07$^{**}$}} & {\textbf{ -0.0018$^{*~}$}} & {\textbf{  0.09$^{**}$}} & {\textbf{ -0.0017$^{*~}$}} & {\textbf{  0.09$^{**}$}}  \\
    &  3 & { 0.0011$^{~~}$} & { -0.03$^{~~}$} & {\textbf{ -0.0019$^{**}$}} & {\textbf{  0.09$^{**}$}}  & {\textbf{ -0.0019$^{**}$}} & {\textbf{  0.09$^{**}$}} & {\textbf{ -0.0022$^{*~}$}} &  {\textbf{  0.29$^{**}$}} & { $^{~~}$} &  { $^{~~}$}& {\textbf{ -0.0019$^{**}$}} &  {\textbf{  0.08$^{**}$}} & {\textbf{ -0.0018$^{**}$}} &  {\textbf{  0.09$^{**}$}}  & {\textbf{ -0.0017$^{*~}$}} & {\textbf{  0.09$^{**}$}} & {\textbf{ -0.0019$^{**}$}} & {\textbf{  0.10$^{**}$}} & {\textbf{ -0.0017$^{*~}$}} & {\textbf{  0.09$^{**}$}}  \\
    &  4 & { 0.0014$^{~~}$} & { -0.01$^{~~}$} & {\textbf{ -0.0018$^{**}$}} & {\textbf{  0.10$^{**}$}}  & {\textbf{ -0.0018$^{**}$}} & {\textbf{  0.10$^{**}$}} & { -0.0011$^{~~}$} &  {\textbf{  0.30$^{**}$}} & { $^{~~}$} &  { $^{~~}$}& {\textbf{ -0.0019$^{**}$}} &  {\textbf{  0.09$^{**}$}} & {\textbf{ -0.0017$^{**}$}} &  {\textbf{  0.10$^{**}$}}  & {\textbf{ -0.0016$^{*~}$}} & {\textbf{  0.10$^{**}$}} & {\textbf{ -0.0018$^{**}$}} & {\textbf{  0.10$^{**}$}} & {\textbf{ -0.0016$^{*~}$}} & {\textbf{  0.11$^{**}$}}  \\
    &  8 & { 0.0015$^{~~}$} & {  0.01$^{~~}$} & {\textbf{ -0.0017$^{**}$}} & {\textbf{  0.11$^{**}$}}  & {\textbf{ -0.0018$^{**}$}} & {\textbf{  0.11$^{**}$}} & { -0.0011$^{~~}$} &  {\textbf{  0.29$^{**}$}} & { $^{~~}$} &  { $^{~~}$}& {\textbf{ -0.0019$^{**}$}} &  {\textbf{  0.09$^{**}$}} & {\textbf{ -0.0016$^{**}$}} &  {\textbf{  0.11$^{**}$}}  & {\textbf{ -0.0017$^{**}$}} & {\textbf{  0.10$^{**}$}} & {\textbf{ -0.0019$^{**}$}} & {\textbf{  0.11$^{**}$}} & {\textbf{ -0.0015$^{*~}$}} & {\textbf{  0.12$^{**}$}}  \\
    &  13 & { 0.0010$^{~~}$} & {  0.03$^{~~}$} & {\textbf{ -0.0016$^{**}$}} & {\textbf{  0.13$^{**}$}}  & {\textbf{ -0.0017$^{**}$}} & {\textbf{  0.13$^{**}$}} & { -0.0013$^{~~}$} &  {\textbf{  0.26$^{**}$}} & { $^{~~}$} &  { $^{~~}$}& {\textbf{ -0.0018$^{**}$}} &  {\textbf{  0.12$^{**}$}} & {\textbf{ -0.0015$^{**}$}} &  {\textbf{  0.13$^{**}$}}  & {\textbf{ -0.0016$^{**}$}} & {\textbf{  0.13$^{**}$}} & {\textbf{ -0.0017$^{**}$}} & {\textbf{  0.13$^{**}$}} & {\textbf{ -0.0015$^{*~}$}} & {\textbf{  0.14$^{**}$}}  \\
    &  26 & { 0.0000$^{~~}$} & {  0.02$^{~~}$} & {\textbf{ -0.0012$^{*~}$}} & {\textbf{  0.16$^{**}$}}  & {\textbf{ -0.0012$^{**}$}} & {\textbf{  0.16$^{**}$}} & { -0.0005$^{~~}$} &  {\textbf{  0.27$^{**}$}} & { $^{~~}$} &  { $^{~~}$}& {\textbf{ -0.0014$^{**}$}} &  {\textbf{  0.16$^{**}$}} & {\textbf{ -0.0012$^{*~}$}} &  {\textbf{  0.17$^{**}$}}  & {\textbf{ -0.0013$^{*~}$}} & {\textbf{  0.16$^{**}$}} & {\textbf{ -0.0011$^{*~}$}} & {\textbf{  0.17$^{**}$}} & {\textbf{ -0.0011$^{*~}$}} & {\textbf{  0.17$^{**}$}}  \\
    &  52 & { 0.0005$^{~~}$} & {  0.02$^{~~}$} & {\textbf{ -0.0012$^{**}$}} & {\textbf{  0.11$^{**}$}}  & {\textbf{ -0.0012$^{**}$}} & {\textbf{  0.11$^{**}$}} & { -0.0003$^{~~}$} &  {\textbf{  0.20$^{**}$}} & { $^{~~}$} &  { $^{~~}$}& {\textbf{ -0.0014$^{**}$}} &  {\textbf{  0.11$^{**}$}} & {\textbf{ -0.0012$^{*~}$}} &  {\textbf{  0.11$^{**}$}}  & {\textbf{ -0.0013$^{**}$}} & {\textbf{  0.11$^{**}$}} & {\textbf{ -0.0011$^{*~}$}} & {\textbf{  0.12$^{**}$}} & {\textbf{ -0.0011$^{*~}$}} & {\textbf{  0.12$^{**}$}}  \\
   \hline\hline
   \end{tabular}
   \label{TB:Res2Risk:F5:spanning:p1}
\end{table}
\end{landscape}

\begin{landscape}
\begin{table}[!ht]
\setlength\tabcolsep{0pt}
\footnotesize
\caption*{Table \ref{TB:Res2Risk:F5:spanning:p1} (continued): Spanning test for risk-adjusted IMOM portfolios. Augmented FF5F-regression is conducted to examine if the profitability of various risk-based IMOM portfolios could be explained by one of others. Variable X in the first column refers to as the profitability of the X incorporated in the FF5F. Column 2 presents $K$'s value. Columns 3 to 8 report $\alpha$ values of explained variables that are IMOMs based on the rest of risk metrics, and $\beta$ values of explanatory variable X. \cite{Newey-West-1987-Em}'s $t$-statistics are presented in parentheses and the superscripts * and ** denote the significance at 5\% and 1\% levels, respectively. }
\centering
\vspace{-3mm}
   \begin{tabular}{cccccccccccccccccccccc}
   \hline\hline
    Explanatory  & \multirow{2}{*}{$K$} & \multirow{2}{*}{$\alpha$(RR)}  &\multirow{2}{*}{$\beta_X$} & \multirow{2}{*}{$\alpha$(IR)}  &\multirow{2}{*}{$\beta_X$} & \multirow{2}{*}{$\alpha$(IVol)}  &\multirow{2}{*}{$\beta_X$}  &  \multirow{2}{*}{$\alpha$(ISkew)}  &\multirow{2}{*}{$\beta_X$} & \multirow{2}{*}{$\alpha$(IKurt)}  &\multirow{2}{*}{$\beta_X$} & \multirow{2}{*}{$\alpha$(IMD) } &\multirow{2}{*}{$\beta_X$} & \multirow{2}{*}{$\alpha$(IES5)} & \multirow{2}{*}{$\beta_X$ } & \multirow{2}{*}{$\alpha$(IVaR5)} & \multirow{2}{*}{$\beta_X$ } & \multirow{2}{*}{$\alpha$(IES1)} & \multirow{2}{*}{$\beta_X$ } &\multirow{2}{*}{$\alpha$(IVaR1)} & \multirow{2}{*}{$\beta_X$} \\
   variable X  & & & & &  & & &  &  &  &  &  & & & & & & & & & \\
   \hline
   IMD-IMOM &  1 & { 0.0006$^{~~}$} & {\textbf{ -0.38$^{**}$}} & {\textbf{ -0.0007$^{**}$}} & {\textbf{  0.80$^{**}$}}  & {\textbf{ -0.0007$^{**}$}} & {\textbf{  0.81$^{**}$}} & { -0.0019$^{~~}$} &  {\textbf{  0.59$^{**}$}} & { 0.0016$^{~~}$} &  {\textbf{  0.48$^{*~}$}}& { $^{~~}$} &  { $^{~~}$} & { -0.0005$^{~~}$} &  {\textbf{  0.88$^{**}$}}  & { -0.0004$^{~~}$} & {\textbf{  0.90$^{**}$}} & {\textbf{ -0.0008$^{*~}$}} & {\textbf{  0.86$^{**}$}} & { -0.0005$^{~~}$} & {\textbf{  0.89$^{**}$}}  \\
    &  2 & { 0.0005$^{~~}$} & {\textbf{ -0.36$^{**}$}} & {\textbf{ -0.0005$^{*~}$}} & {\textbf{  0.82$^{**}$}}  & {\textbf{ -0.0005$^{*~}$}} & {\textbf{  0.82$^{**}$}} & {\textbf{ -0.0022$^{*~}$}} &  {\textbf{  0.59$^{**}$}} & { 0.0026$^{~~}$} &  {\textbf{  0.57$^{**}$}}& { $^{~~}$} &  { $^{~~}$} & { -0.0002$^{~~}$} &  {\textbf{  0.89$^{**}$}}  & { -0.0003$^{~~}$} & {\textbf{  0.91$^{**}$}} & { -0.0005$^{~~}$} & {\textbf{  0.86$^{**}$}} & { -0.0002$^{~~}$} & {\textbf{  0.90$^{**}$}}  \\
    &  3 & { 0.0004$^{~~}$} & {\textbf{ -0.35$^{**}$}} & {\textbf{ -0.0004$^{*~}$}} & {\textbf{  0.83$^{**}$}}  & {\textbf{ -0.0004$^{*~}$}} & {\textbf{  0.84$^{**}$}} & {\textbf{ -0.0027$^{**}$}} &  {\textbf{  0.67$^{**}$}} & { 0.0025$^{~~}$} &  {\textbf{  0.57$^{**}$}}& { $^{~~}$} &  { $^{~~}$} & { -0.0002$^{~~}$} &  {\textbf{  0.90$^{**}$}}  & { -0.0002$^{~~}$} & {\textbf{  0.92$^{**}$}} & { -0.0003$^{~~}$} & {\textbf{  0.87$^{**}$}} & { -0.0001$^{~~}$} & {\textbf{  0.91$^{**}$}}  \\
    &  4 & { 0.0004$^{~~}$} & {\textbf{ -0.33$^{**}$}} & { -0.0003$^{~~}$} & {\textbf{  0.83$^{**}$}}  & { -0.0003$^{~~}$} & {\textbf{  0.85$^{**}$}} & { -0.0017$^{~~}$} &  {\textbf{  0.67$^{**}$}} & { 0.0011$^{~~}$} &  {\textbf{  0.52$^{**}$}}& { $^{~~}$} &  { $^{~~}$} & { -0.0000$^{~~}$} &  {\textbf{  0.91$^{**}$}}  & { -0.0001$^{~~}$} & {\textbf{  0.93$^{**}$}} & { -0.0002$^{~~}$} & {\textbf{  0.88$^{**}$}} & { 0.0000$^{~~}$} & {\textbf{  0.92$^{**}$}}  \\
    &  8 & { 0.0002$^{~~}$} & {\textbf{ -0.31$^{**}$}} & { -0.0001$^{~~}$} & {\textbf{  0.86$^{**}$}}  & { -0.0002$^{~~}$} & {\textbf{  0.87$^{**}$}} & { -0.0006$^{~~}$} &  {\textbf{  0.61$^{**}$}} & { 0.0011$^{~~}$} &  {\textbf{  0.50$^{**}$}}& { $^{~~}$} &  { $^{~~}$} & { 0.0001$^{~~}$} &  {\textbf{  0.93$^{**}$}}  & { -0.0000$^{~~}$} & {\textbf{  0.95$^{**}$}} & { -0.0001$^{~~}$} & {\textbf{  0.89$^{**}$}} & { 0.0001$^{~~}$} & {\textbf{  0.95$^{**}$}}  \\
    &  13 & { -0.0002$^{~~}$} & {\textbf{ -0.27$^{**}$}} & { -0.0001$^{~~}$} & {\textbf{  0.87$^{**}$}}  & { -0.0001$^{~~}$} & {\textbf{  0.88$^{**}$}} & { 0.0004$^{~~}$} &  {\textbf{  0.65$^{**}$}} & { 0.0010$^{~~}$} &  {\textbf{  0.58$^{**}$}}& { $^{~~}$} &  { $^{~~}$} & { 0.0001$^{~~}$} &  {\textbf{  0.94$^{**}$}}  & { 0.0000$^{~~}$} & {\textbf{  0.97$^{**}$}} & { -0.0001$^{~~}$} & {\textbf{  0.92$^{**}$}} & { 0.0001$^{~~}$} & {\textbf{  0.96$^{**}$}}  \\
    &  26 & { -0.0005$^{~~}$} & {\textbf{ -0.22$^{**}$}} & { 0.0000$^{~~}$} & {\textbf{  0.89$^{**}$}}  & { 0.0000$^{~~}$} & {\textbf{  0.90$^{**}$}} & { 0.0002$^{~~}$} &  {\textbf{  0.60$^{**}$}} & { -0.0003$^{~~}$} &  {\textbf{  0.69$^{**}$}}& { $^{~~}$} &  { $^{~~}$} & { 0.0001$^{~~}$} &  {\textbf{  0.97$^{**}$}}  & { 0.0001$^{~~}$} & {\textbf{  0.99$^{**}$}} & { 0.0001$^{~~}$} & {\textbf{  0.96$^{**}$}} & { 0.0002$^{~~}$} & {\textbf{  0.99$^{**}$}}  \\
    &  52 & { 0.0002$^{~~}$} & {\textbf{ -0.14$^{*~}$}} & { 0.0000$^{~~}$} & {\textbf{  0.90$^{**}$}}  & { 0.0000$^{~~}$} & {\textbf{  0.92$^{**}$}} & { 0.0002$^{~~}$} &  {\textbf{  0.63$^{**}$}} & { -0.0006$^{~~}$} &  {\textbf{  0.55$^{**}$}}& { $^{~~}$} &  { $^{~~}$} & {\textbf{ 0.0002$^{*~}$}} &  {\textbf{  0.98$^{**}$}}  & { 0.0001$^{~~}$} & {\textbf{  1.00$^{**}$}} & { 0.0002$^{~~}$} & {\textbf{  0.95$^{**}$}} & { 0.0002$^{~~}$} & {\textbf{  0.99$^{**}$}}
      \smallskip\\
   IES5-IMOM &  1 & { 0.0004$^{~~}$} & {\textbf{ -0.40$^{**}$}} & {\textbf{ -0.0004$^{*~}$}} & {\textbf{  0.86$^{**}$}}  & {\textbf{ -0.0004$^{**}$}} & {\textbf{  0.87$^{**}$}} & { -0.0019$^{~~}$} &  {\textbf{  0.52$^{**}$}} & { 0.0018$^{~~}$} &  {\textbf{  0.57$^{**}$}}& { -0.0000$^{~~}$} &  {\textbf{  0.96$^{**}$}} & { $^{~~}$} &  { $^{~~}$}  & { -0.0000$^{~~}$} & {\textbf{  0.98$^{**}$}} & { -0.0003$^{~~}$} & {\textbf{  0.97$^{**}$}} & { -0.0000$^{~~}$} & {\textbf{  1.00$^{**}$}}  \\
    &  2 & { 0.0004$^{~~}$} & {\textbf{ -0.38$^{**}$}} & {\textbf{ -0.0004$^{*~}$}} & {\textbf{  0.87$^{**}$}}  & {\textbf{ -0.0004$^{**}$}} & {\textbf{  0.88$^{**}$}} & {\textbf{ -0.0024$^{*~}$}} &  {\textbf{  0.51$^{**}$}} & { 0.0027$^{~~}$} &  {\textbf{  0.67$^{**}$}}& { -0.0002$^{~~}$} &  {\textbf{  0.97$^{**}$}} & { $^{~~}$} &  { $^{~~}$}  & { -0.0002$^{~~}$} & {\textbf{  0.99$^{**}$}} & { -0.0002$^{~~}$} & {\textbf{  0.97$^{**}$}} & { 0.0000$^{~~}$} & {\textbf{  1.01$^{**}$}}  \\
    &  3 & { 0.0005$^{~~}$} & {\textbf{ -0.36$^{**}$}} & {\textbf{ -0.0003$^{*~}$}} & {\textbf{  0.88$^{**}$}}  & {\textbf{ -0.0004$^{*~}$}} & {\textbf{  0.89$^{**}$}} & {\textbf{ -0.0030$^{**}$}} &  {\textbf{  0.61$^{**}$}} & { 0.0026$^{~~}$} &  {\textbf{  0.69$^{**}$}}& { -0.0002$^{~~}$} &  {\textbf{  0.98$^{**}$}} & { $^{~~}$} &  { $^{~~}$}  & { -0.0001$^{~~}$} & {\textbf{  0.99$^{**}$}} & { -0.0002$^{~~}$} & {\textbf{  0.98$^{**}$}} & { 0.0001$^{~~}$} & {\textbf{  1.01$^{**}$}}  \\
    &  4 & { 0.0005$^{~~}$} & {\textbf{ -0.33$^{**}$}} & {\textbf{ -0.0003$^{*~}$}} & {\textbf{  0.88$^{**}$}}  & {\textbf{ -0.0003$^{*~}$}} & {\textbf{  0.89$^{**}$}} & {\textbf{ -0.0020$^{*~}$}} &  {\textbf{  0.61$^{**}$}} & { 0.0013$^{~~}$} &  {\textbf{  0.68$^{**}$}}& { -0.0003$^{~~}$} &  {\textbf{  0.97$^{**}$}} & { $^{~~}$} &  { $^{~~}$}  & { -0.0001$^{~~}$} & {\textbf{  0.99$^{**}$}} & { -0.0002$^{~~}$} & {\textbf{  0.97$^{**}$}} & { 0.0001$^{~~}$} & {\textbf{  1.02$^{**}$}}  \\
    &  8 & { 0.0003$^{~~}$} & {\textbf{ -0.30$^{**}$}} & {\textbf{ -0.0003$^{*~}$}} & {\textbf{  0.89$^{**}$}}  & {\textbf{ -0.0003$^{*~}$}} & {\textbf{  0.90$^{**}$}} & { -0.0009$^{~~}$} &  {\textbf{  0.56$^{**}$}} & { 0.0011$^{~~}$} &  {\textbf{  0.63$^{**}$}}& { -0.0003$^{~~}$} &  {\textbf{  0.97$^{**}$}} & { $^{~~}$} &  { $^{~~}$}  & { -0.0002$^{~~}$} & {\textbf{  0.99$^{**}$}} & { -0.0002$^{~~}$} & {\textbf{  0.97$^{**}$}} & { 0.0000$^{~~}$} & {\textbf{  1.02$^{**}$}}  \\
    &  13 & { -0.0001$^{~~}$} & {\textbf{ -0.26$^{**}$}} & { -0.0002$^{~~}$} & {\textbf{  0.89$^{**}$}}  & {\textbf{ -0.0002$^{*~}$}} & {\textbf{  0.90$^{**}$}} & { 0.0001$^{~~}$} &  {\textbf{  0.61$^{**}$}} & { 0.0010$^{~~}$} &  {\textbf{  0.68$^{**}$}}& { -0.0003$^{~~}$} &  {\textbf{  0.97$^{**}$}} & { $^{~~}$} &  { $^{~~}$}  & { -0.0001$^{~~}$} & {\textbf{  1.00$^{**}$}} & { -0.0001$^{~~}$} & {\textbf{  0.98$^{**}$}} & { 0.0000$^{~~}$} & {\textbf{  1.02$^{**}$}}  \\
    &  26 & { -0.0004$^{~~}$} & {\textbf{ -0.20$^{**}$}} & { -0.0001$^{~~}$} & {\textbf{  0.88$^{**}$}}  & { -0.0002$^{~~}$} & {\textbf{  0.89$^{**}$}} & { 0.0000$^{~~}$} &  {\textbf{  0.55$^{**}$}} & { -0.0005$^{~~}$} &  {\textbf{  0.71$^{**}$}}& {\textbf{ -0.0003$^{*~}$}} &  {\textbf{  0.96$^{**}$}} & { $^{~~}$} &  { $^{~~}$}  & { -0.0001$^{~~}$} & {\textbf{  0.99$^{**}$}} & { 0.0000$^{~~}$} & {\textbf{  0.99$^{**}$}} & { 0.0000$^{~~}$} & {\textbf{  1.02$^{**}$}}  \\
    &  52 & { 0.0003$^{~~}$} & { -0.11$^{~~}$} & { -0.0002$^{~~}$} & {\textbf{  0.89$^{**}$}}  & {\textbf{ -0.0002$^{*~}$}} & {\textbf{  0.91$^{**}$}} & { 0.0000$^{~~}$} &  {\textbf{  0.58$^{**}$}} & { -0.0007$^{~~}$} &  {\textbf{  0.56$^{**}$}}& {\textbf{ -0.0003$^{**}$}} &  {\textbf{  0.96$^{**}$}} & { $^{~~}$} &  { $^{~~}$}  & { -0.0001$^{~~}$} & {\textbf{  1.00$^{**}$}} & { -0.0000$^{~~}$} & {\textbf{  0.98$^{**}$}} & { 0.0000$^{~~}$} & {\textbf{  1.01$^{**}$}}
      \smallskip\\
   IVaR5-IMOM &  1 & { 0.0006$^{~~}$} & {\textbf{ -0.35$^{**}$}} & {\textbf{ -0.0004$^{**}$}} & {\textbf{  0.86$^{**}$}}  & {\textbf{ -0.0005$^{**}$}} & {\textbf{  0.86$^{**}$}} & { -0.0019$^{~~}$} &  {\textbf{  0.55$^{**}$}} & { 0.0015$^{~~}$} &  {\textbf{  0.48$^{*~}$}}& { -0.0001$^{~~}$} &  {\textbf{  0.95$^{**}$}} & { -0.0002$^{~~}$} &  {\textbf{  0.95$^{**}$}}  & { $^{~~}$} & { $^{~~}$} & { -0.0005$^{~~}$} & {\textbf{  0.91$^{**}$}} & { -0.0002$^{~~}$} & {\textbf{  0.95$^{**}$}}  \\
    &  2 & { 0.0006$^{~~}$} & {\textbf{ -0.32$^{**}$}} & {\textbf{ -0.0003$^{*~}$}} & {\textbf{  0.87$^{**}$}}  & {\textbf{ -0.0003$^{*~}$}} & {\textbf{  0.88$^{**}$}} & {\textbf{ -0.0023$^{*~}$}} &  {\textbf{  0.54$^{**}$}} & { 0.0025$^{~~}$} &  {\textbf{  0.62$^{**}$}}& { -0.0001$^{~~}$} &  {\textbf{  0.96$^{**}$}} & { -0.0000$^{~~}$} &  {\textbf{  0.96$^{**}$}}  & { $^{~~}$} & { $^{~~}$} & { -0.0003$^{~~}$} & {\textbf{  0.92$^{**}$}} & { -0.0000$^{~~}$} & {\textbf{  0.96$^{**}$}}  \\
    &  3 & { 0.0006$^{~~}$} & {\textbf{ -0.32$^{**}$}} & {\textbf{ -0.0003$^{*~}$}} & {\textbf{  0.87$^{**}$}}  & {\textbf{ -0.0003$^{**}$}} & {\textbf{  0.88$^{**}$}} & {\textbf{ -0.0028$^{**}$}} &  {\textbf{  0.63$^{**}$}} & { 0.0025$^{~~}$} &  {\textbf{  0.66$^{**}$}}& { -0.0002$^{~~}$} &  {\textbf{  0.97$^{**}$}} & { -0.0001$^{~~}$} &  {\textbf{  0.96$^{**}$}}  & { $^{~~}$} & { $^{~~}$} & { -0.0002$^{~~}$} & {\textbf{  0.92$^{**}$}} & { 0.0000$^{~~}$} & {\textbf{  0.97$^{**}$}}  \\
    &  4 & { 0.0006$^{~~}$} & {\textbf{ -0.29$^{**}$}} & {\textbf{ -0.0003$^{*~}$}} & {\textbf{  0.88$^{**}$}}  & {\textbf{ -0.0003$^{*~}$}} & {\textbf{  0.89$^{**}$}} & { -0.0019$^{~~}$} &  {\textbf{  0.64$^{**}$}} & { 0.0011$^{~~}$} &  {\textbf{  0.64$^{**}$}}& { -0.0002$^{~~}$} &  {\textbf{  0.97$^{**}$}} & { -0.0000$^{~~}$} &  {\textbf{  0.96$^{**}$}}  & { $^{~~}$} & { $^{~~}$} & { -0.0002$^{~~}$} & {\textbf{  0.92$^{**}$}} & { 0.0000$^{~~}$} & {\textbf{  0.97$^{**}$}}  \\
    &  8 & { 0.0003$^{~~}$} & {\textbf{ -0.28$^{**}$}} & { -0.0002$^{~~}$} & {\textbf{  0.88$^{**}$}}  & { -0.0002$^{~~}$} & {\textbf{  0.89$^{**}$}} & { -0.0007$^{~~}$} &  {\textbf{  0.58$^{**}$}} & { 0.0011$^{~~}$} &  {\textbf{  0.57$^{**}$}}& { -0.0002$^{~~}$} &  {\textbf{  0.97$^{**}$}} & { 0.0001$^{~~}$} &  {\textbf{  0.96$^{**}$}}  & { $^{~~}$} & { $^{~~}$} & { -0.0002$^{~~}$} & {\textbf{  0.92$^{**}$}} & { 0.0001$^{~~}$} & {\textbf{  0.98$^{**}$}}  \\
    &  13 & { -0.0001$^{~~}$} & {\textbf{ -0.25$^{**}$}} & { -0.0001$^{~~}$} & {\textbf{  0.89$^{**}$}}  & { -0.0001$^{~~}$} & {\textbf{  0.90$^{**}$}} & { 0.0002$^{~~}$} &  {\textbf{  0.62$^{**}$}} & { 0.0011$^{~~}$} &  {\textbf{  0.63$^{**}$}}& { -0.0002$^{~~}$} &  {\textbf{  0.97$^{**}$}} & { 0.0000$^{~~}$} &  {\textbf{  0.97$^{**}$}}  & { $^{~~}$} & { $^{~~}$} & { -0.0001$^{~~}$} & {\textbf{  0.93$^{**}$}} & { 0.0000$^{~~}$} & {\textbf{  0.98$^{**}$}}  \\
    &  26 & { -0.0004$^{~~}$} & {\textbf{ -0.20$^{**}$}} & { -0.0001$^{~~}$} & {\textbf{  0.88$^{**}$}}  & { -0.0001$^{~~}$} & {\textbf{  0.90$^{**}$}} & { 0.0001$^{~~}$} &  {\textbf{  0.56$^{**}$}} & { -0.0004$^{~~}$} &  {\textbf{  0.67$^{**}$}}& { -0.0002$^{~~}$} &  {\textbf{  0.96$^{**}$}} & { 0.0001$^{~~}$} &  {\textbf{  0.98$^{**}$}}  & { $^{~~}$} & { $^{~~}$} & { 0.0000$^{~~}$} & {\textbf{  0.95$^{**}$}} & { 0.0001$^{~~}$} & {\textbf{  0.99$^{**}$}}  \\
    &  52 & { 0.0003$^{~~}$} & {\textbf{ -0.12$^{*~}$}} & { -0.0001$^{~~}$} & {\textbf{  0.89$^{**}$}}  & { -0.0001$^{~~}$} & {\textbf{  0.90$^{**}$}} & { 0.0001$^{~~}$} &  {\textbf{  0.59$^{**}$}} & { -0.0007$^{~~}$} &  {\textbf{  0.53$^{**}$}}& {\textbf{ -0.0002$^{*~}$}} &  {\textbf{  0.96$^{**}$}} & { 0.0001$^{~~}$} &  {\textbf{  0.97$^{**}$}}  & { $^{~~}$} & { $^{~~}$} & { 0.0000$^{~~}$} & {\textbf{  0.93$^{**}$}} & { 0.0001$^{~~}$} & {\textbf{  0.98$^{**}$}}
      \smallskip\\
   IES1-IMOM &  1 & { 0.0004$^{~~}$} & {\textbf{ -0.40$^{**}$}} & { -0.0004$^{~~}$} & {\textbf{  0.81$^{**}$}}  & { -0.0005$^{~~}$} & {\textbf{  0.82$^{**}$}} & { -0.0019$^{~~}$} &  {\textbf{  0.50$^{**}$}} & { 0.0020$^{~~}$} &  {\textbf{  0.66$^{**}$}}& { -0.0000$^{~~}$} &  {\textbf{  0.91$^{**}$}} & { -0.0000$^{~~}$} &  {\textbf{  0.94$^{**}$}}  & { -0.0001$^{~~}$} & {\textbf{  0.91$^{**}$}} & { $^{~~}$} & { $^{~~}$} & { 0.0000$^{~~}$} & {\textbf{  0.97$^{**}$}}  \\
    &  2 & { 0.0004$^{~~}$} & {\textbf{ -0.38$^{**}$}} & { -0.0004$^{~~}$} & {\textbf{  0.82$^{**}$}}  & {\textbf{ -0.0005$^{*~}$}} & {\textbf{  0.83$^{**}$}} & {\textbf{ -0.0024$^{*~}$}} &  {\textbf{  0.50$^{**}$}} & { 0.0027$^{~~}$} &  {\textbf{  0.70$^{**}$}}& { -0.0003$^{~~}$} &  {\textbf{  0.91$^{**}$}} & { -0.0001$^{~~}$} &  {\textbf{  0.94$^{**}$}}  & { -0.0003$^{~~}$} & {\textbf{  0.92$^{**}$}} & { $^{~~}$} & { $^{~~}$} & { 0.0000$^{~~}$} & {\textbf{  0.97$^{**}$}}  \\
    &  3 & { 0.0005$^{~~}$} & {\textbf{ -0.35$^{**}$}} & { -0.0004$^{~~}$} & {\textbf{  0.83$^{**}$}}  & { -0.0004$^{~~}$} & {\textbf{  0.84$^{**}$}} & {\textbf{ -0.0030$^{**}$}} &  {\textbf{  0.58$^{**}$}} & { 0.0026$^{~~}$} &  {\textbf{  0.68$^{**}$}}& { -0.0003$^{~~}$} &  {\textbf{  0.92$^{**}$}} & { -0.0001$^{~~}$} &  {\textbf{  0.95$^{**}$}}  & { -0.0002$^{~~}$} & {\textbf{  0.92$^{**}$}} & { $^{~~}$} & { $^{~~}$} & { 0.0001$^{~~}$} & {\textbf{  0.98$^{**}$}}  \\
    &  4 & { 0.0005$^{~~}$} & {\textbf{ -0.32$^{**}$}} & { -0.0004$^{~~}$} & {\textbf{  0.83$^{**}$}}  & { -0.0004$^{~~}$} & {\textbf{  0.85$^{**}$}} & {\textbf{ -0.0020$^{*~}$}} &  {\textbf{  0.59$^{**}$}} & { 0.0013$^{~~}$} &  {\textbf{  0.65$^{**}$}}& { -0.0003$^{~~}$} &  {\textbf{  0.92$^{**}$}} & { -0.0000$^{~~}$} &  {\textbf{  0.95$^{**}$}}  & { -0.0002$^{~~}$} & {\textbf{  0.93$^{**}$}} & { $^{~~}$} & { $^{~~}$} & { 0.0001$^{~~}$} & {\textbf{  0.99$^{**}$}}  \\
    &  8 & { 0.0003$^{~~}$} & {\textbf{ -0.29$^{**}$}} & { -0.0003$^{~~}$} & {\textbf{  0.85$^{**}$}}  & { -0.0003$^{~~}$} & {\textbf{  0.86$^{**}$}} & { -0.0008$^{~~}$} &  {\textbf{  0.55$^{**}$}} & { 0.0012$^{~~}$} &  {\textbf{  0.60$^{**}$}}& { -0.0003$^{~~}$} &  {\textbf{  0.93$^{**}$}} & { 0.0000$^{~~}$} &  {\textbf{  0.96$^{**}$}}  & { -0.0002$^{~~}$} & {\textbf{  0.94$^{**}$}} & { $^{~~}$} & { $^{~~}$} & { 0.0001$^{~~}$} & {\textbf{  1.00$^{**}$}}  \\
    &  13 & { -0.0001$^{~~}$} & {\textbf{ -0.24$^{**}$}} & { -0.0002$^{~~}$} & {\textbf{  0.85$^{**}$}}  & { -0.0003$^{~~}$} & {\textbf{  0.86$^{**}$}} & { 0.0002$^{~~}$} &  {\textbf{  0.60$^{**}$}} & { 0.0011$^{~~}$} &  {\textbf{  0.66$^{**}$}}& { -0.0003$^{~~}$} &  {\textbf{  0.92$^{**}$}} & { -0.0000$^{~~}$} &  {\textbf{  0.96$^{**}$}}  & { -0.0002$^{~~}$} & {\textbf{  0.94$^{**}$}} & { $^{~~}$} & { $^{~~}$} & { 0.0000$^{~~}$} & {\textbf{  0.99$^{**}$}}  \\
    &  26 & { -0.0003$^{~~}$} & {\textbf{ -0.17$^{*~}$}} & { -0.0002$^{~~}$} & {\textbf{  0.84$^{**}$}}  & { -0.0003$^{~~}$} & {\textbf{  0.85$^{**}$}} & { -0.0000$^{~~}$} &  {\textbf{  0.53$^{**}$}} & { -0.0005$^{~~}$} &  {\textbf{  0.73$^{**}$}}& {\textbf{ -0.0004$^{*~}$}} &  {\textbf{  0.92$^{**}$}} & { -0.0001$^{~~}$} &  {\textbf{  0.96$^{**}$}}  & { -0.0002$^{~~}$} & {\textbf{  0.94$^{**}$}} & { $^{~~}$} & { $^{~~}$} & { -0.0000$^{~~}$} & {\textbf{  1.00$^{**}$}}  \\
    &  52 & { 0.0003$^{~~}$} & { -0.10$^{~~}$} & { -0.0002$^{~~}$} & {\textbf{  0.86$^{**}$}}  & { -0.0002$^{~~}$} & {\textbf{  0.88$^{**}$}} & { 0.0000$^{~~}$} &  {\textbf{  0.57$^{**}$}} & { -0.0007$^{~~}$} &  {\textbf{  0.58$^{**}$}}& {\textbf{ -0.0003$^{*~}$}} &  {\textbf{  0.93$^{**}$}} & { -0.0000$^{~~}$} &  {\textbf{  0.98$^{**}$}}  & { -0.0002$^{~~}$} & {\textbf{  0.96$^{**}$}} & { $^{~~}$} & { $^{~~}$} & { 0.0000$^{~~}$} & {\textbf{  1.00$^{**}$}}
      \smallskip\\
   IVaR1-IMOM &  1 & { 0.0004$^{~~}$} & {\textbf{ -0.40$^{**}$}} & {\textbf{ -0.0005$^{*~}$}} & {\textbf{  0.81$^{**}$}}  & {\textbf{ -0.0006$^{**}$}} & {\textbf{  0.82$^{**}$}} & { -0.0022$^{~~}$} &  {\textbf{  0.46$^{**}$}} & { 0.0018$^{~~}$} &  {\textbf{  0.55$^{**}$}}& { -0.0002$^{~~}$} &  {\textbf{  0.90$^{**}$}} & { -0.0002$^{~~}$} &  {\textbf{  0.94$^{**}$}}  & { -0.0002$^{~~}$} & {\textbf{  0.92$^{**}$}} & { -0.0004$^{~~}$} & {\textbf{  0.93$^{**}$}} & { $^{~~}$} & { $^{~~}$}  \\
    &  2 & { 0.0005$^{~~}$} & {\textbf{ -0.37$^{**}$}} & {\textbf{ -0.0005$^{**}$}} & {\textbf{  0.82$^{**}$}}  & {\textbf{ -0.0006$^{**}$}} & {\textbf{  0.83$^{**}$}} & {\textbf{ -0.0025$^{*~}$}} &  {\textbf{  0.48$^{**}$}} & { 0.0026$^{~~}$} &  {\textbf{  0.63$^{**}$}}& { -0.0004$^{~~}$} &  {\textbf{  0.92$^{**}$}} & { -0.0002$^{~~}$} &  {\textbf{  0.94$^{**}$}}  & { -0.0003$^{~~}$} & {\textbf{  0.92$^{**}$}} & { -0.0003$^{~~}$} & {\textbf{  0.94$^{**}$}} & { $^{~~}$} & { $^{~~}$}  \\
    &  3 & { 0.0005$^{~~}$} & {\textbf{ -0.35$^{**}$}} & {\textbf{ -0.0005$^{**}$}} & {\textbf{  0.83$^{**}$}}  & {\textbf{ -0.0005$^{**}$}} & {\textbf{  0.84$^{**}$}} & {\textbf{ -0.0031$^{**}$}} &  {\textbf{  0.58$^{**}$}} & { 0.0025$^{~~}$} &  {\textbf{  0.64$^{**}$}}& { -0.0004$^{~~}$} &  {\textbf{  0.92$^{**}$}} & { -0.0002$^{~~}$} &  {\textbf{  0.94$^{**}$}}  & { -0.0003$^{~~}$} & {\textbf{  0.93$^{**}$}} & { -0.0003$^{~~}$} & {\textbf{  0.94$^{**}$}} & { $^{~~}$} & { $^{~~}$}  \\
    &  4 & { 0.0006$^{~~}$} & {\textbf{ -0.32$^{**}$}} & {\textbf{ -0.0005$^{**}$}} & {\textbf{  0.83$^{**}$}}  & {\textbf{ -0.0005$^{**}$}} & {\textbf{  0.84$^{**}$}} & {\textbf{ -0.0021$^{*~}$}} &  {\textbf{  0.58$^{**}$}} & { 0.0011$^{~~}$} &  {\textbf{  0.63$^{**}$}}& {\textbf{ -0.0005$^{*~}$}} &  {\textbf{  0.92$^{**}$}} & { -0.0002$^{~~}$} &  {\textbf{  0.94$^{**}$}}  & { -0.0003$^{~~}$} & {\textbf{  0.93$^{**}$}} & { -0.0003$^{~~}$} & {\textbf{  0.94$^{**}$}} & { $^{~~}$} & { $^{~~}$}  \\
    &  8 & { 0.0003$^{~~}$} & {\textbf{ -0.29$^{**}$}} & {\textbf{ -0.0004$^{*~}$}} & {\textbf{  0.84$^{**}$}}  & {\textbf{ -0.0004$^{*~}$}} & {\textbf{  0.85$^{**}$}} & { -0.0009$^{~~}$} &  {\textbf{  0.54$^{**}$}} & { 0.0010$^{~~}$} &  {\textbf{  0.61$^{**}$}}& {\textbf{ -0.0004$^{*~}$}} &  {\textbf{  0.92$^{**}$}} & { -0.0001$^{~~}$} &  {\textbf{  0.95$^{**}$}}  & { -0.0003$^{~~}$} & {\textbf{  0.94$^{**}$}} & { -0.0003$^{~~}$} & {\textbf{  0.94$^{**}$}} & { $^{~~}$} & { $^{~~}$}  \\
    &  13 & { -0.0001$^{~~}$} & {\textbf{ -0.25$^{**}$}} & { -0.0003$^{~~}$} & {\textbf{  0.85$^{**}$}}  & { -0.0003$^{~~}$} & {\textbf{  0.86$^{**}$}} & { 0.0001$^{~~}$} &  {\textbf{  0.58$^{**}$}} & { 0.0010$^{~~}$} &  {\textbf{  0.66$^{**}$}}& { -0.0004$^{~~}$} &  {\textbf{  0.92$^{**}$}} & { -0.0001$^{~~}$} &  {\textbf{  0.96$^{**}$}}  & { -0.0002$^{~~}$} & {\textbf{  0.95$^{**}$}} & { -0.0002$^{~~}$} & {\textbf{  0.95$^{**}$}} & { $^{~~}$} & { $^{~~}$}  \\
    &  26 & { -0.0004$^{~~}$} & {\textbf{ -0.18$^{**}$}} & { -0.0002$^{~~}$} & {\textbf{  0.84$^{**}$}}  & { -0.0003$^{~~}$} & {\textbf{  0.85$^{**}$}} & { -0.0001$^{~~}$} &  {\textbf{  0.52$^{**}$}} & { -0.0005$^{~~}$} &  {\textbf{  0.69$^{**}$}}& {\textbf{ -0.0003$^{*~}$}} &  {\textbf{  0.92$^{**}$}} & { -0.0001$^{~~}$} &  {\textbf{  0.96$^{**}$}}  & { -0.0002$^{~~}$} & {\textbf{  0.94$^{**}$}} & { -0.0001$^{~~}$} & {\textbf{  0.96$^{**}$}} & { $^{~~}$} & { $^{~~}$}  \\
    &  52 & { 0.0003$^{~~}$} & { -0.09$^{~~}$} & { -0.0002$^{~~}$} & {\textbf{  0.85$^{**}$}}  & { -0.0003$^{~~}$} & {\textbf{  0.86$^{**}$}} & { -0.0000$^{~~}$} &  {\textbf{  0.55$^{**}$}} & { -0.0007$^{~~}$} &  {\textbf{  0.55$^{**}$}}& {\textbf{ -0.0004$^{**}$}} &  {\textbf{  0.91$^{**}$}} & { -0.0001$^{~~}$} &  {\textbf{  0.96$^{**}$}}  & { -0.0002$^{~~}$} & {\textbf{  0.95$^{**}$}} & { -0.0001$^{~~}$} & {\textbf{  0.95$^{**}$}} & { $^{~~}$} & { $^{~~}$}  \\
   \hline\hline
   \end{tabular}
   \label{TB:Res2Risk:F5:spanning:p2}
\end{table}
\end{landscape}

In addition, liquidity factor plays an important role in asset pricing. In fact, a body of literature has concentrated on the liquidity factor \citep{Avramov-Cheng-Hameed-2016-JFQA}, whose importance has been highlighted after the Global Financial Tsunami \citep{Jiang-Zhou-Sornette-Woodard-Bastiaensen-Cauwels-2010-JEBO,Han-Xie-Xiong-Zhang-Zhou-2017-FNL}. In our work, we follow \cite{Amihud-2002-JFinM} to construct illiquidity factor as the measure of market liquidity state. We first construct a daily illiquidity measure for each individual stock by calculating the ratio of absolute daily return to its trading volume in RMB:
\begin{equation}
  ILLIQ_{i,t,d}=\frac{|R_{i,t,d}|}{VOLD_{i,t,d}},
\label{Eq:Illiq:daily}
\end{equation}
where $R_{i,t,d}$ is the return of the stock $i$ on day $d$ of week $t$ and $VOLD_{i,t,d}$ is the respective daily trading volume in RMB.This measure is averaged over all trading days
\begin{equation}
  ILLIQ_{i,t}=\frac{1}{D_{i,t}}\sum_{j=1}^{D_{i,t}}ILLIQ_{i,t,d},
\label{Eq:Illiq:weekly}
\end{equation}
 and then over all stocks available in the same week $t$:
\begin{equation}
  AILLIQ_{t}=\frac{1}{N_{t}}\sum_{j=1}^{N_{t}}ILLIQ_{i,t},
  \label{Eq:Illiq:All}
\end{equation}
where $D_{i,t}$ is the number of trading days and $N_{t}$ is the number of available stocks.
Market illiquidity would be at high (low) level when the illiquidity at week $t$ is above (below) its median value for the whole sample period. In addition, we consider the periods with illiquidity extremes, including the periods with illiquidity value above its 80 percentile (Top20) as well as the periods with illiquidity value below its 20 percentile (Bottom20).

Lastly, we consider behaviorally associated factors, which have gained much more attention in recent years. In fact, many abnormal phenomena in the market become reasonable within the framework of behavioral finance, and the newly proposed behavioral three-factor model indeed works well in explaining most of the anomalies \citep{Daniel-Hirshleifer-Sun-2020-RFS}.
More importantly, the momentum effect has been found to be closely linked to investor sentiment \citep{Antoniou-Doukas-Subrahmanyam-2013-JFQA}.
In our work, we adopt the method proposed by \cite{Baker-Wurgler-2006-JF} (BW, hereafter) to construct the measure of investor sentiment. Specifically, the BW sentiment index is the first component obtained from the principle components analysis on six proxies of sentiment.%
\footnote{The sentiment index in our work is simplified. We omit the lagged effect of the six proxies, and the index is not orthogonalized with respect to the set of macroeconomic variables. Thus, the sentiment index conveys more information that is not limited to the stock market. }
We obtain the six factor loadings associated with the first component:
\begin{equation}
Sentiment_{t}  =  0.55CEFD_{t} - 0.07NIPO_{t} + 0.45RIPO_{t} + 0.19P_{t}^{D-ND} + 0.57S_{t} + 0.36TURN_{t},
\label{Eq:sentiment}
\end{equation}
where $CEFD_t$ denotes the closed-end fund discount at the end of week $t$, which is the average difference between the net asset values of closed-end stock fund shares and their market prices; $NIPO_t$ denotes the number of IPOs at week $t$ and $RIPO_t$ denotes their average first-day returns; $P^{D-ND}$ denotes the dividend premium, which is the log difference of the average market-to-book ratios of payers and nonpayers; $S_t$ denotes the equity share, which is the gross equity issuance divided by gross equity plus the gross long-term debt issuance;%
\footnote{ Note that the data records of $P^{D-ND}$ and $S_t$ are on the yearly basis. In our work, we assign yearly observation to each week within the same year. In other words, the observations for each week are identical within the same year. }
and lastly, $TURN_t$ denotes the turnover rate at week $t$, which is defined as the ratio of share volume for the A-share market to its all outstanding shares.
Likewise, the high (low) level of investor sentiment is defined as the periods when the sentiment measure at week $t$ is above (below) its median for the whole sample period.
The extremely high (low) level of sentiment is also taken into account, which is defined as the periods with the sentiment value being greater (less) than its 80 (20) percentile, denoted as Top20 (Bottom20).
In addition, we also consider the change in sentiment measured by the first-order difference of the sentiment index, which is argued to influence the momentum performance \citep{Moskowitz-Ooi-Pedersen-2012-JFE}. And its extremes are taken into account as well.
It is noted that the sentiment index formed in our work spans from January 1999 to December 2017, because the data of close-end fund discounts are not available before 1999.

\begin{landscape}
\begin{table}[!ht]
\setlength\tabcolsep{1pt}
\footnotesize
\caption{Conditional performance test. This table reports the weekly average returns of MOMs, IMOMs, IVol-IMOMs and IMD-IMOMs conditionally in the periods with different levels of market state, illiquidity, sentiment and change in sentiment. Following \cite{Cooper-Gutierrez-Hameed-2004-JF}, the upside (downside) market state at week $t$ is defined as the periods with the positive (negative) cumulative return of Shanghai Composite Index (SHCI) over the past $N$ trading weeks. The results are presented in the columns of Market state ($N$) and $N$ is assigned to 26 and 52. Up (Down) denotes the periods with upside (downside) market state. Following \cite{Amihud-2002-JFinM}, market illiquidity measure is constructed according to Eq.~(\ref{Eq:Illiq:All}). Market illiquidity is in the high (low) level when the illiquidity at week $t$ is above (below) its median value for the whole sample period. The periods with liquidity extremes refer to the periods with the illiquidity above than its 80 percentile (Top20) and the periods with illiquidity value below than its 20 percentile (Bottom20).
Following \cite{Baker-Wurgler-2006-JF}, the sentiment index is constructed according to the measure of investor sentiment in Eq.~(\ref{Eq:sentiment}). High (low) level of investor sentiment is defined as the periods when the sentiment index at week $t$ is above (below) its median for the whole sample period. The extremely high (low) level of sentiment is defined as the periods with sentiment being higher (lower) than its 80 (20) percentile, denoted as Top20 (Bottom20). Lastly, change in sentiment measured by the first-order difference of sentiment index as well as its extremes are considered, which follow the similar definitions mentioned above. \cite{Newey-West-1987-Em}'s $t$-statistics are obtained and the superscripts * and ** denote the significance at 5\% and 1\% levels, respectively. }
\centering
\vspace{-3mm}
   \begin{tabular}{ccccccccccccccccccccccc}
   \hline\hline
      & & & \multicolumn{2}{c}{Market state(26)} && \multicolumn{2}{c}{Market state(52)} && \multicolumn{4}{c}{Illiquidity} && \multicolumn{4}{c}{Sentiment} && \multicolumn{4}{c}{Change in sentiment}\\
     \cline{4-5} \cline{7-8}  \cline{10-13} \cline{15-18}  \cline{20-23}
      & $K$ && Up & Down && Up & Down && High & Low & Top20 & Bottom20&& High & Low & Top20 & Bottom20 && High & Low & Top20 & Bottom20  \\
   \hline
   MOM &  1 && { -0.0013$^{~~}$} & { -0.0009$^{~~}$} && { -0.0013$^{~~}$} &  { -0.0010$^{~~}$} && { 0.0016$^{~~}$} &  {\textbf{ -0.0029$^{*~}$}}& { 0.0021$^{~~}$} &  {\textbf{ -0.0033$^{*~}$}} && { -0.0008$^{~~}$} &  { -0.0017$^{~~}$}  & { 0.0010$^{~~}$} & { 0.0006$^{~~}$} && { -0.0027$^{~~}$} & { 0.0001$^{~~}$} & { -0.0000$^{~~}$} & { -0.0017$^{~~}$}  \\
    &  2 && { -0.0012$^{~~}$} & { -0.0007$^{~~}$} && { -0.0012$^{~~}$} &  { -0.0007$^{~~}$} && { 0.0012$^{~~}$} &  {\textbf{ -0.0023$^{*~}$}}& { 0.0027$^{~~}$} &  {\textbf{ -0.0028$^{*~}$}} && { -0.0010$^{~~}$} &  { -0.0011$^{~~}$}  & { 0.0005$^{~~}$} & { 0.0008$^{~~}$} && { -0.0026$^{~~}$} & { 0.0004$^{~~}$} & { -0.0005$^{~~}$} & { -0.0013$^{~~}$}  \\
    &  3 && { -0.0011$^{~~}$} & { -0.0006$^{~~}$} && { -0.0011$^{~~}$} &  { -0.0007$^{~~}$} && { 0.0010$^{~~}$} &  {\textbf{ -0.0021$^{*~}$}}& { 0.0033$^{~~}$} &  { -0.0026$^{~~}$} && { -0.0011$^{~~}$} &  { -0.0010$^{~~}$}  & { 0.0002$^{~~}$} & { 0.0007$^{~~}$} && {\textbf{ -0.0026$^{*~}$}} & { 0.0005$^{~~}$} & { -0.0005$^{~~}$} & { -0.0013$^{~~}$}  \\
    &  4 && { -0.0011$^{~~}$} & { -0.0005$^{~~}$} && { -0.0011$^{~~}$} &  { -0.0005$^{~~}$} && { 0.0008$^{~~}$} &  { -0.0019$^{~~}$}& { 0.0036$^{~~}$} &  { -0.0024$^{~~}$} && { -0.0010$^{~~}$} &  { -0.0008$^{~~}$}  & { 0.0003$^{~~}$} & { 0.0005$^{~~}$} && {\textbf{ -0.0025$^{*~}$}} & { 0.0006$^{~~}$} & { -0.0005$^{~~}$} & { -0.0011$^{~~}$}  \\
    &  8 && { -0.0010$^{~~}$} & { 0.0002$^{~~}$} && { -0.0009$^{~~}$} &  { 0.0001$^{~~}$} && { 0.0010$^{~~}$} &  { -0.0014$^{~~}$}& { 0.0042$^{~~}$} &  { -0.0017$^{~~}$} && { -0.0005$^{~~}$} &  { -0.0004$^{~~}$}  & { 0.0002$^{~~}$} & { 0.0002$^{~~}$} && { -0.0016$^{~~}$} & { 0.0007$^{~~}$} & { -0.0004$^{~~}$} & { -0.0002$^{~~}$}  \\
    &  13 && { -0.0008$^{~~}$} & { 0.0009$^{~~}$} && { -0.0006$^{~~}$} &  { 0.0006$^{~~}$} && { 0.0012$^{~~}$} &  { -0.0009$^{~~}$}& { 0.0053$^{~~}$} &  { -0.0007$^{~~}$} && { -0.0000$^{~~}$} &  { 0.0001$^{~~}$}  & { 0.0000$^{~~}$} & { 0.0002$^{~~}$} && { -0.0009$^{~~}$} & { 0.0010$^{~~}$} & { -0.0004$^{~~}$} & { 0.0003$^{~~}$}  \\
    &  26 && { -0.0006$^{~~}$} & { 0.0014$^{~~}$} && { -0.0002$^{~~}$} &  { 0.0010$^{~~}$} && { 0.0019$^{~~}$} &  { -0.0006$^{~~}$}& {\textbf{ 0.0069$^{*~}$}} &  { -0.0005$^{~~}$} && { 0.0006$^{~~}$} &  { 0.0005$^{~~}$}  & { 0.0005$^{~~}$} & { 0.0006$^{~~}$} && { -0.0002$^{~~}$} & { 0.0012$^{~~}$} & { -0.0003$^{~~}$} & { 0.0007$^{~~}$}  \\
    &  52 && { -0.0011$^{~~}$} & { 0.0007$^{~~}$} && { -0.0008$^{~~}$} &  { 0.0003$^{~~}$} && { 0.0011$^{~~}$} &  {\textbf{ -0.0012$^{*~}$}}& {\textbf{ 0.0059$^{**}$}} &  { -0.0011$^{~~}$} && { -0.0001$^{~~}$} &  { 0.0001$^{~~}$}  & { -0.0003$^{~~}$} & { -0.0002$^{~~}$} && { -0.0004$^{~~}$} & { 0.0003$^{~~}$} & { -0.0000$^{~~}$} & { 0.0002$^{~~}$}
    \smallskip\\
   IMOM &  1 && {\textbf{ 0.0034$^{**}$}} & { 0.0012$^{~~}$} && {\textbf{ 0.0032$^{**}$}} &  { 0.0015$^{~~}$} && { 0.0002$^{~~}$} &  {\textbf{ 0.0038$^{**}$}}& { 0.0025$^{~~}$} &  {\textbf{ 0.0054$^{**}$}} && {\textbf{ 0.0043$^{**}$}} &  { 0.0015$^{~~}$}  & {\textbf{ 0.0035$^{*~}$}} & {\textbf{ 0.0046$^{**}$}} && {\textbf{ 0.0036$^{**}$}} & {\textbf{ 0.0021$^{*~}$}} & { 0.0006$^{~~}$} & {\textbf{ 0.0041$^{*~}$}}  \\
    &  2 && {\textbf{ 0.0031$^{**}$}} & { 0.0012$^{~~}$} && {\textbf{ 0.0030$^{**}$}} &  { 0.0013$^{~~}$} && { 0.0004$^{~~}$} &  {\textbf{ 0.0034$^{**}$}}& { 0.0024$^{~~}$} &  {\textbf{ 0.0050$^{**}$}} && {\textbf{ 0.0039$^{**}$}} &  { 0.0014$^{~~}$}  & {\textbf{ 0.0033$^{*~}$}} & {\textbf{ 0.0042$^{**}$}} && {\textbf{ 0.0033$^{**}$}} & { 0.0020$^{~~}$} & { 0.0006$^{~~}$} & {\textbf{ 0.0035$^{*~}$}}  \\
    &  3 && {\textbf{ 0.0028$^{**}$}} & { 0.0013$^{~~}$} && {\textbf{ 0.0029$^{**}$}} &  { 0.0012$^{~~}$} && { 0.0003$^{~~}$} &  {\textbf{ 0.0032$^{**}$}}& { 0.0024$^{~~}$} &  {\textbf{ 0.0049$^{**}$}} && {\textbf{ 0.0037$^{**}$}} &  { 0.0014$^{~~}$}  & {\textbf{ 0.0031$^{*~}$}} & {\textbf{ 0.0041$^{**}$}} && {\textbf{ 0.0030$^{**}$}} & { 0.0020$^{~~}$} & { 0.0007$^{~~}$} & { 0.0031$^{~~}$}  \\
    &  4 && {\textbf{ 0.0025$^{**}$}} & { 0.0013$^{~~}$} && {\textbf{ 0.0027$^{**}$}} &  { 0.0012$^{~~}$} && { 0.0003$^{~~}$} &  {\textbf{ 0.0031$^{**}$}}& { 0.0025$^{~~}$} &  {\textbf{ 0.0047$^{**}$}} && {\textbf{ 0.0033$^{**}$}} &  { 0.0014$^{~~}$}  & { 0.0030$^{~~}$} & {\textbf{ 0.0040$^{**}$}} && {\textbf{ 0.0027$^{**}$}} & { 0.0019$^{~~}$} & { 0.0008$^{~~}$} & { 0.0026$^{~~}$}  \\
    &  8 && {\textbf{ 0.0021$^{*~}$}} & { 0.0014$^{~~}$} && {\textbf{ 0.0024$^{*~}$}} &  { 0.0011$^{~~}$} && { 0.0003$^{~~}$} &  {\textbf{ 0.0028$^{**}$}}& { 0.0031$^{~~}$} &  {\textbf{ 0.0044$^{**}$}} && {\textbf{ 0.0027$^{**}$}} &  { 0.0014$^{~~}$}  & { 0.0025$^{~~}$} & {\textbf{ 0.0039$^{**}$}} && {\textbf{ 0.0021$^{*~}$}} & {\textbf{ 0.0020$^{*~}$}} & { 0.0009$^{~~}$} & { 0.0017$^{~~}$}  \\
    &  13 && {\textbf{ 0.0019$^{*~}$}} & { 0.0014$^{~~}$} && {\textbf{ 0.0022$^{*~}$}} &  { 0.0011$^{~~}$} && { 0.0005$^{~~}$} &  {\textbf{ 0.0024$^{**}$}}& { 0.0036$^{~~}$} &  {\textbf{ 0.0040$^{**}$}} && {\textbf{ 0.0024$^{*~}$}} &  { 0.0015$^{~~}$}  & { 0.0025$^{~~}$} & {\textbf{ 0.0037$^{**}$}} && {\textbf{ 0.0018$^{*~}$}} & {\textbf{ 0.0021$^{*~}$}} & { 0.0014$^{~~}$} & { 0.0014$^{~~}$}  \\
    &  26 && { 0.0015$^{~~}$} & { 0.0014$^{~~}$} && {\textbf{ 0.0018$^{*~}$}} &  { 0.0010$^{~~}$} && { 0.0008$^{~~}$} &  {\textbf{ 0.0019$^{*~}$}}& {\textbf{ 0.0037$^{*~}$}} &  {\textbf{ 0.0032$^{*~}$}} && {\textbf{ 0.0020$^{*~}$}} &  { 0.0015$^{~~}$}  & { 0.0024$^{~~}$} & {\textbf{ 0.0032$^{**}$}} && {\textbf{ 0.0017$^{*~}$}} & {\textbf{ 0.0018$^{*~}$}} & { 0.0016$^{~~}$} & { 0.0012$^{~~}$}  \\
    &  52 && { 0.0011$^{~~}$} & { 0.0013$^{~~}$} && { 0.0014$^{~~}$} &  { 0.0010$^{~~}$} && { 0.0010$^{~~}$} &  {\textbf{ 0.0013$^{*~}$}}& {\textbf{ 0.0041$^{**}$}} &  {\textbf{ 0.0024$^{*~}$}} && {\textbf{ 0.0016$^{*~}$}} &  {\textbf{ 0.0014$^{*~}$}}  & {\textbf{ 0.0020$^{*~}$}} & {\textbf{ 0.0026$^{**}$}} && {\textbf{ 0.0015$^{*~}$}} & {\textbf{ 0.0015$^{*~}$}} & { 0.0010$^{~~}$} & { 0.0014$^{~~}$}
        \smallskip\\
   IVol-IMOM &  1 && {\textbf{ 0.0034$^{**}$}} & { 0.0013$^{~~}$} && {\textbf{ 0.0033$^{**}$}} &  { 0.0015$^{~~}$} && { 0.0003$^{~~}$} &  {\textbf{ 0.0039$^{**}$}}& { 0.0025$^{~~}$} &  {\textbf{ 0.0054$^{**}$}} && {\textbf{ 0.0043$^{**}$}} &  { 0.0015$^{~~}$}  & {\textbf{ 0.0035$^{*~}$}} & {\textbf{ 0.0047$^{**}$}} && {\textbf{ 0.0036$^{**}$}} & {\textbf{ 0.0022$^{*~}$}} & { 0.0006$^{~~}$} & {\textbf{ 0.0041$^{*~}$}}  \\
    &  2 && {\textbf{ 0.0031$^{**}$}} & { 0.0012$^{~~}$} && {\textbf{ 0.0031$^{**}$}} &  { 0.0013$^{~~}$} && { 0.0004$^{~~}$} &  {\textbf{ 0.0035$^{**}$}}& { 0.0023$^{~~}$} &  {\textbf{ 0.0049$^{**}$}} && {\textbf{ 0.0039$^{**}$}} &  { 0.0014$^{~~}$}  & {\textbf{ 0.0032$^{*~}$}} & {\textbf{ 0.0044$^{**}$}} && {\textbf{ 0.0033$^{**}$}} & {\textbf{ 0.0020$^{*~}$}} & { 0.0006$^{~~}$} & {\textbf{ 0.0035$^{*~}$}}  \\
    &  3 && {\textbf{ 0.0028$^{**}$}} & { 0.0013$^{~~}$} && {\textbf{ 0.0029$^{**}$}} &  { 0.0012$^{~~}$} && { 0.0003$^{~~}$} &  {\textbf{ 0.0033$^{**}$}}& { 0.0023$^{~~}$} &  {\textbf{ 0.0049$^{**}$}} && {\textbf{ 0.0036$^{**}$}} &  { 0.0014$^{~~}$}  & {\textbf{ 0.0031$^{*~}$}} & {\textbf{ 0.0042$^{**}$}} && {\textbf{ 0.0030$^{**}$}} & { 0.0020$^{~~}$} & { 0.0007$^{~~}$} & { 0.0031$^{~~}$}  \\
    &  4 && {\textbf{ 0.0025$^{**}$}} & { 0.0014$^{~~}$} && {\textbf{ 0.0027$^{**}$}} &  { 0.0012$^{~~}$} && { 0.0003$^{~~}$} &  {\textbf{ 0.0031$^{**}$}}& { 0.0025$^{~~}$} &  {\textbf{ 0.0047$^{**}$}} && {\textbf{ 0.0033$^{**}$}} &  { 0.0014$^{~~}$}  & { 0.0030$^{~~}$} & {\textbf{ 0.0042$^{**}$}} && {\textbf{ 0.0027$^{**}$}} & {\textbf{ 0.0020$^{*~}$}} & { 0.0007$^{~~}$} & { 0.0025$^{~~}$}  \\
    &  8 && {\textbf{ 0.0021$^{*~}$}} & { 0.0014$^{~~}$} && {\textbf{ 0.0024$^{*~}$}} &  { 0.0011$^{~~}$} && { 0.0003$^{~~}$} &  {\textbf{ 0.0028$^{**}$}}& { 0.0031$^{~~}$} &  {\textbf{ 0.0044$^{**}$}} && {\textbf{ 0.0027$^{**}$}} &  { 0.0015$^{~~}$}  & { 0.0025$^{~~}$} & {\textbf{ 0.0040$^{**}$}} && {\textbf{ 0.0021$^{*~}$}} & {\textbf{ 0.0021$^{*~}$}} & { 0.0009$^{~~}$} & { 0.0017$^{~~}$}  \\
    &  13 && {\textbf{ 0.0020$^{*~}$}} & { 0.0014$^{~~}$} && {\textbf{ 0.0022$^{*~}$}} &  { 0.0012$^{~~}$} && { 0.0005$^{~~}$} &  {\textbf{ 0.0025$^{**}$}}& {\textbf{ 0.0036$^{*~}$}} &  {\textbf{ 0.0040$^{**}$}} && {\textbf{ 0.0024$^{*~}$}} &  { 0.0015$^{~~}$}  & { 0.0025$^{~~}$} & {\textbf{ 0.0038$^{**}$}} && {\textbf{ 0.0018$^{*~}$}} & {\textbf{ 0.0021$^{*~}$}} & { 0.0014$^{~~}$} & { 0.0013$^{~~}$}  \\
    &  26 && { 0.0015$^{~~}$} & { 0.0014$^{~~}$} && {\textbf{ 0.0018$^{*~}$}} &  { 0.0011$^{~~}$} && { 0.0008$^{~~}$} &  {\textbf{ 0.0019$^{*~}$}}& {\textbf{ 0.0038$^{*~}$}} &  {\textbf{ 0.0033$^{*~}$}} && {\textbf{ 0.0020$^{*~}$}} &  { 0.0015$^{~~}$}  & { 0.0025$^{~~}$} & {\textbf{ 0.0033$^{**}$}} && {\textbf{ 0.0017$^{*~}$}} & {\textbf{ 0.0018$^{*~}$}} & { 0.0016$^{~~}$} & { 0.0012$^{~~}$}  \\
    &  52 && { 0.0011$^{~~}$} & { 0.0014$^{~~}$} && { 0.0014$^{~~}$} &  { 0.0010$^{~~}$} && { 0.0011$^{~~}$} &  {\textbf{ 0.0014$^{*~}$}}& {\textbf{ 0.0042$^{**}$}} &  {\textbf{ 0.0024$^{*~}$}} && {\textbf{ 0.0016$^{*~}$}} &  {\textbf{ 0.0015$^{*~}$}}  & {\textbf{ 0.0020$^{*~}$}} & {\textbf{ 0.0026$^{**}$}} && {\textbf{ 0.0015$^{*~}$}} & {\textbf{ 0.0015$^{*~}$}} & { 0.0010$^{~~}$} & { 0.0014$^{~~}$}
        \smallskip\\
   IMD-IMOM &  1 && {\textbf{ 0.0034$^{**}$}} & { 0.0012$^{~~}$} && {\textbf{ 0.0032$^{**}$}} &  { 0.0015$^{~~}$} && { -0.0003$^{~~}$} &  {\textbf{ 0.0042$^{**}$}}& { 0.0021$^{~~}$} &  {\textbf{ 0.0062$^{**}$}} && {\textbf{ 0.0043$^{**}$}} &  { 0.0016$^{~~}$}  & { 0.0029$^{~~}$} & {\textbf{ 0.0049$^{**}$}} && {\textbf{ 0.0041$^{**}$}} & { 0.0017$^{~~}$} & { 0.0002$^{~~}$} & { 0.0034$^{~~}$}  \\
    &  2 && {\textbf{ 0.0033$^{**}$}} & { 0.0013$^{~~}$} && {\textbf{ 0.0032$^{**}$}} &  { 0.0014$^{~~}$} && { -0.0000$^{~~}$} &  {\textbf{ 0.0039$^{**}$}}& { 0.0019$^{~~}$} &  {\textbf{ 0.0058$^{**}$}} && {\textbf{ 0.0042$^{**}$}} &  { 0.0016$^{~~}$}  & { 0.0031$^{~~}$} & {\textbf{ 0.0048$^{**}$}} && {\textbf{ 0.0038$^{**}$}} & { 0.0019$^{~~}$} & { 0.0004$^{~~}$} & { 0.0032$^{~~}$}  \\
    &  3 && {\textbf{ 0.0030$^{**}$}} & { 0.0014$^{~~}$} && {\textbf{ 0.0031$^{**}$}} &  { 0.0014$^{~~}$} && { 0.0001$^{~~}$} &  {\textbf{ 0.0037$^{**}$}}& { 0.0022$^{~~}$} &  {\textbf{ 0.0057$^{**}$}} && {\textbf{ 0.0039$^{**}$}} &  { 0.0016$^{~~}$}  & { 0.0028$^{~~}$} & {\textbf{ 0.0047$^{**}$}} && {\textbf{ 0.0035$^{**}$}} & { 0.0019$^{~~}$} & { 0.0004$^{~~}$} & { 0.0030$^{~~}$}  \\
    &  4 && {\textbf{ 0.0028$^{**}$}} & { 0.0016$^{~~}$} && {\textbf{ 0.0029$^{**}$}} &  { 0.0015$^{~~}$} && { 0.0002$^{~~}$} &  {\textbf{ 0.0036$^{**}$}}& { 0.0026$^{~~}$} &  {\textbf{ 0.0054$^{**}$}} && {\textbf{ 0.0037$^{**}$}} &  { 0.0016$^{~~}$}  & { 0.0029$^{~~}$} & {\textbf{ 0.0046$^{**}$}} && {\textbf{ 0.0032$^{**}$}} & { 0.0021$^{~~}$} & { 0.0005$^{~~}$} & { 0.0026$^{~~}$}  \\
    &  8 && {\textbf{ 0.0024$^{*~}$}} & { 0.0016$^{~~}$} && {\textbf{ 0.0027$^{*~}$}} &  { 0.0014$^{~~}$} && { 0.0004$^{~~}$} &  {\textbf{ 0.0031$^{**}$}}& { 0.0033$^{~~}$} &  {\textbf{ 0.0048$^{**}$}} && {\textbf{ 0.0028$^{**}$}} &  { 0.0018$^{~~}$}  & { 0.0022$^{~~}$} & {\textbf{ 0.0044$^{**}$}} && {\textbf{ 0.0022$^{*~}$}} & {\textbf{ 0.0023$^{*~}$}} & { 0.0006$^{~~}$} & { 0.0017$^{~~}$}  \\
    &  13 && {\textbf{ 0.0022$^{*~}$}} & { 0.0016$^{~~}$} && {\textbf{ 0.0024$^{*~}$}} &  { 0.0015$^{~~}$} && { 0.0006$^{~~}$} &  {\textbf{ 0.0029$^{**}$}}& { 0.0034$^{~~}$} &  {\textbf{ 0.0045$^{**}$}} && {\textbf{ 0.0024$^{*~}$}} &  {\textbf{ 0.0019$^{*~}$}}  & { 0.0021$^{~~}$} & {\textbf{ 0.0042$^{**}$}} && {\textbf{ 0.0019$^{*~}$}} & {\textbf{ 0.0024$^{*~}$}} & { 0.0011$^{~~}$} & { 0.0012$^{~~}$}  \\
    &  26 && {\textbf{ 0.0017$^{*~}$}} & { 0.0016$^{~~}$} && {\textbf{ 0.0019$^{*~}$}} &  { 0.0014$^{~~}$} && { 0.0008$^{~~}$} &  {\textbf{ 0.0023$^{**}$}}& {\textbf{ 0.0035$^{*~}$}} &  {\textbf{ 0.0037$^{**}$}} && {\textbf{ 0.0020$^{*~}$}} &  {\textbf{ 0.0018$^{*~}$}}  & { 0.0023$^{~~}$} & {\textbf{ 0.0035$^{**}$}} && {\textbf{ 0.0018$^{*~}$}} & {\textbf{ 0.0021$^{*~}$}} & { 0.0017$^{~~}$} & { 0.0011$^{~~}$}  \\
    &  52 && { 0.0014$^{~~}$} & { 0.0014$^{~~}$} && {\textbf{ 0.0016$^{*~}$}} &  { 0.0012$^{~~}$} && { 0.0010$^{~~}$} &  {\textbf{ 0.0017$^{*~}$}}& {\textbf{ 0.0038$^{*~}$}} &  {\textbf{ 0.0027$^{*~}$}} && {\textbf{ 0.0018$^{*~}$}} &  {\textbf{ 0.0016$^{*~}$}}  & {\textbf{ 0.0023$^{*~}$}} & {\textbf{ 0.0028$^{**}$}} && {\textbf{ 0.0016$^{*~}$}} & {\textbf{ 0.0018$^{**}$}} & { 0.0011$^{~~}$} & { 0.0015$^{~~}$}  \\
   \hline\hline
   \end{tabular}
   \label{TB:Condition}
\end{table}
\end{landscape}

The results are presented in \autoref{TB:Condition}.
Apparently, the performance regarding the momentum portfolios (including various versions) are highly dependent on the aforementioned factors.
At first, it is evident that more favorable performance is related to upside market states, except that the MOM portfolios remain insignificant in periods with either states.
All the statistically significant portfolios regarding the IMOMs appear in the periods with upside market states, while no portfolios are significant in the periods with downside states. In addition, we also observe a considerable increase in return magnitudes in the periods with upside states, particularly in the short term. For instance, pure IMOMs have weekly average returns about 0.24\% when $K = 1$ week in \autoref{TB:IMOM:detail:F5}, while it increases up to 0.34\% in the periods with upside states. In fact, in the short term, returns of most portfolios increase to the new highs by about 10 basis points on average.
\textcolor{black}{
Our finding above can be explained by combining different theories.
Due to the limit-to-arbitrage, such as short-sale constraints featuring the Chinese market,
stocks that are less accurately priced would be more overvalued during the period of upside market state.
This further results in more asymmetric arbitrage, and thereby the more pronouned return predictability based on the IVol \citep{Gu-Kang-Xu-2018-JBF}.
Another feasible explanation is shed light by investor sentiment.
Upside market state is typically related to high level of investor sentiment, thus leading to high level of comovement \citep{Chue-Gul-Mian-2019-JBF}. In this context, investors generally allocate more attention on market-level information, while overlooking firm-specific information. In doing so, investor underreaction to idiosyncratic information would be intensified.
Accordingly, the IMOMs, IVol-IMOMs and IMD-IMOMs would be more profitable during the period of upside market state, as shown in the \autoref{TB:Condition}.
}

Our results also unveil that market liquidity does influence momentum performance significantly. We readily observe that all of the portfolios with respect to the IMOM are statistically significant in the periods with both high and extremely high levels of market liquidity. Specifically, portfolios regarding the IMOM achieve even higher returns than those in periods with upside market states. This is highlighted by the fact that, in the periods with extremely high level of liquidity, the associated returns are twice as their raw returns.
\textcolor{black}{It is noted that our findings are in accordance with those associated with momentum portfolios \citep{Avramov-Cheng-Hameed-2016-JFQA}. As pointed out in \cite{Avramov-Cheng-Hameed-2016-JFQA}, within the framework of limit-to-arbitrage, the profitability of arbitrage portfolios would be weakened (enhanced) when the market is liquid (illiquid).
In other words, arbitrage cost would be lower (higher) during the period with lower (higher) level of market illiquidity.
Following this logic, the positive relation is expected to exist between illiquidity level and profitability regarding the IMOMs and their risk-adjusted versions, which is exactly as opposed to our findings.
Therefore, further in-depth study is needed, which will be one of focal points in our future research.}

The results of sentiment are also of interest. As for the sentiment, it is evident that all of the portfolios regarding the IMOM are significant in the periods with high sentiment levels.
As documented by \cite{Antoniou-Doukas-Subrahmanyam-2013-JFQA}, ``bad
(good) news among loser (winner) stocks will diffuse slowly when sentiment is
optimistic (pessimistic)''. Because of cognitive dissonance and costly short-selling, there would be an asymmetric momentum that mainly depends on the negative returns drift of loser portfolios.
Accordingly, a negative abnormal return drift of loser portfolio mainly contributes to the idiosyncratic momentum effect in periods with high levels of sentiment, while in the periods with low levels of sentiment, a positive abnormal return drift of winner portfolios mainly contributes to the idiosyncratic momentum effect.
Unfortunately, we obtain the results that loser portfolios have smaller magnitudes of abnormal returns than those of winners at high levels of sentiment,%
\footnote{ The results could be provided upon the request. }
which implies that further theoretical analysis is needed.
Additionally, we also observe that the portfolios regarding the IMOM are significant in the periods with extremely low level of sentiment, which is not very counter-intuitive. Extremely low level of sentiment reflects the pessimism about future.
Investors are likely to demonstrate more inattention in this situation, and have no incentive to enter into the positions of equities. Despite of the cognitive dissonance, this would also lead to the much slower diffusion of good or bad news, thus giving rise to more pronounced IMOM effects, as unveiled by our results.

We additionally find that returns are statistical significant mainly in periods with high levels of sentiment change, which is also consistent with the arguments mentioned above.
In comparison, the returns of a few of portfolios are statistically significant in the periods with low levels of sentiment change, which indicates the asymmetric momentum associated with lower costs of buying the winners.

In summary, we conclude that the better performance with respect to the IMOM is closely related to the upside market state, the high levels of market liquidity and the high levels of investor sentiment.

\section{Conclusion}
\label{S1:Conclusion}

This paper concentrates on the short-term idiosyncratic momentum (IMOM) as well as its risk-adjusted versions with respect to various idiosyncratic risk metrics. Specifically, we attempt to evaluate the performance of short-term IMOM effects and explore more profitable risk-adjusted IMOMs.

Taking the A-share individual stocks in the Chinese market as data sample from January 1997 to December 2017, we first evaluate the performance of weekly momentum and idiosyncratic momentum based on raw returns and idiosyncratic returns, respectively.
We find a more prevailing contrarian effect and a pronounced IMOM in the whole sample period, which implies the influence of time-varying common factors to pure momentum portfolios.

After that, statistical tests are conducted to investigate the predictability with respect to various idiosyncratic risk metrics. The univariate portfolio analysis confirms the negative relation between most of the idiosyncratic risk metrics and cross-sectional returns, which is consistent with behavioral finance theories.
More importantly, better performance is found to be related to IVol and IMD. Spanning tests also provide evidence for the better explanatory power of IVol and IMD upon others. In addition, our results also unveil that the price limits mechanism might function well in alleviating the extreme pessimism or optimism of investors in China.

Based on the preceding tests, we further conduct a comparison study on the performance of the IMOM portfolios with respect to various risk metrics. Different risk adjustment methods are employed and we obtain the robust results revealing the outperformance of IVol-IMOMs and IMD-IMOMs, particularly in the bivariate portfolio analysis. Spanning tests demonstrate that IVol-IMOMs and IMD-IMOMs exhibit a more powerful explanation to other risk-based IMOM portfolios as well.

Finally, the study is conducted to explore the possible explanations to IMOMs as well as risk-adjusted IMOMs, including IVol-IMOMs and IMD-IMOMs. We find that the performance of portfolios regarding the IMOMs is closely linked to market states, illiquidity and sentiment. Specifically, upside market state, high levels of liquidity and high levels of investor sentiment give rise to the higher profitability of the IMOMs and risk-adjusted IMOMs.

\section*{Acknowledgements}

This work was partly supported by Humanities and Social Sciences Fund of Ministry of Education of China [grant number 20YJC790113], the Project of Philosophy and Social Science Research in Colleges and Universities in Jiangsu Province [grant number 2019SJA0156], the Startup Foundation for Introducing Talent of NUIST, and the Fundamental Research Funds for the Central Universities.

%

\newpage
\appendix
\section{Additional tables}

\setcounter{table}{0}
\begin{table}[!ht]
\small
\caption{Average weekly returns of momentum portfolios. This table reports the average weekly returns of momentum portfolios for different $J-K$ portfolios. The whole sample period is January 1997 to December 2017. The results for whole period and three sub-periods are reported in Panel A, B, C and D.  \cite{Newey-West-1987-Em}'s $t$-statistics are obtained and the superscripts * and ** denote the significance at 5\% and 1\% levels, respectively.}
\centering
\vspace{-3mm}
   \begin{tabular}{ccccccccc}
   \hline
     $J$ & $K=1$ & $2$ & $3$ & $4$ &  $8$ & $13$ & $26$ & $52$ \\
   \hline
 \multicolumn{9}{l}{Panel A: whole period from 1997 to 2017} \\
   2 &  {\textbf{ -0.0020$^{*~}$}} & { -0.0006$^{~~}$}  & { 0.0006$^{~~}$} & {\textbf{ 0.0013$^{*~}$}} & {\textbf{ 0.0010$^{**}$}} & {\textbf{ 0.0009$^{**}$}} & { 0.0002$^{~~}$} & { 0.0000$^{~~}$}  \\
   3 &  { -0.0002$^{~~}$} & { 0.0010$^{~~}$}  & {\textbf{ 0.0019$^{**}$}} & {\textbf{ 0.0022$^{**}$}} & {\textbf{ 0.0015$^{**}$}} & {\textbf{ 0.0013$^{**}$}} & { 0.0003$^{~~}$} & { 0.0001$^{~~}$}  \\
   4 &  { 0.0007$^{~~}$} & {\textbf{ 0.0019$^{*~}$}}  & {\textbf{ 0.0023$^{**}$}} & {\textbf{ 0.0024$^{**}$}} & {\textbf{ 0.0016$^{**}$}} & {\textbf{ 0.0013$^{**}$}} & { 0.0002$^{~~}$} & { 0.0001$^{~~}$}  \\
   8 &  { 0.0018$^{~~}$} & {\textbf{ 0.0019$^{*~}$}}  & {\textbf{ 0.0021$^{*~}$}} & {\textbf{ 0.0022$^{**}$}} & {\textbf{ 0.0017$^{**}$}} & {\textbf{ 0.0013$^{*~}$}} & { -0.0000$^{~~}$} & { 0.0001$^{~~}$}  \\
   13 &  {\textbf{ 0.0022$^{*~}$}} & {\textbf{ 0.0024$^{*~}$}}  & {\textbf{ 0.0023$^{*~}$}} & {\textbf{ 0.0022$^{*~}$}} & { 0.0014$^{~~}$} & { 0.0009$^{~~}$} & { -0.0003$^{~~}$} & { 0.0001$^{~~}$}  \\
   26 &  { 0.0011$^{~~}$} & { 0.0009$^{~~}$}  & { 0.0009$^{~~}$} & { 0.0008$^{~~}$} & { 0.0003$^{~~}$} & { 0.0000$^{~~}$} & { -0.0004$^{~~}$} & { 0.0003$^{~~}$}  \\
   52 &  { 0.0005$^{~~}$} & { 0.0005$^{~~}$}  & { 0.0005$^{~~}$} & { 0.0006$^{~~}$} & { 0.0003$^{~~}$} & { 0.0004$^{~~}$} & { 0.0002$^{~~}$} & { 0.0007$^{~~}$}  \\
     & & & & & & & & \\
 \multicolumn{9}{l}{Panel B: sub-period from 1997 to 2003} \\
   2 &  {\textbf{ -0.0049$^{**}$}} & {\textbf{ -0.0035$^{**}$}}  & { -0.0017$^{~~}$} & { -0.0006$^{~~}$} & { 0.0006$^{~~}$} & { 0.0009$^{~~}$} & { -0.0003$^{~~}$} & { -0.0000$^{~~}$}  \\
   3 &  {\textbf{ -0.0037$^{**}$}} & { -0.0022$^{~~}$}  & { -0.0007$^{~~}$} & { 0.0003$^{~~}$} & { 0.0011$^{~~}$} & { 0.0013$^{~~}$} & { -0.0002$^{~~}$} & { 0.0001$^{~~}$}  \\
   4 &  {\textbf{ -0.0032$^{*~}$}} & { -0.0016$^{~~}$}  & { -0.0004$^{~~}$} & { 0.0005$^{~~}$} & { 0.0013$^{~~}$} & { 0.0013$^{~~}$} & { -0.0003$^{~~}$} & { 0.0000$^{~~}$}  \\
   8 &  { -0.0005$^{~~}$} & { 0.0002$^{~~}$}  & { 0.0009$^{~~}$} & { 0.0016$^{~~}$} & { 0.0022$^{~~}$} & { 0.0014$^{~~}$} & { -0.0005$^{~~}$} & { -0.0000$^{~~}$}  \\
   13 &  { -0.0001$^{~~}$} & { 0.0008$^{~~}$}  & { 0.0014$^{~~}$} & { 0.0016$^{~~}$} & { 0.0014$^{~~}$} & { 0.0004$^{~~}$} & { -0.0009$^{~~}$} & { -0.0001$^{~~}$}  \\
   26 &  { -0.0013$^{~~}$} & { -0.0012$^{~~}$}  & { -0.0008$^{~~}$} & { -0.0006$^{~~}$} & { -0.0005$^{~~}$} & { -0.0005$^{~~}$} & { -0.0006$^{~~}$} & { 0.0002$^{~~}$}  \\
   52 &  { -0.0016$^{~~}$} & { -0.0011$^{~~}$}  & { -0.0006$^{~~}$} & { -0.0003$^{~~}$} & { 0.0000$^{~~}$} & { 0.0000$^{~~}$} & { -0.0000$^{~~}$} & { 0.0008$^{~~}$}  \\
     & & & & & & & & \\
 \multicolumn{9}{l}{Panel C: sub-period from 2004 to 2010} \\
   2 &  {\textbf{ -0.0037$^{**}$}} & { -0.0010$^{~~}$}  & { 0.0008$^{~~}$} & { 0.0020$^{~~}$} & { 0.0014$^{~~}$} & { 0.0008$^{~~}$} & { 0.0004$^{~~}$} & { 0.0001$^{~~}$}  \\
   3 &  { -0.0005$^{~~}$} & { 0.0018$^{~~}$}  & {\textbf{ 0.0027$^{*~}$}} & {\textbf{ 0.0034$^{**}$}} & {\textbf{ 0.0020$^{*~}$}} & { 0.0013$^{~~}$} & { 0.0007$^{~~}$} & { 0.0003$^{~~}$}  \\
   4 &  { 0.0011$^{~~}$} & {\textbf{ 0.0029$^{*~}$}}  & {\textbf{ 0.0034$^{*~}$}} & {\textbf{ 0.0037$^{**}$}} & {\textbf{ 0.0021$^{*~}$}} & { 0.0013$^{~~}$} & { 0.0007$^{~~}$} & { 0.0003$^{~~}$}  \\
   8 &  { 0.0016$^{~~}$} & { 0.0023$^{~~}$}  & { 0.0023$^{~~}$} & { 0.0026$^{~~}$} & { 0.0014$^{~~}$} & { 0.0012$^{~~}$} & { 0.0005$^{~~}$} & { 0.0004$^{~~}$}  \\
   13 &  { 0.0020$^{~~}$} & { 0.0024$^{~~}$}  & { 0.0020$^{~~}$} & { 0.0020$^{~~}$} & { 0.0012$^{~~}$} & { 0.0011$^{~~}$} & { 0.0004$^{~~}$} & { 0.0005$^{~~}$}  \\
   26 &  { 0.0015$^{~~}$} & { 0.0017$^{~~}$}  & { 0.0013$^{~~}$} & { 0.0014$^{~~}$} & { 0.0010$^{~~}$} & { 0.0006$^{~~}$} & { 0.0001$^{~~}$} & { 0.0005$^{~~}$}  \\
   52 &  { 0.0007$^{~~}$} & { 0.0010$^{~~}$}  & { 0.0010$^{~~}$} & { 0.0012$^{~~}$} & { 0.0010$^{~~}$} & { 0.0012$^{~~}$} & { 0.0010$^{~~}$} & { 0.0005$^{~~}$}  \\
     & & & & & & & & \\
 \multicolumn{9}{l}{Panel D: sub-period from 2011 to 2017} \\
   2 &  { 0.0025$^{~~}$} & {\textbf{ 0.0027$^{*~}$}}  & {\textbf{ 0.0027$^{*~}$}} & {\textbf{ 0.0027$^{**}$}} & { 0.0012$^{~~}$} & {\textbf{ 0.0012$^{*~}$}} & { 0.0004$^{~~}$} & { 0.0002$^{~~}$}  \\
   3 &  {\textbf{ 0.0035$^{*~}$}} & {\textbf{ 0.0033$^{*~}$}}  & {\textbf{ 0.0037$^{**}$}} & {\textbf{ 0.0032$^{**}$}} & {\textbf{ 0.0016$^{*~}$}} & {\textbf{ 0.0015$^{*~}$}} & { 0.0005$^{~~}$} & { 0.0002$^{~~}$}  \\
   4 &  {\textbf{ 0.0042$^{*~}$}} & {\textbf{ 0.0042$^{**}$}}  & {\textbf{ 0.0040$^{**}$}} & {\textbf{ 0.0033$^{**}$}} & {\textbf{ 0.0017$^{*~}$}} & {\textbf{ 0.0015$^{*~}$}} & { 0.0004$^{~~}$} & { 0.0002$^{~~}$}  \\
   8 &  {\textbf{ 0.0040$^{*~}$}} & {\textbf{ 0.0032$^{*~}$}}  & {\textbf{ 0.0030$^{*~}$}} & {\textbf{ 0.0027$^{*~}$}} & { 0.0020$^{~~}$} & { 0.0015$^{~~}$} & { 0.0002$^{~~}$} & { 0.0003$^{~~}$}  \\
   13 &  {\textbf{ 0.0048$^{**}$}} & {\textbf{ 0.0041$^{*~}$}}  & {\textbf{ 0.0036$^{*~}$}} & {\textbf{ 0.0032$^{*~}$}} & { 0.0019$^{~~}$} & { 0.0016$^{~~}$} & { -0.0001$^{~~}$} & { 0.0004$^{~~}$}  \\
   26 &  {\textbf{ 0.0031$^{*~}$}} & { 0.0023$^{~~}$}  & { 0.0022$^{~~}$} & { 0.0019$^{~~}$} & { 0.0008$^{~~}$} & { 0.0003$^{~~}$} & { -0.0006$^{~~}$} & { 0.0004$^{~~}$}  \\
   52 &  { 0.0022$^{~~}$} & { 0.0015$^{~~}$}  & { 0.0013$^{~~}$} & { 0.0012$^{~~}$} & { 0.0005$^{~~}$} & { 0.0008$^{~~}$} & { 0.0004$^{~~}$} & { 0.0010$^{~~}$}  \\
   \hline
   \end{tabular}
   \label{TB:MOM:subperiod}
\end{table}

\begin{table}[!ht]
\small
\caption{Average weekly returns of idiosyncratic momentum portfolios. This table reports the average weekly returns of idiosyncratic momentum portfolios for different $J-K$ portfolios. The whole sample period is January 1997 to December 2017. The results for whole period and three sub-periods are reported in Panel A, B, C and D.  \cite{Newey-West-1987-Em}'s $t$-statistics are obtained and the superscripts * and ** denote the significance at 5\% and 1\% levels, respectively.}
\centering
\vspace{-3mm}
   \begin{tabular}{ccccccccc}
   \hline
     $J$ & $K=1$ & $2$ & $3$ & $4$ &  $8$ & $13$ & $26$ & $52$ \\
   \hline
 \multicolumn{9}{l}{Panel A: whole period from 1997 to 2017} \\
   2 &  {\textbf{ 0.0031$^{**}$}} & {\textbf{ 0.0029$^{**}$}}  & {\textbf{ 0.0028$^{**}$}} & {\textbf{ 0.0025$^{**}$}} & {\textbf{ 0.0016$^{**}$}} & {\textbf{ 0.0013$^{**}$}} & {\textbf{ 0.0008$^{**}$}} & {\textbf{ 0.0007$^{**}$}}  \\
   3 &  {\textbf{ 0.0036$^{**}$}} & {\textbf{ 0.0034$^{**}$}}  & {\textbf{ 0.0032$^{**}$}} & {\textbf{ 0.0028$^{**}$}} & {\textbf{ 0.0019$^{**}$}} & {\textbf{ 0.0015$^{**}$}} & {\textbf{ 0.0010$^{*~}$}} & {\textbf{ 0.0008$^{*~}$}}  \\
   4 &  {\textbf{ 0.0038$^{**}$}} & {\textbf{ 0.0034$^{**}$}}  & {\textbf{ 0.0032$^{**}$}} & {\textbf{ 0.0028$^{**}$}} & {\textbf{ 0.0020$^{**}$}} & {\textbf{ 0.0016$^{**}$}} & {\textbf{ 0.0010$^{*~}$}} & {\textbf{ 0.0008$^{*~}$}}  \\
   8 &  {\textbf{ 0.0033$^{**}$}} & {\textbf{ 0.0030$^{**}$}}  & {\textbf{ 0.0028$^{**}$}} & {\textbf{ 0.0026$^{**}$}} & {\textbf{ 0.0021$^{**}$}} & {\textbf{ 0.0017$^{**}$}} & {\textbf{ 0.0012$^{*~}$}} & {\textbf{ 0.0010$^{*~}$}}  \\
   13 &  {\textbf{ 0.0030$^{**}$}} & {\textbf{ 0.0028$^{**}$}}  & {\textbf{ 0.0026$^{**}$}} & {\textbf{ 0.0024$^{**}$}} & {\textbf{ 0.0020$^{**}$}} & {\textbf{ 0.0017$^{**}$}} & {\textbf{ 0.0012$^{*~}$}} & {\textbf{ 0.0011$^{*~}$}}  \\
   26 &  {\textbf{ 0.0024$^{**}$}} & {\textbf{ 0.0022$^{**}$}}  & {\textbf{ 0.0021$^{**}$}} & {\textbf{ 0.0020$^{**}$}} & {\textbf{ 0.0018$^{**}$}} & {\textbf{ 0.0016$^{*~}$}} & {\textbf{ 0.0014$^{*~}$}} & {\textbf{ 0.0012$^{*~}$}}  \\
   52 &  {\textbf{ 0.0019$^{**}$}} & {\textbf{ 0.0018$^{**}$}}  & {\textbf{ 0.0018$^{**}$}} & {\textbf{ 0.0017$^{*~}$}} & {\textbf{ 0.0016$^{*~}$}} & {\textbf{ 0.0016$^{*~}$}} & {\textbf{ 0.0013$^{*~}$}} & {\textbf{ 0.0011$^{*~}$}}  \\
     & & & & & & & & \\
 \multicolumn{9}{l}{Panel B: sub-period from 1997 to 2003} \\
   2 &  { 0.0010$^{~~}$} & { 0.0011$^{~~}$}  & { 0.0011$^{~~}$} & { 0.0009$^{~~}$} & { 0.0006$^{~~}$} & { 0.0006$^{~~}$} & { 0.0004$^{~~}$} & { 0.0003$^{~~}$}  \\
   3 &  { 0.0015$^{~~}$} & { 0.0015$^{~~}$}  & { 0.0015$^{~~}$} & { 0.0013$^{~~}$} & { 0.0008$^{~~}$} & { 0.0007$^{~~}$} & { 0.0005$^{~~}$} & { 0.0004$^{~~}$}  \\
   4 &  { 0.0016$^{~~}$} & { 0.0016$^{~~}$}  & { 0.0015$^{~~}$} & { 0.0013$^{~~}$} & { 0.0008$^{~~}$} & { 0.0007$^{~~}$} & { 0.0004$^{~~}$} & { 0.0003$^{~~}$}  \\
   8 &  { 0.0019$^{~~}$} & { 0.0018$^{~~}$}  & { 0.0018$^{~~}$} & { 0.0017$^{~~}$} & { 0.0013$^{~~}$} & { 0.0011$^{~~}$} & { 0.0008$^{~~}$} & { 0.0006$^{~~}$}  \\
   13 &  {\textbf{ 0.0023$^{*~}$}} & {\textbf{ 0.0023$^{*~}$}}  & { 0.0021$^{~~}$} & { 0.0019$^{~~}$} & { 0.0014$^{~~}$} & { 0.0013$^{~~}$} & { 0.0009$^{~~}$} & { 0.0007$^{~~}$}  \\
   26 &  { 0.0021$^{~~}$} & { 0.0021$^{~~}$}  & { 0.0019$^{~~}$} & { 0.0018$^{~~}$} & { 0.0015$^{~~}$} & { 0.0014$^{~~}$} & { 0.0013$^{~~}$} & { 0.0009$^{~~}$}  \\
   52 &  { 0.0020$^{~~}$} & { 0.0020$^{~~}$}  & { 0.0019$^{~~}$} & { 0.0018$^{~~}$} & { 0.0014$^{~~}$} & { 0.0013$^{~~}$} & { 0.0011$^{~~}$} & { 0.0009$^{~~}$}  \\
     & & & & & & & & \\
 \multicolumn{9}{l}{Panel C: sub-period from 2004 to 2010} \\
   2 &  {\textbf{ 0.0035$^{**}$}} & {\textbf{ 0.0036$^{**}$}}  & {\textbf{ 0.0034$^{**}$}} & {\textbf{ 0.0032$^{**}$}} & {\textbf{ 0.0019$^{**}$}} & {\textbf{ 0.0015$^{**}$}} & {\textbf{ 0.0009$^{*~}$}} & {\textbf{ 0.0008$^{*~}$}}  \\
   3 &  {\textbf{ 0.0042$^{**}$}} & {\textbf{ 0.0044$^{**}$}}  & {\textbf{ 0.0041$^{**}$}} & {\textbf{ 0.0037$^{**}$}} & {\textbf{ 0.0022$^{**}$}} & {\textbf{ 0.0017$^{**}$}} & { 0.0009$^{~~}$} & { 0.0008$^{~~}$}  \\
   4 &  {\textbf{ 0.0045$^{**}$}} & {\textbf{ 0.0044$^{**}$}}  & {\textbf{ 0.0041$^{**}$}} & {\textbf{ 0.0037$^{**}$}} & {\textbf{ 0.0025$^{**}$}} & {\textbf{ 0.0019$^{**}$}} & { 0.0010$^{~~}$} & { 0.0008$^{~~}$}  \\
   8 &  {\textbf{ 0.0041$^{**}$}} & {\textbf{ 0.0039$^{**}$}}  & {\textbf{ 0.0036$^{**}$}} & {\textbf{ 0.0033$^{**}$}} & {\textbf{ 0.0022$^{*~}$}} & {\textbf{ 0.0018$^{*~}$}} & { 0.0011$^{~~}$} & { 0.0010$^{~~}$}  \\
   13 &  {\textbf{ 0.0034$^{**}$}} & {\textbf{ 0.0032$^{**}$}}  & {\textbf{ 0.0030$^{**}$}} & {\textbf{ 0.0028$^{**}$}} & {\textbf{ 0.0020$^{*~}$}} & { 0.0015$^{~~}$} & { 0.0009$^{~~}$} & { 0.0010$^{~~}$}  \\
   26 &  { 0.0021$^{~~}$} & {\textbf{ 0.0021$^{*~}$}}  & { 0.0019$^{~~}$} & { 0.0016$^{~~}$} & { 0.0012$^{~~}$} & { 0.0010$^{~~}$} & { 0.0009$^{~~}$} & { 0.0010$^{~~}$}  \\
   52 &  { 0.0015$^{~~}$} & { 0.0013$^{~~}$}  & { 0.0013$^{~~}$} & { 0.0012$^{~~}$} & { 0.0010$^{~~}$} & { 0.0010$^{~~}$} & { 0.0009$^{~~}$} & { 0.0010$^{~~}$}  \\
     & & & & & & & & \\
 \multicolumn{9}{l}{Panel D: sub-period from 2011 to 2017} \\
   2 &  {\textbf{ 0.0046$^{**}$}} & {\textbf{ 0.0041$^{**}$}}  & {\textbf{ 0.0039$^{**}$}} & {\textbf{ 0.0035$^{**}$}} & {\textbf{ 0.0022$^{**}$}} & {\textbf{ 0.0018$^{**}$}} & { 0.0012$^{~~}$} & {\textbf{ 0.0012$^{*~}$}}  \\
   3 &  {\textbf{ 0.0050$^{**}$}} & {\textbf{ 0.0042$^{**}$}}  & {\textbf{ 0.0040$^{**}$}} & {\textbf{ 0.0035$^{**}$}} & {\textbf{ 0.0023$^{**}$}} & {\textbf{ 0.0019$^{*~}$}} & { 0.0014$^{~~}$} & {\textbf{ 0.0014$^{*~}$}}  \\
   4 &  {\textbf{ 0.0052$^{**}$}} & {\textbf{ 0.0043$^{**}$}}  & {\textbf{ 0.0039$^{**}$}} & {\textbf{ 0.0035$^{**}$}} & {\textbf{ 0.0026$^{**}$}} & {\textbf{ 0.0020$^{*~}$}} & { 0.0015$^{~~}$} & {\textbf{ 0.0015$^{*~}$}}  \\
   8 &  {\textbf{ 0.0040$^{**}$}} & {\textbf{ 0.0033$^{*~}$}}  & {\textbf{ 0.0031$^{*~}$}} & {\textbf{ 0.0029$^{*~}$}} & {\textbf{ 0.0024$^{*~}$}} & { 0.0019$^{~~}$} & { 0.0015$^{~~}$} & { 0.0016$^{~~}$}  \\
   13 &  {\textbf{ 0.0033$^{*~}$}} & {\textbf{ 0.0028$^{*~}$}}  & {\textbf{ 0.0027$^{*~}$}} & {\textbf{ 0.0026$^{*~}$}} & { 0.0022$^{~~}$} & { 0.0018$^{~~}$} & { 0.0016$^{~~}$} & { 0.0017$^{~~}$}  \\
   26 &  { 0.0029$^{~~}$} & { 0.0026$^{~~}$}  & { 0.0026$^{~~}$} & { 0.0026$^{~~}$} & { 0.0023$^{~~}$} & { 0.0021$^{~~}$} & { 0.0018$^{~~}$} & { 0.0017$^{~~}$}  \\
   52 &  { 0.0023$^{~~}$} & { 0.0021$^{~~}$}  & { 0.0022$^{~~}$} & { 0.0022$^{~~}$} & { 0.0021$^{~~}$} & { 0.0019$^{~~}$} & { 0.0017$^{~~}$} & { 0.0015$^{~~}$}  \\
   \hline
   \end{tabular}
   \label{TB:IMOM:subperiod:F5}
\end{table}

\begin{table}[!ht]
\small
\caption{Univariate portfolio analysis based on various idiosyncratic risk metrics. This table reports the results associated with univariate portfolios formed according to various idiosyncratic risk metrics in \autoref{TB:risks}. Average weekly returns (Raw), FF5F-$\alpha$ values ($\alpha$),  annualized Sharpe ratios (SR) and maximum drawdowns (MD) are presented in each panel. The sample period is from January 1997 to December 2017. At the beginning of each week, each risk metric is calculated using idiosyncratic returns over past 130 trading days for individual stocks, by which they can be sorted into ten decile groups. The zero-cost arbitrage portfolios can be constructed by buying the stocks from the decile group with lowest risk and selling stocks from group with highest risk. Portfolios would be held for $K$ weeks, and calendar-time method is applied to obtain the average weekly return. \cite{Newey-West-1987-Em}'s $t$-statistics are obtained and the superscripts * and ** denote the significance at 5\% and 1\% levels, respectively. }
\centering
\vspace{-3mm}
   \begin{tabular}{ccccccccc}
   \hline\hline
      & $K=1$ & $2$ & $3$ & $4$ &  $8$ & $13$ & $26$ & $52$ \\
   \hline
 \multicolumn{9}{l}{Panel A: Idiosyncratic volatility} \\
   Raw &  {\textbf{ 0.0024$^{**}$}} & {\textbf{ 0.0022$^{**}$}}  & {\textbf{ 0.0021$^{**}$}} & {\textbf{ 0.0020$^{**}$}} & {\textbf{ 0.0018$^{**}$}} & {\textbf{ 0.0017$^{**}$}} & {\textbf{ 0.0014$^{*~}$}} & {\textbf{ 0.0012$^{*~}$}}  \\
   $\alpha$ &  {\textbf{ 0.0034$^{**}$}} & {\textbf{ 0.0031$^{**}$}}  & {\textbf{ 0.0029$^{**}$}} & {\textbf{ 0.0027$^{**}$}} & {\textbf{ 0.0024$^{**}$}} & {\textbf{ 0.0021$^{**}$}} & {\textbf{ 0.0017$^{**}$}} & {\textbf{ 0.0014$^{**}$}}  \\
   SR &  { 0.7449$^{~~}$} & { 0.7020$^{~~}$}  & { 0.6753$^{~~}$} & { 0.6566$^{~~}$} & { 0.6431$^{~~}$} & { 0.6551$^{~~}$} & { 0.6530$^{~~}$} & { 0.6902$^{~~}$}  \\
   MD &  { 0.3176$^{~~}$} & { 0.3291$^{~~}$}  & { 0.3041$^{~~}$} & { 0.2855$^{~~}$} & { 0.2801$^{~~}$} & { 0.2605$^{~~}$} & { 0.2259$^{~~}$} & { 0.2323$^{~~}$}
       \smallskip\\
 \multicolumn{9}{l}{Panel B: Idiosyncratic skewness} \\
   Raw &  {\textbf{ 0.0009$^{*~}$}} & {\textbf{ 0.0009$^{*~}$}}  & {\textbf{ 0.0009$^{*~}$}} & {\textbf{ 0.0007$^{*~}$}} & { 0.0005$^{~~}$} & { 0.0002$^{~~}$} & { -0.0001$^{~~}$} & { -0.0001$^{~~}$}  \\
   $\alpha$ &  {\textbf{ 0.0010$^{*~}$}} & {\textbf{ 0.0010$^{**}$}}  & {\textbf{ 0.0009$^{**}$}} & {\textbf{ 0.0008$^{*~}$}} & { 0.0005$^{~~}$} & { 0.0002$^{~~}$} & { -0.0002$^{~~}$} & { -0.0001$^{~~}$}  \\
   SR &  { 0.5564$^{~~}$} & { 0.6037$^{~~}$}  & { 0.5975$^{~~}$} & { 0.5323$^{~~}$} & { 0.3774$^{~~}$} & { 0.1896$^{~~}$} & { -0.1365$^{~~}$} & { -0.1684$^{~~}$}  \\
   MD &  { 0.3733$^{~~}$} & { 0.3601$^{~~}$}  & { 0.3600$^{~~}$} & { 0.3489$^{~~}$} & { 0.3625$^{~~}$} & { 0.4250$^{~~}$} & { 0.4701$^{~~}$} & { 0.3887$^{~~}$}
      \smallskip\\
 \multicolumn{9}{l}{Panel C: Idiosyncratic kurtosis} \\
   Raw &  { 0.0007$^{~~}$} & { 0.0007$^{~~}$}  & { 0.0006$^{~~}$} & { 0.0005$^{~~}$} & { 0.0002$^{~~}$} & { -0.0000$^{~~}$} & { -0.0002$^{~~}$} & { -0.0003$^{~~}$}  \\
   $\alpha$ &  { 0.0008$^{~~}$} & { 0.0007$^{~~}$}  & { 0.0006$^{~~}$} & { 0.0006$^{~~}$} & { 0.0002$^{~~}$} & { 0.0000$^{~~}$} & { -0.0002$^{~~}$} & { -0.0003$^{~~}$}  \\
   SR &  { 0.3787$^{~~}$} & { 0.3638$^{~~}$}  & { 0.3283$^{~~}$} & { 0.2986$^{~~}$} & { 0.0991$^{~~}$} & { -0.0109$^{~~}$} & { -0.1491$^{~~}$} & { -0.3131$^{~~}$}  \\
   MD &  { 0.4371$^{~~}$} & { 0.4125$^{~~}$}  & { 0.3914$^{~~}$} & { 0.3797$^{~~}$} & { 0.4175$^{~~}$} & { 0.4161$^{~~}$} & { 0.4296$^{~~}$} & { 0.4681$^{~~}$}
       \smallskip\\
 \multicolumn{9}{l}{Panel D: Idiosyncratic maximum drawdown} \\
   Raw &  {\textbf{ 0.0015$^{*~}$}} & {\textbf{ 0.0017$^{**}$}}  & {\textbf{ 0.0017$^{**}$}} & {\textbf{ 0.0018$^{**}$}} & {\textbf{ 0.0017$^{**}$}} & {\textbf{ 0.0016$^{**}$}} & {\textbf{ 0.0014$^{**}$}} & {\textbf{ 0.0012$^{**}$}}  \\
   $\alpha$ &  {\textbf{ 0.0020$^{**}$}} & {\textbf{ 0.0022$^{**}$}}  & {\textbf{ 0.0022$^{**}$}} & {\textbf{ 0.0022$^{**}$}} & {\textbf{ 0.0021$^{**}$}} & {\textbf{ 0.0019$^{**}$}} & {\textbf{ 0.0016$^{**}$}} & {\textbf{ 0.0013$^{**}$}}  \\
   SR &  { 0.5586$^{~~}$} & { 0.6537$^{~~}$}  & { 0.6871$^{~~}$} & { 0.7155$^{~~}$} & { 0.7512$^{~~}$} & { 0.7796$^{~~}$} & { 0.8141$^{~~}$} & { 0.8758$^{~~}$}  \\
   MD &  { 0.3079$^{~~}$} & { 0.3132$^{~~}$}  & { 0.3052$^{~~}$} & { 0.2965$^{~~}$} & { 0.2786$^{~~}$} & { 0.2498$^{~~}$} & { 0.2138$^{~~}$} & { 0.2164$^{~~}$}
       \smallskip\\
 \multicolumn{9}{l}{Panel E: Idiosyncratic ES (5\%)} \\
   Raw &  {\textbf{ 0.0019$^{**}$}} & {\textbf{ 0.0016$^{*~}$}}  & {\textbf{ 0.0015$^{*~}$}} & {\textbf{ 0.0014$^{*~}$}} & {\textbf{ 0.0013$^{*~}$}} & {\textbf{ 0.0013$^{*~}$}} & {\textbf{ 0.0012$^{*~}$}} & { 0.0009$^{~~}$}  \\
   $\alpha$ &  {\textbf{ 0.0029$^{**}$}} & {\textbf{ 0.0025$^{**}$}}  & {\textbf{ 0.0023$^{**}$}} & {\textbf{ 0.0022$^{**}$}} & {\textbf{ 0.0020$^{**}$}} & {\textbf{ 0.0018$^{**}$}} & {\textbf{ 0.0015$^{**}$}} & {\textbf{ 0.0011$^{**}$}}  \\
   SR &  { 0.6101$^{~~}$} & { 0.5487$^{~~}$}  & { 0.5224$^{~~}$} & { 0.5145$^{~~}$} & { 0.5119$^{~~}$} & { 0.5533$^{~~}$} & { 0.5804$^{~~}$} & { 0.5771$^{~~}$}  \\
   MD &  { 0.3226$^{~~}$} & { 0.3398$^{~~}$}  & { 0.3145$^{~~}$} & { 0.2937$^{~~}$} & { 0.2872$^{~~}$} & { 0.2727$^{~~}$} & { 0.2194$^{~~}$} & { 0.2301$^{~~}$}
       \smallskip\\
 \multicolumn{9}{l}{Panel F: Idiosyncratic VaR (5\%)} \\
   Raw &  {\textbf{ 0.0017$^{*~}$}} & {\textbf{ 0.0016$^{*~}$}}  & {\textbf{ 0.0015$^{*~}$}} & {\textbf{ 0.0014$^{*~}$}} & {\textbf{ 0.0013$^{*~}$}} & {\textbf{ 0.0013$^{*~}$}} & {\textbf{ 0.0012$^{*~}$}} & { 0.0010$^{~~}$}  \\
   $\alpha$ &  {\textbf{ 0.0025$^{**}$}} & {\textbf{ 0.0023$^{**}$}}  & {\textbf{ 0.0022$^{**}$}} & {\textbf{ 0.0021$^{**}$}} & {\textbf{ 0.0019$^{**}$}} & {\textbf{ 0.0018$^{**}$}} & {\textbf{ 0.0015$^{**}$}} & {\textbf{ 0.0012$^{**}$}}  \\
   SR &  { 0.5422$^{~~}$} & { 0.5096$^{~~}$}  & { 0.4935$^{~~}$} & { 0.4848$^{~~}$} & { 0.4919$^{~~}$} & { 0.5393$^{~~}$} & { 0.5759$^{~~}$} & { 0.5896$^{~~}$}  \\
   MD &  { 0.3619$^{~~}$} & { 0.3713$^{~~}$}  & { 0.3448$^{~~}$} & { 0.3186$^{~~}$} & { 0.3016$^{~~}$} & { 0.2737$^{~~}$} & { 0.2316$^{~~}$} & { 0.2405$^{~~}$}
       \smallskip\\
 \multicolumn{9}{l}{Panel G: Idiosyncratic ES (1\%)} \\
   Raw &  {\textbf{ 0.0016$^{**}$}} & {\textbf{ 0.0014$^{*~}$}}  & {\textbf{ 0.0012$^{*~}$}} & {\textbf{ 0.0012$^{*~}$}} & {\textbf{ 0.0011$^{*~}$}} & {\textbf{ 0.0010$^{*~}$}} & { 0.0009$^{~~}$} & { 0.0007$^{~~}$}  \\
   $\alpha$ &  {\textbf{ 0.0024$^{**}$}} & {\textbf{ 0.0021$^{**}$}}  & {\textbf{ 0.0019$^{**}$}} & {\textbf{ 0.0018$^{**}$}} & {\textbf{ 0.0016$^{**}$}} & {\textbf{ 0.0015$^{**}$}} & {\textbf{ 0.0011$^{**}$}} & {\textbf{ 0.0009$^{*~}$}}  \\
   SR &  { 0.6189$^{~~}$} & { 0.5493$^{~~}$}  & { 0.5181$^{~~}$} & { 0.5026$^{~~}$} & { 0.5008$^{~~}$} & { 0.5138$^{~~}$} & { 0.5101$^{~~}$} & { 0.5327$^{~~}$}  \\
   MD &  { 0.2823$^{~~}$} & { 0.2960$^{~~}$}  & { 0.2793$^{~~}$} & { 0.2610$^{~~}$} & { 0.2460$^{~~}$} & { 0.2367$^{~~}$} & { 0.1951$^{~~}$} & { 0.2021$^{~~}$}
       \smallskip\\
 \multicolumn{9}{l}{Panel H: Idiosyncratic VaR (1\%)} \\
   Raw &  {\textbf{ 0.0016$^{*~}$}} & {\textbf{ 0.0014$^{*~}$}}  & {\textbf{ 0.0012$^{*~}$}} & { 0.0012$^{~~}$} & { 0.0010$^{~~}$} & { 0.0011$^{~~}$} & { 0.0010$^{~~}$} & { 0.0008$^{~~}$}  \\
   $\alpha$ &  {\textbf{ 0.0025$^{**}$}} & {\textbf{ 0.0022$^{**}$}}  & {\textbf{ 0.0020$^{**}$}} & {\textbf{ 0.0019$^{**}$}} & {\textbf{ 0.0016$^{**}$}} & {\textbf{ 0.0015$^{**}$}} & {\textbf{ 0.0013$^{**}$}} & {\textbf{ 0.0010$^{*~}$}}  \\
   SR &  { 0.5754$^{~~}$} & { 0.5088$^{~~}$}  & { 0.4716$^{~~}$} & { 0.4506$^{~~}$} & { 0.4300$^{~~}$} & { 0.4782$^{~~}$} & { 0.5194$^{~~}$} & { 0.5237$^{~~}$}  \\
   MD &  { 0.2965$^{~~}$} & { 0.3153$^{~~}$}  & { 0.2926$^{~~}$} & { 0.2719$^{~~}$} & { 0.2650$^{~~}$} & { 0.2497$^{~~}$} & { 0.2273$^{~~}$} & { 0.2161$^{~~}$}  \\
   \hline\hline
   \end{tabular}
   \label{TB:Risk:F5}
\end{table}

\begin{table}[!ht]
\small
\caption{Univariate portfolio analysis based on various idiosyncratic risk metrics. This table reports the results associated with long sides for univariate portfolios formed according to various idiosyncratic risk metrics in \autoref{TB:risks}. Average weekly returns (Raw), FF5F-$\alpha$s ($\alpha$), annualized Sharpe ratios (SR) and maximum drawdowns (MD) are presented in each panel. The sample period is from January 1997 to December 2017. At the beginning of each week, each risk metric is calculated using idiosyncratic returns over past 130 trading days for individual stocks, by which they can be sorted into ten decile groups. The long sides for portfolios can be constructed by buying the stocks from the decile group with lowest risk. Portfolios would be held for $K$ weeks, and calendar-time method is applied to obtain the average weekly return. \cite{Newey-West-1987-Em}'s $t$-statistics are obtained and the superscripts * and ** denote the significance at 5\% and 1\% levels, respectively.}
\centering
\vspace{-3mm}
   \begin{tabular}{ccccccccc}
   \hline
      & $K=1$ & $2$ & $3$ & $4$ &  $8$ & $13$ & $26$ & $52$ \\
   \hline
 \multicolumn{9}{l}{Panel A: Idiosyncratic volatility} \\
   Raw &  {\textbf{ 0.0033$^{*~}$}} & {\textbf{ 0.0033$^{*~}$}}  & {\textbf{ 0.0033$^{*~}$}} & {\textbf{ 0.0033$^{*~}$}} & {\textbf{ 0.0032$^{*~}$}} & {\textbf{ 0.0033$^{*~}$}} & {\textbf{ 0.0031$^{*~}$}} & {\textbf{ 0.0032$^{*~}$}}  \\
   $\alpha$ &  { 0.0006$^{~~}$} & { 0.0007$^{~~}$}  & {\textbf{ 0.0009$^{*~}$}} & {\textbf{ 0.0010$^{*~}$}} & {\textbf{ 0.0013$^{*~}$}} & {\textbf{ 0.0017$^{**}$}} & {\textbf{ 0.0021$^{**}$}} & {\textbf{ 0.0025$^{**}$}}  \\
   SR &  { 0.5380$^{~~}$} & { 0.5471$^{~~}$}  & { 0.5545$^{~~}$} & { 0.5639$^{~~}$} & { 0.5642$^{~~}$} & { 0.5917$^{~~}$} & { 0.6056$^{~~}$} & { 0.6822$^{~~}$}  \\
   MD &  { 0.6454$^{~~}$} & { 0.6502$^{~~}$}  & { 0.6407$^{~~}$} & { 0.6379$^{~~}$} & { 0.6400$^{~~}$} & { 0.6390$^{~~}$} & { 0.6328$^{~~}$} & { 0.6096$^{~~}$}  \\
     & & & & & & & & \\
 \multicolumn{9}{l}{Panel B: Idiosyncratic skewness} \\
   Raw &  {\textbf{ 0.0036$^{*~}$}} & {\textbf{ 0.0036$^{*~}$}}  & {\textbf{ 0.0036$^{*~}$}} & {\textbf{ 0.0036$^{*~}$}} & {\textbf{ 0.0034$^{*~}$}} & {\textbf{ 0.0033$^{*~}$}} & {\textbf{ 0.0029$^{*~}$}} & {\textbf{ 0.0030$^{*~}$}}  \\
   $\alpha$ &  { 0.0005$^{~~}$} & { 0.0007$^{~~}$}  & {\textbf{ 0.0008$^{*~}$}} & {\textbf{ 0.0009$^{*~}$}} & {\textbf{ 0.0011$^{*~}$}} & {\textbf{ 0.0014$^{*~}$}} & {\textbf{ 0.0018$^{*~}$}} & {\textbf{ 0.0022$^{*~}$}}  \\
   SR &  { 0.5202$^{~~}$} & { 0.5411$^{~~}$}  & { 0.5498$^{~~}$} & { 0.5537$^{~~}$} & { 0.5308$^{~~}$} & { 0.5313$^{~~}$} & { 0.5171$^{~~}$} & { 0.5950$^{~~}$}  \\
   MD &  { 0.6980$^{~~}$} & { 0.6814$^{~~}$}  & { 0.6720$^{~~}$} & { 0.6763$^{~~}$} & { 0.6828$^{~~}$} & { 0.6836$^{~~}$} & { 0.6728$^{~~}$} & { 0.6412$^{~~}$}  \\
     & & & & & & & & \\
 \multicolumn{9}{l}{Panel C: Idiosyncratic kurtosis} \\
   Raw &  {\textbf{ 0.0036$^{*~}$}} & {\textbf{ 0.0035$^{*~}$}}  & {\textbf{ 0.0035$^{*~}$}} & {\textbf{ 0.0035$^{*~}$}} & {\textbf{ 0.0032$^{*~}$}} & {\textbf{ 0.0031$^{*~}$}} & {\textbf{ 0.0030$^{*~}$}} & {\textbf{ 0.0028$^{*~}$}}  \\
   $\alpha$ &  { 0.0004$^{~~}$} & { 0.0005$^{~~}$}  & { 0.0007$^{~~}$} & { 0.0008$^{~~}$} & { 0.0010$^{~~}$} & {\textbf{ 0.0013$^{*~}$}} & {\textbf{ 0.0017$^{*~}$}} & {\textbf{ 0.0019$^{*~}$}}  \\
   SR &  { 0.5068$^{~~}$} & { 0.5190$^{~~}$}  & { 0.5164$^{~~}$} & { 0.5321$^{~~}$} & { 0.5013$^{~~}$} & { 0.5045$^{~~}$} & { 0.5132$^{~~}$} & { 0.5443$^{~~}$}  \\
   MD &  { 0.7208$^{~~}$} & { 0.7197$^{~~}$}  & { 0.7085$^{~~}$} & { 0.7085$^{~~}$} & { 0.7098$^{~~}$} & { 0.7074$^{~~}$} & { 0.6915$^{~~}$} & { 0.6585$^{~~}$}  \\
     & & & & & & & & \\
 \multicolumn{9}{l}{Panel D: Idiosyncratic maximum drawdown} \\
   Raw &  {\textbf{ 0.0034$^{*~}$}} & {\textbf{ 0.0035$^{*~}$}}  & {\textbf{ 0.0035$^{*~}$}} & {\textbf{ 0.0035$^{**}$}} & {\textbf{ 0.0034$^{*~}$}} & {\textbf{ 0.0035$^{**}$}} & {\textbf{ 0.0033$^{*~}$}} & {\textbf{ 0.0033$^{**}$}}  \\
   $\alpha$ &  { 0.0005$^{~~}$} & {\textbf{ 0.0008$^{*~}$}}  & {\textbf{ 0.0009$^{*~}$}} & {\textbf{ 0.0011$^{*~}$}} & {\textbf{ 0.0014$^{**}$}} & {\textbf{ 0.0018$^{**}$}} & {\textbf{ 0.0022$^{**}$}} & {\textbf{ 0.0026$^{**}$}}  \\
   SR &  { 0.5394$^{~~}$} & { 0.5707$^{~~}$}  & { 0.5804$^{~~}$} & { 0.6003$^{~~}$} & { 0.5965$^{~~}$} & { 0.6200$^{~~}$} & { 0.6318$^{~~}$} & { 0.7037$^{~~}$}  \\
   MD &  { 0.6746$^{~~}$} & { 0.6569$^{~~}$}  & { 0.6501$^{~~}$} & { 0.6341$^{~~}$} & { 0.6390$^{~~}$} & { 0.6355$^{~~}$} & { 0.6273$^{~~}$} & { 0.6098$^{~~}$}  \\
     & & & & & & & & \\
 \multicolumn{9}{l}{Panel E: Idiosyncratic ES (5\%)} \\
   Raw &  {\textbf{ 0.0033$^{*~}$}} & {\textbf{ 0.0032$^{*~}$}}  & {\textbf{ 0.0032$^{*~}$}} & {\textbf{ 0.0032$^{*~}$}} & {\textbf{ 0.0031$^{*~}$}} & {\textbf{ 0.0032$^{*~}$}} & {\textbf{ 0.0031$^{*~}$}} & {\textbf{ 0.0032$^{*~}$}}  \\
   $\alpha$ &  { 0.0007$^{~~}$} & { 0.0008$^{~~}$}  & {\textbf{ 0.0009$^{*~}$}} & {\textbf{ 0.0010$^{*~}$}} & {\textbf{ 0.0013$^{*~}$}} & {\textbf{ 0.0018$^{**}$}} & {\textbf{ 0.0022$^{**}$}} & {\textbf{ 0.0025$^{**}$}}  \\
   SR &  { 0.5391$^{~~}$} & { 0.5338$^{~~}$}  & { 0.5365$^{~~}$} & { 0.5491$^{~~}$} & { 0.5455$^{~~}$} & { 0.5835$^{~~}$} & { 0.6039$^{~~}$} & { 0.6799$^{~~}$}  \\
   MD &  { 0.6759$^{~~}$} & { 0.6669$^{~~}$}  & { 0.6643$^{~~}$} & { 0.6516$^{~~}$} & { 0.6403$^{~~}$} & { 0.6335$^{~~}$} & { 0.6283$^{~~}$} & { 0.6049$^{~~}$}  \\
     & & & & & & & & \\
 \multicolumn{9}{l}{Panel F: Idiosyncratic VaR (5\%)} \\
   Raw &  {\textbf{ 0.0031$^{*~}$}} & {\textbf{ 0.0030$^{*~}$}}  & {\textbf{ 0.0030$^{*~}$}} & {\textbf{ 0.0031$^{*~}$}} & {\textbf{ 0.0030$^{*~}$}} & {\textbf{ 0.0032$^{*~}$}} & {\textbf{ 0.0031$^{*~}$}} & {\textbf{ 0.0031$^{*~}$}}  \\
   $\alpha$ &  { 0.0004$^{~~}$} & { 0.0004$^{~~}$}  & { 0.0006$^{~~}$} & { 0.0008$^{~~}$} & {\textbf{ 0.0011$^{*~}$}} & {\textbf{ 0.0016$^{**}$}} & {\textbf{ 0.0021$^{**}$}} & {\textbf{ 0.0024$^{**}$}}  \\
   SR &  { 0.4926$^{~~}$} & { 0.4942$^{~~}$}  & { 0.5033$^{~~}$} & { 0.5200$^{~~}$} & { 0.5226$^{~~}$} & { 0.5617$^{~~}$} & { 0.5836$^{~~}$} & { 0.6624$^{~~}$}  \\
   MD &  { 0.6853$^{~~}$} & { 0.6692$^{~~}$}  & { 0.6627$^{~~}$} & { 0.6491$^{~~}$} & { 0.6423$^{~~}$} & { 0.6381$^{~~}$} & { 0.6343$^{~~}$} & { 0.6105$^{~~}$}  \\
     & & & & & & & & \\
 \multicolumn{9}{l}{Panel G: Idiosyncratic ES (1\%)} \\
   Raw &  {\textbf{ 0.0035$^{*~}$}} & {\textbf{ 0.0035$^{*~}$}}  & {\textbf{ 0.0035$^{*~}$}} & {\textbf{ 0.0035$^{*~}$}} & {\textbf{ 0.0033$^{*~}$}} & {\textbf{ 0.0034$^{*~}$}} & {\textbf{ 0.0032$^{*~}$}} & {\textbf{ 0.0033$^{*~}$}}  \\
   $\alpha$ &  {\textbf{ 0.0008$^{*~}$}} & {\textbf{ 0.0009$^{*~}$}}  & {\textbf{ 0.0011$^{*~}$}} & {\textbf{ 0.0012$^{**}$}} & {\textbf{ 0.0014$^{**}$}} & {\textbf{ 0.0019$^{**}$}} & {\textbf{ 0.0022$^{**}$}} & {\textbf{ 0.0025$^{**}$}}  \\
   SR &  { 0.5729$^{~~}$} & { 0.5768$^{~~}$}  & { 0.5825$^{~~}$} & { 0.5932$^{~~}$} & { 0.5762$^{~~}$} & { 0.6041$^{~~}$} & { 0.6192$^{~~}$} & { 0.6862$^{~~}$}  \\
   MD &  { 0.6660$^{~~}$} & { 0.6576$^{~~}$}  & { 0.6536$^{~~}$} & { 0.6451$^{~~}$} & { 0.6501$^{~~}$} & { 0.6453$^{~~}$} & { 0.6391$^{~~}$} & { 0.6180$^{~~}$}  \\
     & & & & & & & & \\
 \multicolumn{9}{l}{Panel H: Idiosyncratic VaR (1\%)} \\
   Raw &  {\textbf{ 0.0033$^{*~}$}} & {\textbf{ 0.0033$^{*~}$}}  & {\textbf{ 0.0033$^{*~}$}} & {\textbf{ 0.0033$^{*~}$}} & {\textbf{ 0.0032$^{*~}$}} & {\textbf{ 0.0033$^{*~}$}} & {\textbf{ 0.0032$^{*~}$}} & {\textbf{ 0.0032$^{*~}$}}  \\
   $\alpha$ &  { 0.0007$^{~~}$} & { 0.0008$^{~~}$}  & {\textbf{ 0.0009$^{*~}$}} & {\textbf{ 0.0011$^{*~}$}} & {\textbf{ 0.0013$^{*~}$}} & {\textbf{ 0.0018$^{**}$}} & {\textbf{ 0.0022$^{**}$}} & {\textbf{ 0.0025$^{**}$}}  \\
   SR &  { 0.5437$^{~~}$} & { 0.5374$^{~~}$}  & { 0.5424$^{~~}$} & { 0.5581$^{~~}$} & { 0.5481$^{~~}$} & { 0.5842$^{~~}$} & { 0.6040$^{~~}$} & { 0.6768$^{~~}$}  \\
   MD &  { 0.6865$^{~~}$} & { 0.6757$^{~~}$}  & { 0.6729$^{~~}$} & { 0.6603$^{~~}$} & { 0.6493$^{~~}$} & { 0.6353$^{~~}$} & { 0.6322$^{~~}$} & { 0.6084$^{~~}$}  \\
   \hline
   \end{tabular}
   \label{TB:Risk:F5:long}
\end{table}

\begin{table}[!ht]
\small
\caption{Univariate portfolio analysis based on various idiosyncratic risk metrics. This table reports the results associated with short sides of univariate portfolios formed according to various idiosyncratic risk metrics in \autoref{TB:risks}. Average weekly returns (Raw), FF5F-$\alpha$s ($\alpha$), annualized Sharpe ratios (SR) and maximum drawdowns (MD) are presented in each panel. The sample period is from January 1997 to December 2017. At the beginning of each week, each risk metric is calculated using idiosyncratic returns over past 130 trading days for individual stocks, by which they can be sorted into ten decile groups. The short sides for portfolios can be constructed by buying the stocks from the decile group with highest risk. Portfolios would be held for $K$ weeks, and calendar-time method is applied to obtain the average weekly return. \cite{Newey-West-1987-Em}'s $t$-statistics are obtained and the superscripts * and ** denote the significance at 5\% and 1\% levels, respectively.}
\centering
\vspace{-3mm}
   \begin{tabular}{ccccccccc}
   \hline
      & $K=1$ & $2$ & $3$ & $4$ &  $8$ & $13$ & $26$ & $52$ \\
   \hline
 \multicolumn{9}{l}{Panel A: Idiosyncratic volatility} \\
   Raw &  { 0.0009$^{~~}$} & { 0.0011$^{~~}$}  & { 0.0012$^{~~}$} & { 0.0013$^{~~}$} & { 0.0014$^{~~}$} & { 0.0016$^{~~}$} & { 0.0017$^{~~}$} & { 0.0020$^{~~}$}  \\
   $\alpha$ &  {\textbf{ -0.0028$^{**}$}} & {\textbf{ -0.0023$^{**}$}}  & {\textbf{ -0.0020$^{**}$}} & {\textbf{ -0.0017$^{**}$}} & { -0.0011$^{~~}$} & { -0.0004$^{~~}$} & { 0.0004$^{~~}$} & { 0.0011$^{~~}$}  \\
   SR &  { 0.0493$^{~~}$} & { 0.0819$^{~~}$}  & { 0.1031$^{~~}$} & { 0.1230$^{~~}$} & { 0.1431$^{~~}$} & { 0.1757$^{~~}$} & { 0.2141$^{~~}$} & { 0.3071$^{~~}$}  \\
   MD &  { 0.8528$^{~~}$} & { 0.8424$^{~~}$}  & { 0.8318$^{~~}$} & { 0.8171$^{~~}$} & { 0.7906$^{~~}$} & { 0.7714$^{~~}$} & { 0.7565$^{~~}$} & { 0.7372$^{~~}$}  \\
     & & & & & & & & \\
 \multicolumn{9}{l}{Panel B: Idiosyncratic skewness} \\
   Raw &  { 0.0027$^{~~}$} & { 0.0027$^{~~}$}  & { 0.0027$^{~~}$} & {\textbf{ 0.0028$^{*~}$}} & {\textbf{ 0.0029$^{*~}$}} & {\textbf{ 0.0030$^{*~}$}} & {\textbf{ 0.0031$^{*~}$}} & {\textbf{ 0.0031$^{*~}$}}  \\
   $\alpha$ &  { -0.0005$^{~~}$} & { -0.0003$^{~~}$}  & { -0.0001$^{~~}$} & { 0.0001$^{~~}$} & { 0.0006$^{~~}$} & { 0.0012$^{~~}$} & {\textbf{ 0.0019$^{*~}$}} & {\textbf{ 0.0023$^{*~}$}}  \\
   SR &  { 0.3714$^{~~}$} & { 0.3836$^{~~}$}  & { 0.3962$^{~~}$} & { 0.4194$^{~~}$} & { 0.4380$^{~~}$} & { 0.4804$^{~~}$} & { 0.5357$^{~~}$} & { 0.6068$^{~~}$}  \\
   MD &  { 0.7783$^{~~}$} & { 0.7686$^{~~}$}  & { 0.7624$^{~~}$} & { 0.7466$^{~~}$} & { 0.7218$^{~~}$} & { 0.6967$^{~~}$} & { 0.6805$^{~~}$} & { 0.6661$^{~~}$}  \\
     & & & & & & & & \\
 \multicolumn{9}{l}{Panel C: Idiosyncratic kurtosis} \\
   Raw &  { 0.0028$^{~~}$} & {\textbf{ 0.0029$^{*~}$}}  & {\textbf{ 0.0029$^{*~}$}} & {\textbf{ 0.0030$^{*~}$}} & {\textbf{ 0.0031$^{*~}$}} & {\textbf{ 0.0032$^{*~}$}} & {\textbf{ 0.0031$^{*~}$}} & {\textbf{ 0.0031$^{*~}$}}  \\
   $\alpha$ &  { -0.0004$^{~~}$} & { -0.0002$^{~~}$}  & { 0.0000$^{~~}$} & { 0.0003$^{~~}$} & { 0.0007$^{~~}$} & {\textbf{ 0.0013$^{*~}$}} & {\textbf{ 0.0019$^{*~}$}} & {\textbf{ 0.0022$^{*~}$}}  \\
   SR &  { 0.3997$^{~~}$} & { 0.4181$^{~~}$}  & { 0.4274$^{~~}$} & { 0.4520$^{~~}$} & { 0.4795$^{~~}$} & { 0.5139$^{~~}$} & { 0.5549$^{~~}$} & { 0.6119$^{~~}$}  \\
   MD &  { 0.7579$^{~~}$} & { 0.7467$^{~~}$}  & { 0.7401$^{~~}$} & { 0.7226$^{~~}$} & { 0.6974$^{~~}$} & { 0.6780$^{~~}$} & { 0.6735$^{~~}$} & { 0.6599$^{~~}$}  \\
     & & & & & & & & \\
 \multicolumn{9}{l}{Panel D: Idiosyncratic maximum drawdown} \\
   Raw &  { 0.0019$^{~~}$} & { 0.0018$^{~~}$}  & { 0.0017$^{~~}$} & { 0.0018$^{~~}$} & { 0.0017$^{~~}$} & { 0.0018$^{~~}$} & { 0.0019$^{~~}$} & { 0.0021$^{~~}$}  \\
   $\alpha$ &  {\textbf{ -0.0015$^{**}$}} & {\textbf{ -0.0014$^{**}$}}  & {\textbf{ -0.0013$^{**}$}} & {\textbf{ -0.0011$^{*~}$}} & { -0.0007$^{~~}$} & { -0.0001$^{~~}$} & { 0.0006$^{~~}$} & { 0.0012$^{~~}$}  \\
   SR &  { 0.2035$^{~~}$} & { 0.1961$^{~~}$}  & { 0.1911$^{~~}$} & { 0.1998$^{~~}$} & { 0.2004$^{~~}$} & { 0.2285$^{~~}$} & { 0.2571$^{~~}$} & { 0.3442$^{~~}$}  \\
   MD &  { 0.8256$^{~~}$} & { 0.8230$^{~~}$}  & { 0.8158$^{~~}$} & { 0.8023$^{~~}$} & { 0.7789$^{~~}$} & { 0.7542$^{~~}$} & { 0.7423$^{~~}$} & { 0.7248$^{~~}$}  \\
     & & & & & & & & \\
 \multicolumn{9}{l}{Panel E: Idiosyncratic ES (5\%)} \\
   Raw &  { 0.0014$^{~~}$} & { 0.0016$^{~~}$}  & { 0.0017$^{~~}$} & { 0.0018$^{~~}$} & { 0.0018$^{~~}$} & { 0.0019$^{~~}$} & { 0.0019$^{~~}$} & { 0.0023$^{~~}$}  \\
   $\alpha$ &  {\textbf{ -0.0021$^{**}$}} & {\textbf{ -0.0018$^{**}$}}  & {\textbf{ -0.0015$^{**}$}} & {\textbf{ -0.0012$^{*~}$}} & { -0.0007$^{~~}$} & { -0.0001$^{~~}$} & { 0.0007$^{~~}$} & { 0.0014$^{~~}$}  \\
   SR &  { 0.1293$^{~~}$} & { 0.1614$^{~~}$}  & { 0.1798$^{~~}$} & { 0.1975$^{~~}$} & { 0.2103$^{~~}$} & { 0.2363$^{~~}$} & { 0.2658$^{~~}$} & { 0.3720$^{~~}$}  \\
   MD &  { 0.8238$^{~~}$} & { 0.8139$^{~~}$}  & { 0.8049$^{~~}$} & { 0.7911$^{~~}$} & { 0.7744$^{~~}$} & { 0.7594$^{~~}$} & { 0.7507$^{~~}$} & { 0.7255$^{~~}$}  \\
     & & & & & & & & \\
 \multicolumn{9}{l}{Panel F: Idiosyncratic VaR (5\%)} \\
   Raw &  { 0.0014$^{~~}$} & { 0.0015$^{~~}$}  & { 0.0016$^{~~}$} & { 0.0017$^{~~}$} & { 0.0017$^{~~}$} & { 0.0018$^{~~}$} & { 0.0018$^{~~}$} & { 0.0021$^{~~}$}  \\
   $\alpha$ &  {\textbf{ -0.0022$^{**}$}} & {\textbf{ -0.0019$^{**}$}}  & {\textbf{ -0.0016$^{**}$}} & {\textbf{ -0.0013$^{*~}$}} & { -0.0008$^{~~}$} & { -0.0002$^{~~}$} & { 0.0006$^{~~}$} & { 0.0012$^{~~}$}  \\
   SR &  { 0.1237$^{~~}$} & { 0.1451$^{~~}$}  & { 0.1628$^{~~}$} & { 0.1838$^{~~}$} & { 0.1951$^{~~}$} & { 0.2200$^{~~}$} & { 0.2454$^{~~}$} & { 0.3429$^{~~}$}  \\
   MD &  { 0.8427$^{~~}$} & { 0.8309$^{~~}$}  & { 0.8221$^{~~}$} & { 0.8069$^{~~}$} & { 0.7865$^{~~}$} & { 0.7679$^{~~}$} & { 0.7531$^{~~}$} & { 0.7311$^{~~}$}  \\
     & & & & & & & & \\
 \multicolumn{9}{l}{Panel G: Idiosyncratic ES (1\%)} \\
   Raw &  { 0.0019$^{~~}$} & { 0.0021$^{~~}$}  & { 0.0022$^{~~}$} & { 0.0023$^{~~}$} & { 0.0022$^{~~}$} & { 0.0023$^{~~}$} & { 0.0023$^{~~}$} & { 0.0025$^{~~}$}  \\
   $\alpha$ &  {\textbf{ -0.0015$^{**}$}} & {\textbf{ -0.0012$^{**}$}}  & { -0.0009$^{~~}$} & { -0.0006$^{~~}$} & { -0.0002$^{~~}$} & { 0.0004$^{~~}$} & { 0.0011$^{~~}$} & { 0.0016$^{~~}$}  \\
   SR &  { 0.2155$^{~~}$} & { 0.2524$^{~~}$}  & { 0.2724$^{~~}$} & { 0.2901$^{~~}$} & { 0.2888$^{~~}$} & { 0.3161$^{~~}$} & { 0.3488$^{~~}$} & { 0.4352$^{~~}$}  \\
   MD &  { 0.8105$^{~~}$} & { 0.7984$^{~~}$}  & { 0.7870$^{~~}$} & { 0.7700$^{~~}$} & { 0.7520$^{~~}$} & { 0.7338$^{~~}$} & { 0.7301$^{~~}$} & { 0.7052$^{~~}$}  \\
     & & & & & & & & \\
 \multicolumn{9}{l}{Panel H: Idiosyncratic VaR (1\%)} \\
   Raw &  { 0.0017$^{~~}$} & { 0.0019$^{~~}$}  & { 0.0020$^{~~}$} & { 0.0021$^{~~}$} & { 0.0022$^{~~}$} & { 0.0022$^{~~}$} & { 0.0022$^{~~}$} & { 0.0024$^{~~}$}  \\
   $\alpha$ &  {\textbf{ -0.0018$^{**}$}} & {\textbf{ -0.0014$^{**}$}}  & {\textbf{ -0.0011$^{*~}$}} & { -0.0008$^{~~}$} & { -0.0003$^{~~}$} & { 0.0003$^{~~}$} & { 0.0009$^{~~}$} & { 0.0015$^{~~}$}  \\
   SR &  { 0.1790$^{~~}$} & { 0.2093$^{~~}$}  & { 0.2336$^{~~}$} & { 0.2585$^{~~}$} & { 0.2721$^{~~}$} & { 0.2939$^{~~}$} & { 0.3161$^{~~}$} & { 0.4135$^{~~}$}  \\
   MD &  { 0.8190$^{~~}$} & { 0.8079$^{~~}$}  & { 0.7979$^{~~}$} & { 0.7836$^{~~}$} & { 0.7674$^{~~}$} & { 0.7516$^{~~}$} & { 0.7409$^{~~}$} & { 0.7184$^{~~}$}  \\
   \hline
   \end{tabular}
   \label{TB:Risk:F5:short}
\end{table}

\begin{landscape}
\begin{table}[!ht]
\setlength\tabcolsep{1pt}
\small
\caption{Spanning test for arbitrage portfolios with respect to idiosyncratic risk metrics. Augmented FF5F-regression is conducted to examine if the profitability of various risk-based portfolios could be explained by one of others. Variable X in the first column refers to as the profitability of X-based portfolios incorporated in the FF5F model, derived from \autoref{TB:Risk:F5}. Column 2 reports $K$'s value. Columns 3 to 8 report the $\alpha$ values of the augmented FF5F model, and $\beta$ values of variable X. \cite{Newey-West-1987-Em}'s $t$-statistics are adopted and the superscripts * and ** denote the significance at 5\% and 1\% levels, respectively. }
\centering
\vspace{-3mm}
   \begin{tabular}{cccccccccccccccccc}
   \hline\hline
    Explanatory  & \multirow{2}{*}{$K$} & \multirow{2}{*}{$\alpha$(IVol)}  &\multirow{2}{*}{$\beta_X$}  &  \multirow{2}{*}{$\alpha$(ISkew)}  &\multirow{2}{*}{$\beta_X$} & \multirow{2}{*}{$\alpha$(IKurt)}  &\multirow{2}{*}{$\beta_X$} & \multirow{2}{*}{$\alpha$(IMD) } &\multirow{2}{*}{$\beta_X$} & \multirow{2}{*}{$\alpha$(IES5)} & \multirow{2}{*}{$\beta_X$ } & \multirow{2}{*}{$\alpha$(IVaR5)} & \multirow{2}{*}{$\beta_X$ } & \multirow{2}{*}{$\alpha$(IES1)} & \multirow{2}{*}{$\beta_X$ } &\multirow{2}{*}{$\alpha$(IVaR1)} & \multirow{2}{*}{$\beta_X$} \\
   variable X  &  &  &  &  &  &  &  &  & & & & & & & & & \\
   \hline
   IVol &  1 & { $^{~~}$} & { $^{~~}$} & {\textbf{ 0.0014$^{**}$}} &  {\textbf{ -0.12$^{**}$}} & {\textbf{ 0.0021$^{**}$}} &  {\textbf{ -0.37$^{**}$}}& {\textbf{ -0.0006$^{*~}$}} &  {\textbf{  0.79$^{**}$}} & { -0.0001$^{~~}$} &  {\textbf{  0.90$^{**}$}}  & {\textbf{ -0.0007$^{**}$}} & {\textbf{  0.95$^{**}$}} & { 0.0001$^{~~}$} & {\textbf{  0.67$^{**}$}} & { -0.0001$^{~~}$} & {\textbf{  0.80$^{**}$}}  \\
    &  2 & { $^{~~}$} & { $^{~~}$} & {\textbf{ 0.0013$^{**}$}} &  {\textbf{ -0.11$^{**}$}} & {\textbf{ 0.0019$^{**}$}} &  {\textbf{ -0.38$^{**}$}}& { -0.0002$^{~~}$} &  {\textbf{  0.79$^{**}$}} & { -0.0002$^{~~}$} &  {\textbf{  0.89$^{**}$}}  & {\textbf{ -0.0006$^{**}$}} & {\textbf{  0.95$^{**}$}} & { 0.0000$^{~~}$} & {\textbf{  0.67$^{**}$}} & { -0.0002$^{~~}$} & {\textbf{  0.80$^{**}$}}  \\
    &  3 & { $^{~~}$} & { $^{~~}$} & {\textbf{ 0.0012$^{**}$}} &  {\textbf{ -0.10$^{**}$}} & {\textbf{ 0.0017$^{**}$}} &  {\textbf{ -0.37$^{**}$}}& { -0.0001$^{~~}$} &  {\textbf{  0.79$^{**}$}} & { -0.0002$^{~~}$} &  {\textbf{  0.89$^{**}$}}  & {\textbf{ -0.0005$^{**}$}} & {\textbf{  0.95$^{**}$}} & { -0.0000$^{~~}$} & {\textbf{  0.67$^{**}$}} & { -0.0003$^{~~}$} & {\textbf{  0.80$^{**}$}}  \\
    &  4 & { $^{~~}$} & { $^{~~}$} & {\textbf{ 0.0011$^{**}$}} &  {\textbf{ -0.11$^{**}$}} & {\textbf{ 0.0016$^{**}$}} &  {\textbf{ -0.36$^{**}$}}& { 0.0001$^{~~}$} &  {\textbf{  0.79$^{**}$}} & { -0.0002$^{~~}$} &  {\textbf{  0.89$^{**}$}}  & {\textbf{ -0.0005$^{**}$}} & {\textbf{  0.95$^{**}$}} & { -0.0000$^{~~}$} & {\textbf{  0.67$^{**}$}} & { -0.0003$^{~~}$} & {\textbf{  0.80$^{**}$}}  \\
    &  8 & { $^{~~}$} & { $^{~~}$} & {\textbf{ 0.0008$^{**}$}} &  {\textbf{ -0.12$^{**}$}} & {\textbf{ 0.0011$^{**}$}} &  {\textbf{ -0.37$^{**}$}}& { 0.0002$^{~~}$} &  {\textbf{  0.78$^{**}$}} & { -0.0002$^{~~}$} &  {\textbf{  0.89$^{**}$}}  & {\textbf{ -0.0004$^{**}$}} & {\textbf{  0.95$^{**}$}} & { -0.0000$^{~~}$} & {\textbf{  0.67$^{**}$}} & { -0.0003$^{~~}$} & {\textbf{  0.80$^{**}$}}  \\
    &  13 & { $^{~~}$} & { $^{~~}$} & { 0.0005$^{~~}$} &  {\textbf{ -0.13$^{**}$}} & {\textbf{ 0.0008$^{**}$}} &  {\textbf{ -0.37$^{**}$}}& { 0.0003$^{~~}$} &  {\textbf{  0.78$^{**}$}} & { -0.0001$^{~~}$} &  {\textbf{  0.91$^{**}$}}  & {\textbf{ -0.0003$^{*~}$}} & {\textbf{  0.97$^{**}$}} & { -0.0000$^{~~}$} & {\textbf{  0.68$^{**}$}} & { -0.0002$^{~~}$} & {\textbf{  0.81$^{**}$}}  \\
    &  26 & { $^{~~}$} & { $^{~~}$} & { 0.0001$^{~~}$} &  {\textbf{ -0.13$^{**}$}} & { 0.0004$^{~~}$} &  {\textbf{ -0.36$^{**}$}}& {\textbf{ 0.0003$^{*~}$}} &  {\textbf{  0.78$^{**}$}} & { -0.0001$^{~~}$} &  {\textbf{  0.92$^{**}$}}  & { -0.0002$^{~~}$} & {\textbf{  0.98$^{**}$}} & { -0.0001$^{~~}$} & {\textbf{  0.72$^{**}$}} & { -0.0001$^{~~}$} & {\textbf{  0.82$^{**}$}}  \\
    &  52 & { $^{~~}$} & { $^{~~}$} & { 0.0000$^{~~}$} &  {\textbf{ -0.10$^{**}$}} & { 0.0002$^{~~}$} &  {\textbf{ -0.34$^{**}$}}& {\textbf{ 0.0003$^{*~}$}} &  {\textbf{  0.78$^{**}$}} & { -0.0001$^{~~}$} &  {\textbf{  0.92$^{**}$}}  & { -0.0002$^{~~}$} & {\textbf{  0.97$^{**}$}} & { -0.0001$^{~~}$} & {\textbf{  0.72$^{**}$}} & { -0.0002$^{~~}$} & {\textbf{  0.81$^{**}$}}
        \smallskip\\
   ISkew &  1 & {\textbf{ 0.0036$^{**}$}} & {\textbf{ -0.31$^{**}$}} & { $^{~~}$} &  { $^{~~}$} & { 0.0002$^{~~}$} &  {\textbf{  0.58$^{**}$}}& {\textbf{ 0.0024$^{**}$}} &  {\textbf{ -0.36$^{**}$}} & {\textbf{ 0.0034$^{**}$}} &  {\textbf{ -0.60$^{**}$}}  & {\textbf{ 0.0030$^{**}$}} & {\textbf{ -0.54$^{**}$}} & {\textbf{ 0.0029$^{**}$}} & {\textbf{ -0.54$^{**}$}} & {\textbf{ 0.0031$^{**}$}} & {\textbf{ -0.59$^{**}$}}  \\
    &  2 & {\textbf{ 0.0033$^{**}$}} & {\textbf{ -0.28$^{**}$}} & { $^{~~}$} &  { $^{~~}$} & { 0.0002$^{~~}$} &  {\textbf{  0.56$^{**}$}}& {\textbf{ 0.0026$^{**}$}} &  {\textbf{ -0.38$^{**}$}} & {\textbf{ 0.0031$^{**}$}} &  {\textbf{ -0.59$^{**}$}}  & {\textbf{ 0.0028$^{**}$}} & {\textbf{ -0.51$^{**}$}} & {\textbf{ 0.0026$^{**}$}} & {\textbf{ -0.54$^{**}$}} & {\textbf{ 0.0028$^{**}$}} & {\textbf{ -0.58$^{**}$}}  \\
    &  3 & {\textbf{ 0.0031$^{**}$}} & {\textbf{ -0.28$^{**}$}} & { $^{~~}$} &  { $^{~~}$} & { 0.0001$^{~~}$} &  {\textbf{  0.56$^{**}$}}& {\textbf{ 0.0026$^{**}$}} &  {\textbf{ -0.40$^{**}$}} & {\textbf{ 0.0029$^{**}$}} &  {\textbf{ -0.58$^{**}$}}  & {\textbf{ 0.0027$^{**}$}} & {\textbf{ -0.51$^{**}$}} & {\textbf{ 0.0024$^{**}$}} & {\textbf{ -0.54$^{**}$}} & {\textbf{ 0.0025$^{**}$}} & {\textbf{ -0.58$^{**}$}}  \\
    &  4 & {\textbf{ 0.0030$^{**}$}} & {\textbf{ -0.30$^{**}$}} & { $^{~~}$} &  { $^{~~}$} & { 0.0001$^{~~}$} &  {\textbf{  0.55$^{**}$}}& {\textbf{ 0.0026$^{**}$}} &  {\textbf{ -0.42$^{**}$}} & {\textbf{ 0.0027$^{**}$}} &  {\textbf{ -0.60$^{**}$}}  & {\textbf{ 0.0025$^{**}$}} & {\textbf{ -0.52$^{**}$}} & {\textbf{ 0.0023$^{**}$}} & {\textbf{ -0.56$^{**}$}} & {\textbf{ 0.0024$^{**}$}} & {\textbf{ -0.60$^{**}$}}  \\
    &  8 & {\textbf{ 0.0026$^{**}$}} & {\textbf{ -0.37$^{**}$}} & { $^{~~}$} &  { $^{~~}$} & { -0.0001$^{~~}$} &  {\textbf{  0.58$^{**}$}}& {\textbf{ 0.0023$^{**}$}} &  {\textbf{ -0.47$^{**}$}} & {\textbf{ 0.0023$^{**}$}} &  {\textbf{ -0.65$^{**}$}}  & {\textbf{ 0.0022$^{**}$}} & {\textbf{ -0.57$^{**}$}} & {\textbf{ 0.0019$^{**}$}} & {\textbf{ -0.61$^{**}$}} & {\textbf{ 0.0019$^{**}$}} & {\textbf{ -0.66$^{**}$}}  \\
    &  13 & {\textbf{ 0.0022$^{**}$}} & {\textbf{ -0.41$^{**}$}} & { $^{~~}$} &  { $^{~~}$} & { -0.0001$^{~~}$} &  {\textbf{  0.57$^{**}$}}& {\textbf{ 0.0020$^{**}$}} &  {\textbf{ -0.51$^{**}$}} & {\textbf{ 0.0020$^{**}$}} &  {\textbf{ -0.69$^{**}$}}  & {\textbf{ 0.0019$^{**}$}} & {\textbf{ -0.61$^{**}$}} & {\textbf{ 0.0016$^{**}$}} & {\textbf{ -0.66$^{**}$}} & {\textbf{ 0.0017$^{**}$}} & {\textbf{ -0.70$^{**}$}}  \\
    &  26 & {\textbf{ 0.0016$^{**}$}} & {\textbf{ -0.45$^{**}$}} & { $^{~~}$} &  { $^{~~}$} & { -0.0001$^{~~}$} &  {\textbf{  0.56$^{**}$}}& {\textbf{ 0.0015$^{**}$}} &  {\textbf{ -0.55$^{**}$}} & {\textbf{ 0.0014$^{**}$}} &  {\textbf{ -0.75$^{**}$}}  & {\textbf{ 0.0014$^{**}$}} & {\textbf{ -0.67$^{**}$}} & {\textbf{ 0.0010$^{**}$}} & {\textbf{ -0.72$^{**}$}} & {\textbf{ 0.0011$^{**}$}} & {\textbf{ -0.75$^{**}$}}  \\
    &  52 & {\textbf{ 0.0014$^{**}$}} & {\textbf{ -0.35$^{**}$}} & { $^{~~}$} &  { $^{~~}$} & { -0.0002$^{~~}$} &  {\textbf{  0.52$^{**}$}}& {\textbf{ 0.0013$^{**}$}} &  {\textbf{ -0.49$^{**}$}} & {\textbf{ 0.0011$^{**}$}} &  {\textbf{ -0.69$^{**}$}}  & {\textbf{ 0.0011$^{**}$}} & {\textbf{ -0.59$^{**}$}} & {\textbf{ 0.0008$^{**}$}} & {\textbf{ -0.69$^{**}$}} & {\textbf{ 0.0009$^{*~}$}} & {\textbf{ -0.70$^{**}$}}
        \smallskip\\
   IKurt &  1 & {\textbf{ 0.0039$^{**}$}} & {\textbf{ -0.62$^{**}$}} & { 0.0007$^{~~}$} &  {\textbf{  0.38$^{**}$}} & { $^{~~}$} &  { $^{~~}$}& {\textbf{ 0.0025$^{**}$}} &  {\textbf{ -0.58$^{**}$}} & {\textbf{ 0.0033$^{**}$}} &  {\textbf{ -0.59$^{**}$}}  & {\textbf{ 0.0031$^{**}$}} & {\textbf{ -0.75$^{**}$}} & {\textbf{ 0.0027$^{**}$}} & {\textbf{ -0.34$^{**}$}} & {\textbf{ 0.0029$^{**}$}} & {\textbf{ -0.51$^{**}$}}  \\
    &  2 & {\textbf{ 0.0036$^{**}$}} & {\textbf{ -0.66$^{**}$}} & {\textbf{ 0.0007$^{*~}$}} &  {\textbf{  0.37$^{**}$}} & { $^{~~}$} &  { $^{~~}$}& {\textbf{ 0.0027$^{**}$}} &  {\textbf{ -0.62$^{**}$}} & {\textbf{ 0.0030$^{**}$}} &  {\textbf{ -0.62$^{**}$}}  & {\textbf{ 0.0029$^{**}$}} & {\textbf{ -0.79$^{**}$}} & {\textbf{ 0.0024$^{**}$}} & {\textbf{ -0.35$^{**}$}} & {\textbf{ 0.0026$^{**}$}} & {\textbf{ -0.52$^{**}$}}  \\
    &  3 & {\textbf{ 0.0033$^{**}$}} & {\textbf{ -0.66$^{**}$}} & {\textbf{ 0.0007$^{*~}$}} &  {\textbf{  0.37$^{**}$}} & { $^{~~}$} &  { $^{~~}$}& {\textbf{ 0.0026$^{**}$}} &  {\textbf{ -0.63$^{**}$}} & {\textbf{ 0.0027$^{**}$}} &  {\textbf{ -0.62$^{**}$}}  & {\textbf{ 0.0027$^{**}$}} & {\textbf{ -0.79$^{**}$}} & {\textbf{ 0.0021$^{**}$}} & {\textbf{ -0.35$^{**}$}} & {\textbf{ 0.0023$^{**}$}} & {\textbf{ -0.53$^{**}$}}  \\
    &  4 & {\textbf{ 0.0031$^{**}$}} & {\textbf{ -0.67$^{**}$}} & {\textbf{ 0.0006$^{*~}$}} &  {\textbf{  0.36$^{**}$}} & { $^{~~}$} &  { $^{~~}$}& {\textbf{ 0.0026$^{**}$}} &  {\textbf{ -0.64$^{**}$}} & {\textbf{ 0.0026$^{**}$}} &  {\textbf{ -0.62$^{**}$}}  & {\textbf{ 0.0026$^{**}$}} & {\textbf{ -0.80$^{**}$}} & {\textbf{ 0.0020$^{**}$}} & {\textbf{ -0.36$^{**}$}} & {\textbf{ 0.0022$^{**}$}} & {\textbf{ -0.53$^{**}$}}  \\
    &  8 & {\textbf{ 0.0026$^{**}$}} & {\textbf{ -0.71$^{**}$}} & { 0.0004$^{~~}$} &  {\textbf{  0.37$^{**}$}} & { $^{~~}$} &  { $^{~~}$}& {\textbf{ 0.0022$^{**}$}} &  {\textbf{ -0.67$^{**}$}} & {\textbf{ 0.0021$^{**}$}} &  {\textbf{ -0.65$^{**}$}}  & {\textbf{ 0.0021$^{**}$}} & {\textbf{ -0.83$^{**}$}} & {\textbf{ 0.0017$^{**}$}} & {\textbf{ -0.38$^{**}$}} & {\textbf{ 0.0017$^{**}$}} & {\textbf{ -0.56$^{**}$}}  \\
    &  13 & {\textbf{ 0.0022$^{**}$}} & {\textbf{ -0.76$^{**}$}} & { 0.0002$^{~~}$} &  {\textbf{  0.37$^{**}$}} & { $^{~~}$} &  { $^{~~}$}& {\textbf{ 0.0020$^{**}$}} &  {\textbf{ -0.71$^{**}$}} & {\textbf{ 0.0019$^{**}$}} &  {\textbf{ -0.70$^{**}$}}  & {\textbf{ 0.0018$^{**}$}} & {\textbf{ -0.88$^{**}$}} & {\textbf{ 0.0015$^{**}$}} & {\textbf{ -0.42$^{**}$}} & {\textbf{ 0.0016$^{**}$}} & {\textbf{ -0.60$^{**}$}}  \\
    &  26 & {\textbf{ 0.0016$^{**}$}} & {\textbf{ -0.84$^{**}$}} & { -0.0001$^{~~}$} &  {\textbf{  0.39$^{**}$}} & { $^{~~}$} &  { $^{~~}$}& {\textbf{ 0.0015$^{**}$}} &  {\textbf{ -0.76$^{**}$}} & {\textbf{ 0.0013$^{**}$}} &  {\textbf{ -0.80$^{**}$}}  & {\textbf{ 0.0013$^{**}$}} & {\textbf{ -0.97$^{**}$}} & {\textbf{ 0.0010$^{**}$}} & {\textbf{ -0.51$^{**}$}} & {\textbf{ 0.0011$^{**}$}} & {\textbf{ -0.70$^{**}$}}  \\
    &  52 & {\textbf{ 0.0011$^{**}$}} & {\textbf{ -0.90$^{**}$}} & { 0.0000$^{~~}$} &  {\textbf{  0.39$^{**}$}} & { $^{~~}$} &  { $^{~~}$}& {\textbf{ 0.0011$^{**}$}} &  {\textbf{ -0.82$^{**}$}} & {\textbf{ 0.0009$^{**}$}} &  {\textbf{ -0.87$^{**}$}}  & {\textbf{ 0.0009$^{**}$}} & {\textbf{ -1.03$^{**}$}} & {\textbf{ 0.0007$^{*~}$}} & {\textbf{ -0.56$^{**}$}} & {\textbf{ 0.0007$^{*~}$}} & {\textbf{ -0.76$^{**}$}}
        \smallskip\\
   IMD &  1 & {\textbf{ 0.0014$^{**}$}} & {\textbf{  0.95$^{**}$}} & {\textbf{ 0.0013$^{**}$}} &  {\textbf{ -0.17$^{**}$}} & {\textbf{ 0.0017$^{**}$}} &  {\textbf{ -0.42$^{**}$}}& { $^{~~}$} &  { $^{~~}$} & {\textbf{ 0.0011$^{**}$}} &  {\textbf{  0.88$^{**}$}}  & { 0.0006$^{~~}$} & {\textbf{  0.93$^{**}$}} & {\textbf{ 0.0010$^{**}$}} & {\textbf{  0.66$^{**}$}} & {\textbf{ 0.0009$^{**}$}} & {\textbf{  0.79$^{**}$}}  \\
    &  2 & {\textbf{ 0.0009$^{**}$}} & {\textbf{  0.97$^{**}$}} & {\textbf{ 0.0014$^{**}$}} &  {\textbf{ -0.18$^{**}$}} & {\textbf{ 0.0017$^{**}$}} &  {\textbf{ -0.44$^{**}$}}& { $^{~~}$} &  { $^{~~}$} & { 0.0005$^{~~}$} &  {\textbf{  0.91$^{**}$}}  & { 0.0002$^{~~}$} & {\textbf{  0.97$^{**}$}} & { 0.0006$^{~~}$} & {\textbf{  0.68$^{**}$}} & { 0.0004$^{~~}$} & {\textbf{  0.81$^{**}$}}  \\
    &  3 & {\textbf{ 0.0007$^{*~}$}} & {\textbf{  0.99$^{**}$}} & {\textbf{ 0.0014$^{**}$}} &  {\textbf{ -0.18$^{**}$}} & {\textbf{ 0.0016$^{**}$}} &  {\textbf{ -0.44$^{**}$}}& { $^{~~}$} &  { $^{~~}$} & { 0.0003$^{~~}$} &  {\textbf{  0.93$^{**}$}}  & { 0.0000$^{~~}$} & {\textbf{  0.98$^{**}$}} & { 0.0004$^{~~}$} & {\textbf{  0.70$^{**}$}} & { 0.0001$^{~~}$} & {\textbf{  0.84$^{**}$}}  \\
    &  4 & { 0.0005$^{~~}$} & {\textbf{  1.01$^{**}$}} & {\textbf{ 0.0012$^{**}$}} &  {\textbf{ -0.19$^{**}$}} & {\textbf{ 0.0016$^{**}$}} &  {\textbf{ -0.44$^{**}$}}& { $^{~~}$} &  { $^{~~}$} & { 0.0001$^{~~}$} &  {\textbf{  0.94$^{**}$}}  & { -0.0001$^{~~}$} & {\textbf{  1.00$^{**}$}} & { 0.0002$^{~~}$} & {\textbf{  0.70$^{**}$}} & { -0.0000$^{~~}$} & {\textbf{  0.85$^{**}$}}  \\
    &  8 & { 0.0002$^{~~}$} & {\textbf{  1.05$^{**}$}} & {\textbf{ 0.0009$^{**}$}} &  {\textbf{ -0.21$^{**}$}} & {\textbf{ 0.0012$^{**}$}} &  {\textbf{ -0.47$^{**}$}}& { $^{~~}$} &  { $^{~~}$} & { -0.0001$^{~~}$} &  {\textbf{  0.97$^{**}$}}  & { -0.0003$^{~~}$} & {\textbf{  1.04$^{**}$}} & { 0.0001$^{~~}$} & {\textbf{  0.73$^{**}$}} & { -0.0002$^{~~}$} & {\textbf{  0.87$^{**}$}}  \\
    &  13 & { 0.0001$^{~~}$} & {\textbf{  1.08$^{**}$}} & {\textbf{ 0.0007$^{*~}$}} &  {\textbf{ -0.23$^{**}$}} & {\textbf{ 0.0010$^{**}$}} &  {\textbf{ -0.48$^{**}$}}& { $^{~~}$} &  { $^{~~}$} & { -0.0001$^{~~}$} &  {\textbf{  1.02$^{**}$}}  & { -0.0003$^{~~}$} & {\textbf{  1.09$^{**}$}} & { -0.0000$^{~~}$} & {\textbf{  0.77$^{**}$}} & { -0.0002$^{~~}$} & {\textbf{  0.91$^{**}$}}  \\
    &  26 & { -0.0001$^{~~}$} & {\textbf{  1.12$^{**}$}} & { 0.0002$^{~~}$} &  {\textbf{ -0.24$^{**}$}} & {\textbf{ 0.0006$^{*~}$}} &  {\textbf{ -0.47$^{**}$}}& { $^{~~}$} &  { $^{~~}$} & { -0.0003$^{~~}$} &  {\textbf{  1.08$^{**}$}}  & {\textbf{ -0.0004$^{*~}$}} & {\textbf{  1.14$^{**}$}} & { -0.0002$^{~~}$} & {\textbf{  0.85$^{**}$}} & { -0.0003$^{~~}$} & {\textbf{  0.97$^{**}$}}  \\
    &  52 & { -0.0001$^{~~}$} & {\textbf{  1.14$^{**}$}} & { 0.0002$^{~~}$} &  {\textbf{ -0.20$^{**}$}} & { 0.0003$^{~~}$} &  {\textbf{ -0.46$^{**}$}}& { $^{~~}$} &  { $^{~~}$} & {\textbf{ -0.0003$^{*~}$}} &  {\textbf{  1.10$^{**}$}}  & {\textbf{ -0.0004$^{*~}$}} & {\textbf{  1.16$^{**}$}} & { -0.0003$^{~~}$} & {\textbf{  0.86$^{**}$}} & {\textbf{ -0.0003$^{*~}$}} & {\textbf{  0.97$^{**}$}}  \\
   \hline\hline
   \end{tabular}
   \label{TB:Risk:F5:spanning:p1}
\end{table}
\end{landscape}

\begin{landscape}
\begin{table}[!ht]
\setlength\tabcolsep{1pt}
\small
\caption*{Table \ref{TB:Risk:F5:spanning:p1} (Continued): Spanning test for arbitrage portfolios with respect to idiosyncratic risk metrics. Augmented FF5F-regression is conducted to examine if the profitability of various risk-based portfolios could be explained by one of others. Variable X in the first column refers to as the profitability of X-based portfolios incorporated in the FF5F model, derived from \autoref{TB:Risk:F5}. Column 2 reports $K$'s value. Column 3 to 8 report the $\alpha$ values of the augmented FF5F model, and $\beta$ values of variable X. \cite{Newey-West-1987-Em}'s $t$-statistics are adopted and the superscripts * and ** denote the significance at 5\% and 1\% levels, respectively. }
\centering
\vspace{-3mm}
   \begin{tabular}{cccccccccccccccccc}
   \hline\hline
    Explanatory  & \multirow{2}{*}{$K$} & \multirow{2}{*}{$\alpha$(IVol)}  &\multirow{2}{*}{$\beta_X$}  &  \multirow{2}{*}{$\alpha$(ISkew)}  &\multirow{2}{*}{$\beta_X$} & \multirow{2}{*}{$\alpha$(IKurt)}  &\multirow{2}{*}{$\beta_X$} & \multirow{2}{*}{$\alpha$(IMD) } &\multirow{2}{*}{$\beta_X$} & \multirow{2}{*}{$\alpha$(IES5)} & \multirow{2}{*}{$\beta_X$ } & \multirow{2}{*}{$\alpha$(IVaR5)} & \multirow{2}{*}{$\beta_X$ } & \multirow{2}{*}{$\alpha$(IES1)} & \multirow{2}{*}{$\beta_X$ } &\multirow{2}{*}{$\alpha$(IVaR1)} & \multirow{2}{*}{$\beta_X$} \\
   variable X  &  &  &  &  &  &  &  &  & & & & & & & & & \\
   \hline
   IES5 &  1 & {\textbf{ 0.0006$^{*~}$}} & {\textbf{  0.96$^{**}$}} & {\textbf{ 0.0017$^{**}$}} &  {\textbf{ -0.26$^{**}$}} & {\textbf{ 0.0019$^{**}$}} &  {\textbf{ -0.38$^{**}$}}& { -0.0002$^{~~}$} &  {\textbf{  0.79$^{**}$}} & { $^{~~}$} &  { $^{~~}$}  & { -0.0003$^{~~}$} & {\textbf{  0.98$^{**}$}} & { 0.0002$^{~~}$} & {\textbf{  0.76$^{**}$}} & { -0.0000$^{~~}$} & {\textbf{  0.89$^{**}$}}  \\
    &  2 & {\textbf{ 0.0006$^{**}$}} & {\textbf{  0.97$^{**}$}} & {\textbf{ 0.0016$^{**}$}} &  {\textbf{ -0.24$^{**}$}} & {\textbf{ 0.0017$^{**}$}} &  {\textbf{ -0.39$^{**}$}}& { 0.0002$^{~~}$} &  {\textbf{  0.80$^{**}$}} & { $^{~~}$} &  { $^{~~}$}  & { -0.0002$^{~~}$} & {\textbf{  0.99$^{**}$}} & { 0.0002$^{~~}$} & {\textbf{  0.77$^{**}$}} & { -0.0000$^{~~}$} & {\textbf{  0.90$^{**}$}}  \\
    &  3 & {\textbf{ 0.0006$^{**}$}} & {\textbf{  0.98$^{**}$}} & {\textbf{ 0.0015$^{**}$}} &  {\textbf{ -0.23$^{**}$}} & {\textbf{ 0.0015$^{**}$}} &  {\textbf{ -0.38$^{**}$}}& { 0.0003$^{~~}$} &  {\textbf{  0.82$^{**}$}} & { $^{~~}$} &  { $^{~~}$}  & { -0.0001$^{~~}$} & {\textbf{  0.99$^{**}$}} & { 0.0001$^{~~}$} & {\textbf{  0.77$^{**}$}} & { -0.0001$^{~~}$} & {\textbf{  0.90$^{**}$}}  \\
    &  4 & {\textbf{ 0.0005$^{**}$}} & {\textbf{  0.99$^{**}$}} & {\textbf{ 0.0013$^{**}$}} &  {\textbf{ -0.24$^{**}$}} & {\textbf{ 0.0014$^{**}$}} &  {\textbf{ -0.38$^{**}$}}& { 0.0004$^{~~}$} &  {\textbf{  0.81$^{**}$}} & { $^{~~}$} &  { $^{~~}$}  & { -0.0001$^{~~}$} & {\textbf{  0.99$^{**}$}} & { 0.0001$^{~~}$} & {\textbf{  0.77$^{**}$}} & { -0.0001$^{~~}$} & {\textbf{  0.90$^{**}$}}  \\
    &  8 & {\textbf{ 0.0005$^{*~}$}} & {\textbf{  1.00$^{**}$}} & {\textbf{ 0.0010$^{**}$}} &  {\textbf{ -0.24$^{**}$}} & {\textbf{ 0.0010$^{**}$}} &  {\textbf{ -0.38$^{**}$}}& {\textbf{ 0.0005$^{*~}$}} &  {\textbf{  0.82$^{**}$}} & { $^{~~}$} &  { $^{~~}$}  & { -0.0001$^{~~}$} & {\textbf{  1.01$^{**}$}} & { 0.0001$^{~~}$} & {\textbf{  0.77$^{**}$}} & { -0.0002$^{~~}$} & {\textbf{  0.90$^{**}$}}  \\
    &  13 & { 0.0003$^{~~}$} & {\textbf{  0.99$^{**}$}} & {\textbf{ 0.0007$^{*~}$}} &  {\textbf{ -0.24$^{**}$}} & {\textbf{ 0.0007$^{**}$}} &  {\textbf{ -0.38$^{**}$}}& {\textbf{ 0.0004$^{*~}$}} &  {\textbf{  0.81$^{**}$}} & { $^{~~}$} &  { $^{~~}$}  & { -0.0001$^{~~}$} & {\textbf{  1.01$^{**}$}} & { 0.0000$^{~~}$} & {\textbf{  0.78$^{**}$}} & { -0.0001$^{~~}$} & {\textbf{  0.90$^{**}$}}  \\
    &  26 & { 0.0003$^{~~}$} & {\textbf{  0.98$^{**}$}} & { 0.0002$^{~~}$} &  {\textbf{ -0.24$^{**}$}} & { 0.0004$^{~~}$} &  {\textbf{ -0.37$^{**}$}}& {\textbf{ 0.0004$^{**}$}} &  {\textbf{  0.80$^{**}$}} & { $^{~~}$} &  { $^{~~}$}  & { -0.0000$^{~~}$} & {\textbf{  1.01$^{**}$}} & { -0.0001$^{~~}$} & {\textbf{  0.80$^{**}$}} & { -0.0001$^{~~}$} & {\textbf{  0.90$^{**}$}}  \\
    &  52 & {\textbf{ 0.0003$^{*~}$}} & {\textbf{  0.98$^{**}$}} & { 0.0001$^{~~}$} &  {\textbf{ -0.21$^{**}$}} & { 0.0001$^{~~}$} &  {\textbf{ -0.36$^{**}$}}& {\textbf{ 0.0004$^{**}$}} &  {\textbf{  0.80$^{**}$}} & { $^{~~}$} &  { $^{~~}$}  & { 0.0000$^{~~}$} & {\textbf{  1.01$^{**}$}} & { -0.0000$^{~~}$} & {\textbf{  0.80$^{**}$}} & { -0.0001$^{~~}$} & {\textbf{  0.90$^{**}$}}
        \smallskip\\
   IVaR5 &  1 & {\textbf{ 0.0010$^{**}$}} & {\textbf{  0.94$^{**}$}} & {\textbf{ 0.0015$^{**}$}} &  {\textbf{ -0.21$^{**}$}} & {\textbf{ 0.0019$^{**}$}} &  {\textbf{ -0.45$^{**}$}}& { 0.0001$^{~~}$} &  {\textbf{  0.77$^{**}$}} & {\textbf{ 0.0006$^{**}$}} &  {\textbf{  0.90$^{**}$}}  & { $^{~~}$} & { $^{~~}$} & {\textbf{ 0.0007$^{**}$}} & {\textbf{  0.66$^{**}$}} & {\textbf{ 0.0005$^{*~}$}} & {\textbf{  0.80$^{**}$}}  \\
    &  2 & {\textbf{ 0.0009$^{**}$}} & {\textbf{  0.95$^{**}$}} & {\textbf{ 0.0014$^{**}$}} &  {\textbf{ -0.19$^{**}$}} & {\textbf{ 0.0018$^{**}$}} &  {\textbf{ -0.45$^{**}$}}& { 0.0004$^{~~}$} &  {\textbf{  0.78$^{**}$}} & {\textbf{ 0.0004$^{*~}$}} &  {\textbf{  0.90$^{**}$}}  & { $^{~~}$} & { $^{~~}$} & {\textbf{ 0.0006$^{*~}$}} & {\textbf{  0.66$^{**}$}} & { 0.0003$^{~~}$} & {\textbf{  0.80$^{**}$}}  \\
    &  3 & {\textbf{ 0.0008$^{**}$}} & {\textbf{  0.96$^{**}$}} & {\textbf{ 0.0014$^{**}$}} &  {\textbf{ -0.19$^{**}$}} & {\textbf{ 0.0016$^{**}$}} &  {\textbf{ -0.44$^{**}$}}& {\textbf{ 0.0005$^{*~}$}} &  {\textbf{  0.79$^{**}$}} & { 0.0003$^{~~}$} &  {\textbf{  0.91$^{**}$}}  & { $^{~~}$} & { $^{~~}$} & { 0.0005$^{~~}$} & {\textbf{  0.67$^{**}$}} & { 0.0002$^{~~}$} & {\textbf{  0.81$^{**}$}}  \\
    &  4 & {\textbf{ 0.0007$^{**}$}} & {\textbf{  0.96$^{**}$}} & {\textbf{ 0.0012$^{**}$}} &  {\textbf{ -0.19$^{**}$}} & {\textbf{ 0.0015$^{**}$}} &  {\textbf{ -0.44$^{**}$}}& {\textbf{ 0.0006$^{*~}$}} &  {\textbf{  0.79$^{**}$}} & { 0.0003$^{~~}$} &  {\textbf{  0.91$^{**}$}}  & { $^{~~}$} & { $^{~~}$} & { 0.0004$^{~~}$} & {\textbf{  0.67$^{**}$}} & { 0.0002$^{~~}$} & {\textbf{  0.81$^{**}$}}  \\
    &  8 & {\textbf{ 0.0006$^{**}$}} & {\textbf{  0.97$^{**}$}} & {\textbf{ 0.0009$^{**}$}} &  {\textbf{ -0.19$^{**}$}} & {\textbf{ 0.0011$^{**}$}} &  {\textbf{ -0.44$^{**}$}}& {\textbf{ 0.0006$^{**}$}} &  {\textbf{  0.79$^{**}$}} & { 0.0002$^{~~}$} &  {\textbf{  0.91$^{**}$}}  & { $^{~~}$} & { $^{~~}$} & { 0.0003$^{~~}$} & {\textbf{  0.67$^{**}$}} & { 0.0001$^{~~}$} & {\textbf{  0.81$^{**}$}}  \\
    &  13 & {\textbf{ 0.0004$^{**}$}} & {\textbf{  0.96$^{**}$}} & {\textbf{ 0.0006$^{*~}$}} &  {\textbf{ -0.20$^{**}$}} & {\textbf{ 0.0008$^{**}$}} &  {\textbf{ -0.43$^{**}$}}& {\textbf{ 0.0005$^{**}$}} &  {\textbf{  0.79$^{**}$}} & { 0.0002$^{~~}$} &  {\textbf{  0.92$^{**}$}}  & { $^{~~}$} & { $^{~~}$} & { 0.0002$^{~~}$} & {\textbf{  0.69$^{**}$}} & { 0.0001$^{~~}$} & {\textbf{  0.82$^{**}$}}  \\
    &  26 & {\textbf{ 0.0003$^{*~}$}} & {\textbf{  0.96$^{**}$}} & { 0.0001$^{~~}$} &  {\textbf{ -0.20$^{**}$}} & {\textbf{ 0.0004$^{*~}$}} &  {\textbf{ -0.41$^{**}$}}& {\textbf{ 0.0005$^{**}$}} &  {\textbf{  0.78$^{**}$}} & { 0.0001$^{~~}$} &  {\textbf{  0.93$^{**}$}}  & { $^{~~}$} & { $^{~~}$} & { 0.0001$^{~~}$} & {\textbf{  0.71$^{**}$}} & { 0.0000$^{~~}$} & {\textbf{  0.83$^{**}$}}  \\
    &  52 & {\textbf{ 0.0003$^{*~}$}} & {\textbf{  0.96$^{**}$}} & { 0.0001$^{~~}$} &  {\textbf{ -0.17$^{**}$}} & { 0.0002$^{~~}$} &  {\textbf{ -0.39$^{**}$}}& {\textbf{ 0.0004$^{**}$}} &  {\textbf{  0.78$^{**}$}} & { 0.0000$^{~~}$} &  {\textbf{  0.93$^{**}$}}  & { $^{~~}$} & { $^{~~}$} & { 0.0000$^{~~}$} & {\textbf{  0.72$^{**}$}} & { -0.0000$^{~~}$} & {\textbf{  0.83$^{**}$}}
        \smallskip\\
   IES1 &  1 & {\textbf{ 0.0009$^{**}$}} & {\textbf{  1.04$^{**}$}} & {\textbf{ 0.0018$^{**}$}} &  {\textbf{ -0.33$^{**}$}} & {\textbf{ 0.0015$^{**}$}} &  {\textbf{ -0.31$^{**}$}}& { 0.0000$^{~~}$} &  {\textbf{  0.86$^{**}$}} & { 0.0002$^{~~}$} &  {\textbf{  1.11$^{**}$}}  & { 0.0000$^{~~}$} & {\textbf{  1.04$^{**}$}} & { $^{~~}$} & { $^{~~}$} & { 0.0001$^{~~}$} & {\textbf{  1.02$^{**}$}}  \\
    &  2 & {\textbf{ 0.0009$^{**}$}} & {\textbf{  1.05$^{**}$}} & {\textbf{ 0.0016$^{**}$}} &  {\textbf{ -0.32$^{**}$}} & {\textbf{ 0.0014$^{**}$}} &  {\textbf{ -0.32$^{**}$}}& { 0.0004$^{~~}$} &  {\textbf{  0.87$^{**}$}} & { 0.0002$^{~~}$} &  {\textbf{  1.11$^{**}$}}  & { 0.0001$^{~~}$} & {\textbf{  1.05$^{**}$}} & { $^{~~}$} & { $^{~~}$} & { 0.0001$^{~~}$} & {\textbf{  1.03$^{**}$}}  \\
    &  3 & {\textbf{ 0.0008$^{**}$}} & {\textbf{  1.07$^{**}$}} & {\textbf{ 0.0015$^{**}$}} &  {\textbf{ -0.31$^{**}$}} & {\textbf{ 0.0012$^{**}$}} &  {\textbf{ -0.31$^{**}$}}& { 0.0005$^{~~}$} &  {\textbf{  0.89$^{**}$}} & { 0.0002$^{~~}$} &  {\textbf{  1.12$^{**}$}}  & { 0.0002$^{~~}$} & {\textbf{  1.06$^{**}$}} & { $^{~~}$} & { $^{~~}$} & { 0.0000$^{~~}$} & {\textbf{  1.03$^{**}$}}  \\
    &  4 & {\textbf{ 0.0008$^{**}$}} & {\textbf{  1.08$^{**}$}} & {\textbf{ 0.0014$^{**}$}} &  {\textbf{ -0.32$^{**}$}} & {\textbf{ 0.0011$^{**}$}} &  {\textbf{ -0.32$^{**}$}}& {\textbf{ 0.0006$^{*~}$}} &  {\textbf{  0.89$^{**}$}} & { 0.0002$^{~~}$} &  {\textbf{  1.13$^{**}$}}  & { 0.0002$^{~~}$} & {\textbf{  1.07$^{**}$}} & { $^{~~}$} & { $^{~~}$} & { -0.0000$^{~~}$} & {\textbf{  1.04$^{**}$}}  \\
    &  8 & {\textbf{ 0.0006$^{*~}$}} & {\textbf{  1.10$^{**}$}} & {\textbf{ 0.0010$^{**}$}} &  {\textbf{ -0.33$^{**}$}} & {\textbf{ 0.0008$^{*~}$}} &  {\textbf{ -0.33$^{**}$}}& {\textbf{ 0.0007$^{*~}$}} &  {\textbf{  0.90$^{**}$}} & { 0.0001$^{~~}$} &  {\textbf{  1.14$^{**}$}}  & { 0.0001$^{~~}$} & {\textbf{  1.10$^{**}$}} & { $^{~~}$} & { $^{~~}$} & { -0.0001$^{~~}$} & {\textbf{  1.05$^{**}$}}  \\
    &  13 & {\textbf{ 0.0006$^{*~}$}} & {\textbf{  1.09$^{**}$}} & {\textbf{ 0.0007$^{**}$}} &  {\textbf{ -0.34$^{**}$}} & { 0.0005$^{~~}$} &  {\textbf{ -0.33$^{**}$}}& {\textbf{ 0.0006$^{**}$}} &  {\textbf{  0.89$^{**}$}} & { 0.0002$^{~~}$} &  {\textbf{  1.13$^{**}$}}  & { 0.0002$^{~~}$} & {\textbf{  1.10$^{**}$}} & { $^{~~}$} & { $^{~~}$} & { 0.0000$^{~~}$} & {\textbf{  1.05$^{**}$}}  \\
    &  26 & {\textbf{ 0.0005$^{*~}$}} & {\textbf{  1.08$^{**}$}} & { 0.0002$^{~~}$} &  {\textbf{ -0.33$^{**}$}} & { 0.0002$^{~~}$} &  {\textbf{ -0.33$^{**}$}}& {\textbf{ 0.0006$^{**}$}} &  {\textbf{  0.89$^{**}$}} & { 0.0002$^{~~}$} &  {\textbf{  1.13$^{**}$}}  & { 0.0002$^{~~}$} & {\textbf{  1.10$^{**}$}} & { $^{~~}$} & { $^{~~}$} & { 0.0001$^{~~}$} & {\textbf{  1.04$^{**}$}}  \\
    &  52 & {\textbf{ 0.0004$^{*~}$}} & {\textbf{  1.09$^{**}$}} & { 0.0002$^{~~}$} &  {\textbf{ -0.30$^{**}$}} & { -0.0000$^{~~}$} &  {\textbf{ -0.33$^{**}$}}& {\textbf{ 0.0006$^{**}$}} &  {\textbf{  0.89$^{**}$}} & { 0.0001$^{~~}$} &  {\textbf{  1.14$^{**}$}}  & { 0.0002$^{~~}$} & {\textbf{  1.11$^{**}$}} & { $^{~~}$} & { $^{~~}$} & { 0.0000$^{~~}$} & {\textbf{  1.04$^{**}$}}
        \smallskip\\
   IVaR1 &  1 & {\textbf{ 0.0008$^{**}$}} & {\textbf{  1.00$^{**}$}} & {\textbf{ 0.0017$^{**}$}} &  {\textbf{ -0.29$^{**}$}} & {\textbf{ 0.0018$^{**}$}} &  {\textbf{ -0.38$^{**}$}}& { -0.0001$^{~~}$} &  {\textbf{  0.83$^{**}$}} & { 0.0002$^{~~}$} &  {\textbf{  1.04$^{**}$}}  & { -0.0001$^{~~}$} & {\textbf{  1.01$^{**}$}} & { 0.0003$^{~~}$} & {\textbf{  0.82$^{**}$}} & { $^{~~}$} & { $^{~~}$}  \\
    &  2 & {\textbf{ 0.0008$^{**}$}} & {\textbf{  1.01$^{**}$}} & {\textbf{ 0.0016$^{**}$}} &  {\textbf{ -0.28$^{**}$}} & {\textbf{ 0.0016$^{**}$}} &  {\textbf{ -0.38$^{**}$}}& { 0.0003$^{~~}$} &  {\textbf{  0.84$^{**}$}} & { 0.0002$^{~~}$} &  {\textbf{  1.05$^{**}$}}  & { 0.0001$^{~~}$} & {\textbf{  1.02$^{**}$}} & { 0.0003$^{~~}$} & {\textbf{  0.83$^{**}$}} & { $^{~~}$} & { $^{~~}$}  \\
    &  3 & {\textbf{ 0.0008$^{**}$}} & {\textbf{  1.02$^{**}$}} & {\textbf{ 0.0015$^{**}$}} &  {\textbf{ -0.27$^{**}$}} & {\textbf{ 0.0014$^{**}$}} &  {\textbf{ -0.37$^{**}$}}& { 0.0005$^{~~}$} &  {\textbf{  0.85$^{**}$}} & { 0.0002$^{~~}$} &  {\textbf{  1.05$^{**}$}}  & { 0.0001$^{~~}$} & {\textbf{  1.03$^{**}$}} & { 0.0003$^{~~}$} & {\textbf{  0.83$^{**}$}} & { $^{~~}$} & { $^{~~}$}  \\
    &  4 & {\textbf{ 0.0008$^{**}$}} & {\textbf{  1.03$^{**}$}} & {\textbf{ 0.0013$^{**}$}} &  {\textbf{ -0.27$^{**}$}} & {\textbf{ 0.0013$^{**}$}} &  {\textbf{ -0.37$^{**}$}}& {\textbf{ 0.0006$^{*~}$}} &  {\textbf{  0.85$^{**}$}} & {\textbf{ 0.0003$^{*~}$}} &  {\textbf{  1.05$^{**}$}}  & { 0.0002$^{~~}$} & {\textbf{  1.03$^{**}$}} & { 0.0003$^{~~}$} & {\textbf{  0.83$^{**}$}} & { $^{~~}$} & { $^{~~}$}  \\
    &  8 & {\textbf{ 0.0007$^{**}$}} & {\textbf{  1.04$^{**}$}} & {\textbf{ 0.0010$^{**}$}} &  {\textbf{ -0.28$^{**}$}} & {\textbf{ 0.0008$^{**}$}} &  {\textbf{ -0.38$^{**}$}}& {\textbf{ 0.0007$^{**}$}} &  {\textbf{  0.85$^{**}$}} & {\textbf{ 0.0003$^{*~}$}} &  {\textbf{  1.05$^{**}$}}  & { 0.0002$^{~~}$} & {\textbf{  1.05$^{**}$}} & { 0.0003$^{~~}$} & {\textbf{  0.83$^{**}$}} & { $^{~~}$} & { $^{~~}$}  \\
    &  13 & {\textbf{ 0.0006$^{**}$}} & {\textbf{  1.04$^{**}$}} & {\textbf{ 0.0007$^{**}$}} &  {\textbf{ -0.29$^{**}$}} & {\textbf{ 0.0006$^{*~}$}} &  {\textbf{ -0.38$^{**}$}}& {\textbf{ 0.0006$^{**}$}} &  {\textbf{  0.85$^{**}$}} & {\textbf{ 0.0002$^{*~}$}} &  {\textbf{  1.06$^{**}$}}  & { 0.0002$^{~~}$} & {\textbf{  1.06$^{**}$}} & { 0.0002$^{~~}$} & {\textbf{  0.85$^{**}$}} & { $^{~~}$} & { $^{~~}$}  \\
    &  26 & {\textbf{ 0.0004$^{*~}$}} & {\textbf{  1.03$^{**}$}} & { 0.0002$^{~~}$} &  {\textbf{ -0.28$^{**}$}} & { 0.0003$^{~~}$} &  {\textbf{ -0.38$^{**}$}}& {\textbf{ 0.0005$^{**}$}} &  {\textbf{  0.85$^{**}$}} & { 0.0001$^{~~}$} &  {\textbf{  1.07$^{**}$}}  & { 0.0001$^{~~}$} & {\textbf{  1.07$^{**}$}} & { 0.0000$^{~~}$} & {\textbf{  0.87$^{**}$}} & { $^{~~}$} & { $^{~~}$}  \\
    &  52 & {\textbf{ 0.0004$^{*~}$}} & {\textbf{  1.04$^{**}$}} & { 0.0001$^{~~}$} &  {\textbf{ -0.26$^{**}$}} & { 0.0001$^{~~}$} &  {\textbf{ -0.37$^{**}$}}& {\textbf{ 0.0005$^{**}$}} &  {\textbf{  0.85$^{**}$}} & { 0.0001$^{~~}$} &  {\textbf{  1.07$^{**}$}}  & { 0.0002$^{~~}$} & {\textbf{  1.08$^{**}$}} & { 0.0000$^{~~}$} & {\textbf{  0.88$^{**}$}} & { $^{~~}$} & { $^{~~}$}  \\
   \hline\hline
   \end{tabular}
   \label{TB:Risk:F5:spanning:p2}
\end{table}
\end{landscape}

\begin{table}[!ht]
\small
\caption{The performance of long sides of risk-adjusted IMOM portfolios with indirect adjustment process. This table reports the results associated with the bivariate portfolio analysis, including average weekly raw returns (Raw), FF5F-$\alpha$ values ($\alpha$), annualized Sharpe ratios (SR) and maximum drawdowns (MD). The sample period is January 1997 to December 2017. At the beginning of each week, the idiosyncratic return and its risk metrics can be calculated using idiosyncratic returns over past 26 weeks and 130 trading days for individual stocks, by which the stocks can be separately sorted into decile groups according to cumulatively idiosyncratic return and specific risk metric, respectively. The long sides of the risk-adjusted IMOM portfolios are constructed by buying the stocks from intersected decile groups with lowest risk and highest cumulatively idiosyncratic returns. Portfolios would be held for $K$ weeks, and calendar-time method is applied to obtain the average weekly return. \cite{Newey-West-1987-Em}'s $t$-statistics are obtained and the superscripts * and ** denote the significance at 5\% and 1\% levels, respectively. }
\centering
\vspace{-3mm}
   \begin{tabular}{ccccccccc}
   \hline\hline
      & $K=1$ & $2$ & $3$ & $4$ &  $8$ & $13$ & $26$ & $52$ \\
   \hline
 \multicolumn{9}{l}{Panel A: Idiosyncratic volatility} \\
   Raw &  {\textbf{ 0.0033$^{*~}$}} & {\textbf{ 0.0033$^{*~}$}}  & {\textbf{ 0.0032$^{*~}$}} & {\textbf{ 0.0033$^{*~}$}} & {\textbf{ 0.0032$^{*~}$}} & {\textbf{ 0.0032$^{*~}$}} & {\textbf{ 0.0031$^{*~}$}} & {\textbf{ 0.0032$^{*~}$}}  \\
   $\alpha$ &  { 0.0006$^{~~}$} & { 0.0007$^{~~}$}  & {\textbf{ 0.0009$^{*~}$}} & {\textbf{ 0.0010$^{*~}$}} & {\textbf{ 0.0013$^{*~}$}} & {\textbf{ 0.0017$^{**}$}} & {\textbf{ 0.0021$^{**}$}} & {\textbf{ 0.0025$^{**}$}}  \\
   SR &  { 0.5406$^{~~}$} & { 0.5471$^{~~}$}  & { 0.5526$^{~~}$} & { 0.5626$^{~~}$} & { 0.5622$^{~~}$} & { 0.5921$^{~~}$} & { 0.6051$^{~~}$} & { 0.6824$^{~~}$}  \\
   MD &  { 0.6488$^{~~}$} & { 0.6497$^{~~}$}  & { 0.6409$^{~~}$} & { 0.6382$^{~~}$} & { 0.6401$^{~~}$} & { 0.6384$^{~~}$} & { 0.6322$^{~~}$} & { 0.6094$^{~~}$}  \\
     & & & & & & & & \\
 \multicolumn{9}{l}{Panel B: Idiosyncratic skewness} \\
   Raw &  {\textbf{ 0.0031$^{*~}$}} & {\textbf{ 0.0028$^{*~}$}}  & {\textbf{ 0.0029$^{*~}$}} & {\textbf{ 0.0030$^{*~}$}} & {\textbf{ 0.0031$^{*~}$}} & {\textbf{ 0.0030$^{*~}$}} & {\textbf{ 0.0027$^{*~}$}} & {\textbf{ 0.0029$^{*~}$}}  \\
   $\alpha$ &  { -0.0002$^{~~}$} & { 0.0000$^{~~}$}  & { 0.0003$^{~~}$} & { 0.0005$^{~~}$} & { 0.0008$^{~~}$} & {\textbf{ 0.0013$^{*~}$}} & {\textbf{ 0.0016$^{*~}$}} & {\textbf{ 0.0021$^{*~}$}}  \\
   SR &  { 0.4835$^{~~}$} & { 0.4315$^{~~}$}  & { 0.4552$^{~~}$} & { 0.4806$^{~~}$} & { 0.5002$^{~~}$} & { 0.5242$^{~~}$} & { 0.4978$^{~~}$} & { 0.6097$^{~~}$}  \\
   MD &  { 0.6693$^{~~}$} & { 0.6896$^{~~}$}  & { 0.6938$^{~~}$} & { 0.6814$^{~~}$} & { 0.6743$^{~~}$} & { 0.6720$^{~~}$} & { 0.6858$^{~~}$} & { 0.6423$^{~~}$}  \\
     & & & & & & & & \\
 \multicolumn{9}{l}{Panel C: Idiosyncratic maximum drawdown} \\
   Raw &  {\textbf{ 0.0032$^{*~}$}} & {\textbf{ 0.0032$^{*~}$}}  & {\textbf{ 0.0032$^{*~}$}} & {\textbf{ 0.0032$^{*~}$}} & {\textbf{ 0.0030$^{*~}$}} & {\textbf{ 0.0031$^{*~}$}} & {\textbf{ 0.0030$^{*~}$}} & {\textbf{ 0.0031$^{*~}$}}  \\
   $\alpha$ &  { 0.0005$^{~~}$} & { 0.0007$^{~~}$}  & { 0.0008$^{~~}$} & {\textbf{ 0.0009$^{*~}$}} & {\textbf{ 0.0012$^{*~}$}} & {\textbf{ 0.0016$^{**}$}} & {\textbf{ 0.0020$^{**}$}} & {\textbf{ 0.0024$^{**}$}}  \\
   SR &  { 0.5279$^{~~}$} & { 0.5346$^{~~}$}  & { 0.5402$^{~~}$} & { 0.5502$^{~~}$} & { 0.5368$^{~~}$} & { 0.5668$^{~~}$} & { 0.5845$^{~~}$} & { 0.6699$^{~~}$}  \\
   MD &  { 0.6497$^{~~}$} & { 0.6491$^{~~}$}  & { 0.6430$^{~~}$} & { 0.6373$^{~~}$} & { 0.6396$^{~~}$} & { 0.6346$^{~~}$} & { 0.6288$^{~~}$} & { 0.6069$^{~~}$}  \\
     & & & & & & & & \\
 \multicolumn{9}{l}{Panel D: Idiosyncratic ES (5\%)} \\
   Raw &  {\textbf{ 0.0032$^{*~}$}} & {\textbf{ 0.0031$^{*~}$}}  & {\textbf{ 0.0031$^{*~}$}} & {\textbf{ 0.0031$^{*~}$}} & {\textbf{ 0.0030$^{*~}$}} & {\textbf{ 0.0032$^{*~}$}} & {\textbf{ 0.0031$^{*~}$}} & {\textbf{ 0.0032$^{*~}$}}  \\
   $\alpha$ &  { 0.0007$^{~~}$} & { 0.0007$^{~~}$}  & {\textbf{ 0.0009$^{*~}$}} & {\textbf{ 0.0010$^{*~}$}} & {\textbf{ 0.0012$^{*~}$}} & {\textbf{ 0.0017$^{**}$}} & {\textbf{ 0.0021$^{**}$}} & {\textbf{ 0.0025$^{**}$}}  \\
   SR &  { 0.5337$^{~~}$} & { 0.5254$^{~~}$}  & { 0.5333$^{~~}$} & { 0.5417$^{~~}$} & { 0.5352$^{~~}$} & { 0.5754$^{~~}$} & { 0.5987$^{~~}$} & { 0.6771$^{~~}$}  \\
   MD &  { 0.6642$^{~~}$} & { 0.6594$^{~~}$}  & { 0.6585$^{~~}$} & { 0.6458$^{~~}$} & { 0.6381$^{~~}$} & { 0.6338$^{~~}$} & { 0.6324$^{~~}$} & { 0.6089$^{~~}$}  \\
     & & & & & & & & \\
 \multicolumn{9}{l}{Panel E: Idiosyncratic VaR (5\%)} \\
   Raw &  {\textbf{ 0.0032$^{*~}$}} & {\textbf{ 0.0032$^{*~}$}}  & {\textbf{ 0.0031$^{*~}$}} & {\textbf{ 0.0031$^{*~}$}} & {\textbf{ 0.0031$^{*~}$}} & {\textbf{ 0.0032$^{*~}$}} & {\textbf{ 0.0031$^{*~}$}} & {\textbf{ 0.0031$^{*~}$}}  \\
   $\alpha$ &  { 0.0006$^{~~}$} & { 0.0007$^{~~}$}  & { 0.0008$^{~~}$} & {\textbf{ 0.0009$^{*~}$}} & {\textbf{ 0.0012$^{*~}$}} & {\textbf{ 0.0017$^{**}$}} & {\textbf{ 0.0021$^{**}$}} & {\textbf{ 0.0024$^{**}$}}  \\
   SR &  { 0.5360$^{~~}$} & { 0.5331$^{~~}$}  & { 0.5331$^{~~}$} & { 0.5406$^{~~}$} & { 0.5437$^{~~}$} & { 0.5821$^{~~}$} & { 0.5956$^{~~}$} & { 0.6756$^{~~}$}  \\
   MD &  { 0.6607$^{~~}$} & { 0.6495$^{~~}$}  & { 0.6434$^{~~}$} & { 0.6383$^{~~}$} & { 0.6419$^{~~}$} & { 0.6373$^{~~}$} & { 0.6335$^{~~}$} & { 0.6092$^{~~}$}  \\
     & & & & & & & & \\
 \multicolumn{9}{l}{Panel F: Idiosyncratic ES (1\%)} \\
   Raw &  {\textbf{ 0.0032$^{*~}$}} & {\textbf{ 0.0032$^{*~}$}}  & {\textbf{ 0.0032$^{*~}$}} & {\textbf{ 0.0032$^{*~}$}} & {\textbf{ 0.0031$^{*~}$}} & {\textbf{ 0.0032$^{*~}$}} & {\textbf{ 0.0031$^{*~}$}} & {\textbf{ 0.0031$^{*~}$}}  \\
   $\alpha$ &  { 0.0007$^{~~}$} & {\textbf{ 0.0008$^{*~}$}}  & {\textbf{ 0.0010$^{*~}$}} & {\textbf{ 0.0012$^{*~}$}} & {\textbf{ 0.0014$^{**}$}} & {\textbf{ 0.0018$^{**}$}} & {\textbf{ 0.0022$^{**}$}} & {\textbf{ 0.0025$^{**}$}}  \\
   SR &  { 0.5257$^{~~}$} & { 0.5347$^{~~}$}  & { 0.5475$^{~~}$} & { 0.5637$^{~~}$} & { 0.5547$^{~~}$} & { 0.5872$^{~~}$} & { 0.6010$^{~~}$} & { 0.6714$^{~~}$}  \\
   MD &  { 0.6592$^{~~}$} & { 0.6585$^{~~}$}  & { 0.6469$^{~~}$} & { 0.6416$^{~~}$} & { 0.6459$^{~~}$} & { 0.6419$^{~~}$} & { 0.6402$^{~~}$} & { 0.6191$^{~~}$}  \\
     & & & & & & & & \\
 \multicolumn{9}{l}{Panel G: Idiosyncratic VaR (1\%)} \\
   Raw &  {\textbf{ 0.0032$^{*~}$}} & {\textbf{ 0.0031$^{*~}$}}  & {\textbf{ 0.0031$^{*~}$}} & {\textbf{ 0.0032$^{*~}$}} & {\textbf{ 0.0031$^{*~}$}} & {\textbf{ 0.0032$^{*~}$}} & {\textbf{ 0.0031$^{*~}$}} & {\textbf{ 0.0031$^{*~}$}}  \\
   $\alpha$ &  { 0.0007$^{~~}$} & { 0.0007$^{~~}$}  & {\textbf{ 0.0009$^{*~}$}} & {\textbf{ 0.0011$^{*~}$}} & {\textbf{ 0.0013$^{*~}$}} & {\textbf{ 0.0018$^{**}$}} & {\textbf{ 0.0022$^{**}$}} & {\textbf{ 0.0025$^{**}$}}  \\
   SR &  { 0.5310$^{~~}$} & { 0.5237$^{~~}$}  & { 0.5352$^{~~}$} & { 0.5552$^{~~}$} & { 0.5497$^{~~}$} & { 0.5853$^{~~}$} & { 0.6022$^{~~}$} & { 0.6757$^{~~}$}  \\
   MD &  { 0.6665$^{~~}$} & { 0.6619$^{~~}$}  & { 0.6596$^{~~}$} & { 0.6489$^{~~}$} & { 0.6430$^{~~}$} & { 0.6294$^{~~}$} & { 0.6299$^{~~}$} & { 0.6058$^{~~}$}  \\
   \hline\hline
   \end{tabular}
   \label{TB:Res2Risk:F5:id:long}
\end{table}

\begin{table}[!ht]
\small
\caption{The performance of short sides of risk-adjusted IMOM portfolios with indirect adjustment process. This table reports the results associated with the bivariate portfolio analysis, including average weekly raw returns (Raw), FF5F-$\alpha$s ($\alpha$), annualized Sharpe ratios (SR) and maximum drawdowns (MD). The sample period is January 1997 to December 2017. At the beginning of each week, the idiosyncratic return and its risk metrics can be calculated using idiosyncratic returns over past 26 weeks and 130 trading days for individual stocks, by which the stocks can be separately sorted into decile groups according to cumulatively idiosyncratic return and specific risk metric, respectively. The short sides of the risk-adjusted IMOM portfolios are constructed by buying the stocks from the intersected decile groups with highest risk and lowest cumulatively idiosyncratic returns. Portfolios would be held for $K$ weeks, and calendar-time method is applied to obtain the average weekly return. \cite{Newey-West-1987-Em}'s $t$-statistics are obtained and the superscripts * and ** denote the significance at 5\% and 1\% levels, respectively. }
\centering
\vspace{-3mm}
   \begin{tabular}{ccccccccc}
   \hline\hline
      & $K=1$ & $2$ & $3$ & $4$ &  $8$ & $13$ & $26$ & $52$ \\
   \hline
 \multicolumn{9}{l}{Panel A: Idiosyncratic volatility} \\
   Raw &  { 0.0009$^{~~}$} & { 0.0010$^{~~}$}  & { 0.0012$^{~~}$} & { 0.0013$^{~~}$} & { 0.0014$^{~~}$} & { 0.0015$^{~~}$} & { 0.0016$^{~~}$} & { 0.0019$^{~~}$}  \\
   $\alpha$ &  {\textbf{ -0.0028$^{**}$}} & {\textbf{ -0.0024$^{**}$}}  & {\textbf{ -0.0020$^{**}$}} & {\textbf{ -0.0018$^{**}$}} & { -0.0011$^{~~}$} & { -0.0005$^{~~}$} & { 0.0004$^{~~}$} & { 0.0010$^{~~}$}  \\
   SR &  { 0.0467$^{~~}$} & { 0.0745$^{~~}$}  & { 0.0949$^{~~}$} & { 0.1161$^{~~}$} & { 0.1374$^{~~}$} & { 0.1691$^{~~}$} & { 0.2067$^{~~}$} & { 0.2990$^{~~}$}  \\
   MD &  { 0.8517$^{~~}$} & { 0.8427$^{~~}$}  & { 0.8327$^{~~}$} & { 0.8179$^{~~}$} & { 0.7910$^{~~}$} & { 0.7716$^{~~}$} & { 0.7568$^{~~}$} & { 0.7375$^{~~}$}  \\
     & & & & & & & & \\
 \multicolumn{9}{l}{Panel B: Idiosyncratic skewness} \\
   Raw &  { 0.0004$^{~~}$} & { -0.0003$^{~~}$}  & { -0.0012$^{~~}$} & { -0.0004$^{~~}$} & { 0.0011$^{~~}$} & { 0.0021$^{~~}$} & { 0.0019$^{~~}$} & { 0.0023$^{~~}$}  \\
   $\alpha$ &  {\textbf{ -0.0037$^{**}$}} & {\textbf{ -0.0038$^{**}$}}  & {\textbf{ -0.0043$^{**}$}} & {\textbf{ -0.0033$^{**}$}} & { -0.0015$^{~~}$} & { 0.0001$^{~~}$} & { 0.0006$^{~~}$} & { 0.0013$^{~~}$}  \\
   SR &  { -0.0131$^{~~}$} & { -0.1031$^{~~}$}  & { -0.2227$^{~~}$} & { -0.1201$^{~~}$} & { 0.0819$^{~~}$} & { 0.2293$^{~~}$} & { 0.2314$^{~~}$} & { 0.3604$^{~~}$}  \\
   MD &  { 0.8979$^{~~}$} & { 0.9255$^{~~}$}  & { 0.9589$^{~~}$} & { 0.9033$^{~~}$} & { 0.7947$^{~~}$} & { 0.7820$^{~~}$} & { 0.7508$^{~~}$} & { 0.7402$^{~~}$}  \\
     & & & & & & & & \\
 \multicolumn{9}{l}{Panel C: Idiosyncratic maximum drawdown} \\
   Raw &  { 0.0008$^{~~}$} & { 0.0008$^{~~}$}  & { 0.0009$^{~~}$} & { 0.0009$^{~~}$} & { 0.0010$^{~~}$} & { 0.0012$^{~~}$} & { 0.0013$^{~~}$} & { 0.0017$^{~~}$}  \\
   $\alpha$ &  {\textbf{ -0.0028$^{**}$}} & {\textbf{ -0.0025$^{**}$}}  & {\textbf{ -0.0023$^{**}$}} & {\textbf{ -0.0020$^{**}$}} & {\textbf{ -0.0015$^{*~}$}} & { -0.0008$^{~~}$} & { 0.0001$^{~~}$} & { 0.0008$^{~~}$}  \\
   SR &  { 0.0386$^{~~}$} & { 0.0420$^{~~}$}  & { 0.0521$^{~~}$} & { 0.0621$^{~~}$} & { 0.0749$^{~~}$} & { 0.1057$^{~~}$} & { 0.1432$^{~~}$} & { 0.2438$^{~~}$}  \\
   MD &  { 0.8619$^{~~}$} & { 0.8532$^{~~}$}  & { 0.8415$^{~~}$} & { 0.8297$^{~~}$} & { 0.7989$^{~~}$} & { 0.7763$^{~~}$} & { 0.7680$^{~~}$} & { 0.7463$^{~~}$}  \\
     & & & & & & & & \\
 \multicolumn{9}{l}{Panel D: Idiosyncratic ES (5\%)} \\
   Raw &  { 0.0009$^{~~}$} & { 0.0011$^{~~}$}  & { 0.0012$^{~~}$} & { 0.0013$^{~~}$} & { 0.0014$^{~~}$} & { 0.0015$^{~~}$} & { 0.0016$^{~~}$} & { 0.0020$^{~~}$}  \\
   $\alpha$ &  {\textbf{ -0.0027$^{**}$}} & {\textbf{ -0.0024$^{**}$}}  & {\textbf{ -0.0020$^{**}$}} & {\textbf{ -0.0017$^{**}$}} & { -0.0011$^{~~}$} & { -0.0005$^{~~}$} & { 0.0004$^{~~}$} & { 0.0011$^{~~}$}  \\
   SR &  { 0.0552$^{~~}$} & { 0.0806$^{~~}$}  & { 0.0995$^{~~}$} & { 0.1183$^{~~}$} & { 0.1373$^{~~}$} & { 0.1681$^{~~}$} & { 0.1987$^{~~}$} & { 0.3064$^{~~}$}  \\
   MD &  { 0.8489$^{~~}$} & { 0.8428$^{~~}$}  & { 0.8335$^{~~}$} & { 0.8216$^{~~}$} & { 0.7972$^{~~}$} & { 0.7812$^{~~}$} & { 0.7700$^{~~}$} & { 0.7433$^{~~}$}  \\
     & & & & & & & & \\
 \multicolumn{9}{l}{Panel E: Idiosyncratic VaR (5\%)} \\
   Raw &  { 0.0009$^{~~}$} & { 0.0009$^{~~}$}  & { 0.0011$^{~~}$} & { 0.0012$^{~~}$} & { 0.0012$^{~~}$} & { 0.0014$^{~~}$} & { 0.0015$^{~~}$} & { 0.0019$^{~~}$}  \\
   $\alpha$ &  {\textbf{ -0.0027$^{**}$}} & {\textbf{ -0.0025$^{**}$}}  & {\textbf{ -0.0022$^{**}$}} & {\textbf{ -0.0019$^{**}$}} & {\textbf{ -0.0013$^{*~}$}} & { -0.0006$^{~~}$} & { 0.0002$^{~~}$} & { 0.0010$^{~~}$}  \\
   SR &  { 0.0480$^{~~}$} & { 0.0589$^{~~}$}  & { 0.0816$^{~~}$} & { 0.0989$^{~~}$} & { 0.1138$^{~~}$} & { 0.1493$^{~~}$} & { 0.1789$^{~~}$} & { 0.2816$^{~~}$}  \\
   MD &  { 0.8462$^{~~}$} & { 0.8407$^{~~}$}  & { 0.8304$^{~~}$} & { 0.8160$^{~~}$} & { 0.7938$^{~~}$} & { 0.7739$^{~~}$} & { 0.7596$^{~~}$} & { 0.7389$^{~~}$}  \\
     & & & & & & & & \\
 \multicolumn{9}{l}{Panel F: Idiosyncratic ES (1\%)} \\
   Raw &  { 0.0007$^{~~}$} & { 0.0010$^{~~}$}  & { 0.0012$^{~~}$} & { 0.0013$^{~~}$} & { 0.0014$^{~~}$} & { 0.0015$^{~~}$} & { 0.0016$^{~~}$} & { 0.0020$^{~~}$}  \\
   $\alpha$ &  {\textbf{ -0.0028$^{**}$}} & {\textbf{ -0.0024$^{**}$}}  & {\textbf{ -0.0020$^{**}$}} & {\textbf{ -0.0017$^{**}$}} & { -0.0011$^{~~}$} & { -0.0004$^{~~}$} & { 0.0004$^{~~}$} & { 0.0011$^{~~}$}  \\
   SR &  { 0.0263$^{~~}$} & { 0.0684$^{~~}$}  & { 0.0953$^{~~}$} & { 0.1151$^{~~}$} & { 0.1330$^{~~}$} & { 0.1628$^{~~}$} & { 0.2037$^{~~}$} & { 0.3016$^{~~}$}  \\
   MD &  { 0.8733$^{~~}$} & { 0.8611$^{~~}$}  & { 0.8490$^{~~}$} & { 0.8377$^{~~}$} & { 0.8151$^{~~}$} & { 0.7970$^{~~}$} & { 0.7826$^{~~}$} & { 0.7537$^{~~}$}  \\
     & & & & & & & & \\
 \multicolumn{9}{l}{Panel G: Idiosyncratic VaR (1\%)} \\
   Raw &  { 0.0008$^{~~}$} & { 0.0010$^{~~}$}  & { 0.0012$^{~~}$} & { 0.0013$^{~~}$} & { 0.0014$^{~~}$} & { 0.0015$^{~~}$} & { 0.0016$^{~~}$} & { 0.0020$^{~~}$}  \\
   $\alpha$ &  {\textbf{ -0.0028$^{**}$}} & {\textbf{ -0.0024$^{**}$}}  & {\textbf{ -0.0020$^{**}$}} & {\textbf{ -0.0016$^{**}$}} & { -0.0010$^{~~}$} & { -0.0004$^{~~}$} & { 0.0004$^{~~}$} & { 0.0011$^{~~}$}  \\
   SR &  { 0.0379$^{~~}$} & { 0.0688$^{~~}$}  & { 0.0982$^{~~}$} & { 0.1226$^{~~}$} & { 0.1416$^{~~}$} & { 0.1658$^{~~}$} & { 0.1958$^{~~}$} & { 0.2995$^{~~}$}  \\
   MD &  { 0.8520$^{~~}$} & { 0.8469$^{~~}$}  & { 0.8374$^{~~}$} & { 0.8244$^{~~}$} & { 0.8010$^{~~}$} & { 0.7846$^{~~}$} & { 0.7719$^{~~}$} & { 0.7466$^{~~}$}  \\
   \hline\hline
   \end{tabular}
   \label{TB:Res2Risk:F5:id:short}
\end{table}

\end{document}